**Functional voxel hierarchy and afferent capacity revealed mental state transition on dynamic correlation resting-state fMRI**

Dong Soo Lee[1,2*], Hyun Joo Kim[3], Youngmin Huh[2], Yeon Koo Kang[2], Wonseok Whi[2], Hyekyoung Lee[2,4], Hyejin Kang[2,4*]

[1]Medical Science and Engineering, School of Convergence Science and Technology, Pohang University of Science and Technology, Pohang, Korea
[2]Nuclear Medicine, College of Medicine, Seoul National University, Seoul, Korea
[3]Nuclear Medicine, Korea University Anam Hospital, Seoul, Korea
[4]Biomedical Research Institute, Seoul National University Hospital, Seoul, Korea

*Corresponding Authors

Dong Soo Lee, M.D., Ph.D.
Medical Science and Engineering, POSTECH, Pohang, Korea dsl9322@postech.ac.kr

Hyejin Kang, Ph.D.
Biomedical Research Institute, SNUH, Korea hkang211@snu.ac.kr

**Abstract**

Voxel hierarchy on dynamic brain graphs is produced by k core percolation on functional dynamic amplitude correlation of resting-state fMRI. Directed graphs and their afferent/efferent capacities are produced by Markov modeling of the universal cover of undirected graphs simultaneously with the calculation of volume entropy. Positive and unsigned negative brain graphs were analyzed separately on sliding-window representation to underpin the visualization and quantitation of mental dynamic states with their transitions. Voxel hierarchy animation maps of positive graphs revealed abrupt changes in coreness k and $k_{max}$core, which we called mental state transitions. Afferent voxel capacities of the positive graphs also revealed transient modules composed of dominating voxels/independent components and their exchanges representing mental state transitions. Animation and quantification plots of voxel hierarchy and afferent capacity corroborated each other in underpinning mental state transitions and afferent module exchange on the positive directed functional connectivity graphs. We propose the use of spatiotemporal trajectories of voxels on positive dynamic graphs to construct hierarchical structures by k core percolation and quantified in- and out-flows of information of voxels by volume entropy/directed graphs to subserve diverse resting mental state transitions on resting-state fMRI graphs in normal human individuals.

Keywords: graph node hierarchy, afferent node capacity, information flow, mental state transition, resting-state fMRI, k core percolation, volume entropy

## Introduction

The spatiotemporal trajectory of neurons is designed to represent population codes for subserving behavior and the underlying mental process in humans and animals. The limited spatial and temporal resolution of measuring devices such as functional magnetic resonance imaging (fMRI) has obstructed the scrutiny of human mental processes, which was furthered by computing complexity. Resting-state fMRI might have been widely used clinically, especially considering its ready availability but not, for assessing mental states' fluctuation and/or, if any, transition in humans (1-5). We suppose that another barrier against ready use was the technological one, in that we could not use voxels but regions of interest or principal/independent components. However, improved brain imaging devices and pre- and postprocessing technology with enhanced computing resources (6-11) now allow voxel-based investigations of the independent activities of voxels, which are currently the smallest macrocomplexes of neurons and allies. Dynamic functional structures of time-varying measures, if any and easy to catch, would have been a great help, although voxels contain an average of 100,000 neurons per 1 $mm^3$ (5,11-13). As one good example, time-varying features of instantaneous amplitude-correlation reduced to a few principal components were successfully correlated with dynamic mental states and even personality traits based on their elucidation (11). We now need to expand these pilot investigations to refined intervoxel studies while respecting the identities of voxels and propose proper measures to represent human mental states and their quantified contributions. The literature on intervoxel amplitude correlation has already revealed an equal prevalence of amplitude correlation and anti-correlation on voxel-based approaches (5,14,15), unlike previous investigations that observed mostly positive region-based correlations (14,16-18). Resting-state fMRI releases the output of two immiscible graphs, equally propense graphs of correlation and anti-correlation, awaiting novel approaches on how voxels compete and collaborate in composing resting mental states in humans in dynamic plots as well as in the static state (5).

The parameters of functional brain graphs had been either related to either graph theory or classical many-body pairwise embedding technology. These methods had setbacks in terms of their implicit assumption that graphs are reducible to principal components (6-11) and that understanding the parameter and its distribution (i.e., degree and degree distribution for the rich-club coefficient) would be good for elucidating the hidden structure of intervoxel interactions (19-21). They are correct in terms of searching for the global characteristics of graphs or networks. However, with their unsung assumption, investigators implicitly ignored the identities of voxels. What if the voxels, not neurons as voxels, are already a macrostructure of neuron-glia-vessel complexes and 1 minute is enough time for making ensembles, calculating and acting independently to release emergent behaviors with higher-order interactions? We attempted to restore the identity of voxels by calculating their characteristic contributions to functional hierarchy (5) and in-degree (afferent to the nodes) information flow capacity in dynamic functional graphs. Another limitation of mainstream brain connectivity investigations is their inability to produce dynamic representations of functional brain graphs (1-5) and/or their ignorance of the possibility to make directed weighted graphs using pairwise undirected observables (such as amplitude correlations) of dynamic functional brain graphs (14,22,23). We recently explored these uncharted methods and introduced the following schemes of investigation for using resting-state fMRI to characterize voxel hierarchy (5) and afferent/efferent node capacity, which are supposed to represent the dynamic functional graphs of mental state fluctuation and transition in humans.

In this study, we assumed that 1) dynamic functional brain graphs derived from resting-state fMRI represent the fluctuating and sometimes transiting brain states of human minds at resting-state (4,24,25); 2) sliding-window time-binning of resting-state fMRI (Suppl. Fig. 1) reveal waxing and waning voxel behavior of gathering/dissociating to reveal continuous (nonexplosive) temporal changes in composition (26); 3) waveforms of each voxel, once observed pairwise, act as the simplest representation of their higher-order interaction (27,28), which might reveal the inherent characteristics of many-body intervoxel interactions (29,30); and 4) the above pairwise-observed amplitude correlations, as observables, are the sum of signals (functional connectomic dynamics) and redundancy (including inherent and measurement-related error/noise) (29,31). The first of the above assumptions are not refutable, meaning that this investigation cannot prove or refute it (32); however, the remaining three are to be corroborated or partially proven for some measures or disproved for others by our study results, which we describe in detail below.

Based on our preliminary works, described in previous studies (5,32), the threshold range of dynamic brain graphs was set to guarantee scale freeness of degree distribution (distribution freeness) while cropping a large number of voxels. This thresholding was later found to be necessary for the disclosure of the state transition, if any, of maximum coreness k voxels but not for discovering the abrupt module exchange of afferent node capacity on dynamic directed weighted graphs. The sliding window method (5,32) was adopted to find intervoxel amplitude correlations separately for positive and unsigned negative graphs to determine intervoxel similarity per time bin for further analysis of 1) the production of coreness k animation maps and corollary plots such as glass brain, flagplots, and timepoint plots with trajectory tracing and 2) afferent and efferent node capacity maps on animation. Upon varying thresholds using an exemplary case and the same threshold for all the time bins of individuals from the Human Connectome Project (HCP) (33-35), the total sum of coreness k voxels or temporal progression of afferent node capacity were compared with graphs' sums of the number of edges, and the threshold effect was determined on the basis of the following findings of this investigation.

In this investigation, using resting-state fMRI data from the HCP (33-35) for 180 selected subjects, we performed k core percolation (32,36,37) for dynamic intervoxel amplitude correlations and volume entropy calculations associated with afferent/efferent node capacity estimation (38,39). We observed whether sliding-window-prepared time-binned coreness k maps and their flagplots and $k_{max}$core representation (5,32) would show either the waxing and waning gathering/dissociating characteristics-only pattern or the abrupt explosive changes in composition and/or varying fractions thereof to participate in composing the maximum k core voxels (a.k.a. $k_{max}$core). We found the latter in almost all individual subjects (5). We looked for evidence that the state transitions, observed with the above voxel hierarchy study, were similarly represented in information-flow maps of in-degree (afferent) and out-degree (efferent) voxel capacities of positive and negative graphs. We discovered that the afferent node capacity of the positive graph showed the exchange of modules with various combinations of IC voxels. We re-examined the propensity of modules of different IC compositions on $k_{max}$core fluctuations (on a stacked histogram on the run) and compared the findings between these $k_{max}$core fluctuations and the exchange of modules containing various IC compositions. Finally, the voxels themselves, as separate entities showing diverse temporal trajectories along time bins on the run, could be traced on the quantified maps on animation as well as on timepoint plots, which enabled identification of each voxel on the temporal axis for their characteristic values of coreness k and afferent node capacity on the positive graphs. Unsigned-negative graphs and efferent node capacity were also examined but did not present significant state transitions.

## Results

### Effect of thresholds on the number of edges, the voxel coreness k and the afferent voxel capacity

For the cohort of 180 individuals from the HCP (with 280 time bins each with sliding window methods (Suppl. Fig. 1)), to crop the giant component of at least 85% of voxels included but having a guaranteed scale-free (distribution-free) voxel degree distribution, thresholds were set to 0.65 for the amplitude correlation for positive graphs and 0.50 for negative graphs. Among these individuals, one exemplary case was selected to evaluate the effect of the total number of edges per time bin related to the preset thresholds, and the thresholds were varied from 0.20 to 0.85 for the intervoxel correlations to construct 14 positive graphs and 14 negative graphs (Fig. 1AB). The number of voxels in the graphs varied from 100% (n=5,937) to 5% (n=280) of the total voxels. Obviously, increasing thresholds decreased the number of voxels and the number of edges. The total number of edges showed an almost 1:1 correlation (1.09 $\pm$ 0.02 for 280 time bins with a threshold of 0.65 (voxel n; 5,622−5,174) in the positive network and 1.08 $\pm$ 0.02 for 280 time bins with a threshold of 0.5 (voxel n; 5,937−5,833) in the negative network), with the total sum of coreness k values per graph in both the positive and the negative graph analyses (Fig. 1C). For 180 subjects, this relationship between the total sum of edges and the total sum of coreness k values was consistent in all individual/time bins despite the use of the same thresholds of 0.65 for all the positive graphs and 0.5 for all the negative graphs.

Based on this equivalence of the total number of edges per time bin in every individual, the voxels' coreness k values were divided by the total number of edges to yield edge number-scaled voxel coreness k values. Thus, edge-scaling represents normalization of coreness values of voxels by total number of edges for the graph for each time bin. Voxel coreness k timepoint plots initially showed great similarity in terms of the contours of collective trajectory bundles per independent component (IC), which was due to the effect of varying the total edge number per time bin (Suppl. Fig. 2AB). Edge-scaled coreness k trajectory bundles showed gross similarity with minute differences between ICs. Scale freeness (an almost linear decrease in the log-log plot of the degree distribution) was observed in the positive graphs with thresholds ranging from 0.55 to 0.85 and in the negative graphs with thresholds ranging from 0.50 to 0.85 (Suppl. Fig. 2CD). Voxel number of graphs let disregard the positive graphs with thresholds ≥ 0.7 and the negative graphs with thresholds ≥ 0.65 as the number of voxels of these graphs were less than 85%. Notably, the voxels in the same IC showed heterogeneity of trajectories, contributing to the shape of collective coreness k per IC, meaning that each of the voxels took turns in the collective rise and fall along the time-bin axis. At a threshold ≥ 0.4, $k_{max}$core plots of positive graphs showed state transitions regardless of edge numbers in terms of the voxel composition of the $k_{max}$core voxels and their IC belongings. On positive graphs with thresholds < 0.4, the state transition disappeared, except for the remaining fluctuations (Suppl. Fig. 2A). In contrast, $k_{max}$core plots of negative graphs with thresholds of 0.35−0.55 rarely showed state transitions but depicted grossly similar fluctuating temporal progress (Suppl. Fig. 2B).

Volume entropy, topological invariant of the brain graph using its equivalent of universal cover, representing the unique characteristics of information flow of the functional graph shape, was not related linearly to the total number of edges per time bin. This was scrutinized in this exemplary case. Once the number of edges decreased with or without loss of remaining voxels due to thresholding, the volume entropy decreased proportionally to the number of edges per brain graph (Suppl. Fig. 3). When the lower limit of the total number of nodes was set to ≥ 1,265 (85% of the total 1,489 voxels), the number of graph voxels did not influence volume entropy. In positive graphs with correlation thresholds ≤ 0.4 (from 0.2 to 0.4), the calculated volume entropy was the same for the graphs, while the total number of edges varied from 78 K (average for 280 time bins for a threshold of 0.5) to 330 K (average for a threshold of 0.2). As the threshold increased from 0.45 to 0.85, the total edge number decreased with the initial curvilinear decrease in volume entropy and then decreased linearly from the threshold of 0.65. In negative graphs with thresholds ≥ -0.4 for anti-correlations (or < 0.4 for unsigned anti-correlations), the volume entropy was the same for the graphs, while the total edge numbers varied from 45 K (for a threshold of 0.4) to 193 K (for a threshold of 0.2).

Notably, the volume entropy was the same at thresholds lower than certain values in positive graphs and in negative graphs, as they had the same skeleton structure of information flow on the brain graphs regardless of the thresholds below these thresholds. Extra surplus edges of the graphs with lower thresholds did not affect volume entropy, the topological invariant representing information flow (Suppl. Fig. 3). In other words, unlike the total number of edge-dependent total coreness k values, the volume entropy of a graph, either positive or negative, represents the total amount of information flow capacity of a graph independent of the extra number of edges therein. This finding was reiterated in the pattern of afferent capacity on the directed graphs, especially positive graphs. The modular shapes of the temporal progress of the afferent capacity of the positive graphs were exactly the same when the threshold was $\leq 0.7$ (i.e., from 0.2 to 0.7), and this was also the case when the threshold was $\leq 0.55$ (from 0.2 to 0.55) in the negative graphs. For the afferent capacity, the MRI-overlaid animation maps on the run showed the same pattern of modules and their exchanges, which was also the case for the timepoint plots of the positive graphs regardless of the thresholds when they were $\leq 0.7$.

With these observations and analyses, as the edge number affected the coreness k values, the coreness k values were corrected by dividing them by their total edge number (edge-scaled coreness k) per time bin. We could compare these edge-scaled coreness k values between intertime bins within individuals, between individuals or between positive and negative graphs. In contrast, volume entropy and afferent/efferent node capacity were not affected by the surplus edges when the thresholds were low enough, and we did not perform edge scaling for volume entropy or afferent/efferent capacity.

Optimal thresholds were determined by 1) coreness k (and their accompanying $k_{max}$core), 2) volume entropy (and afferent/efferent node capacity) and 3) scale-freeness of the degree distribution among various thresholds in addition to the number of voxels (85%) (Suppl. Fig. 3). Using the separate thresholds for positive and negative graphs, we further analyzed the state fluctuation and transition of the $k_{max}$core maps with a glass brain representation to determine the dynamic voxel hierarchy. We also analyzed afferent/efferent node capacity overlaid on MRI slices in 180 HCP subjects and the afferent/efferent capacity of voxels on their timepoint plots. With these preset thresholds, we looked for the state fluctuation/transition that was associated with modules and their exchanges of afferent/efferent capacity of voxels/ICs in positive and negative graphs.

**Coreness k voxel-values represent the hierarchical position of voxels in resting-state brain functional graphs**

The intervoxel correlation of the voxel waveform amplitudes exhibited an almost symmetric distribution in the 17 million-edge histogram. We divided the positive and negative edges to construct positive and negative graphs, respectively, which are mutually exclusive and interdependent. Edges with negative correlation (intervoxel anticorrelation) were made to unsigned values; thus, a negative graph is in fact an absolute-valued negative network. All subsequent analyses were performed for positive and unsigned negative networks (Suppl. Fig. 4).

After thresholding the brain graphs with the appropriate correlation values, the brain graphs were distribution-free (descending mostly linearly on the log-log plot in the voxel degree-prevalence distribution), preserved as many voxels (5,000 (85% of the total)) as possible and had a sparse number of edges (0.5%−10% along all the time bins and individuals from the HCP (n=180). Python codes were used to perform K core percolation as previously described (32,36). Coreness k values were annotated to 5,937 voxels and overlaid on 36 slices of MR trans-axial images (Suppl. Fig. 5D). For 280 time bins of 180 individuals from the HCP project, animation movies for coreness k voxel value maps revealed all k values, which varied in total sum and thus waxed and waned on animated displays. The total sum of the coreness k values per graph was equal to the total number of edges for each time bin (Suppl. Fig. 5A), while highly ranked coreness k-valued voxels were shown in scatters or clusters (Suppl. Fig. 5C). According to the theoretical reasoning and the practical observation of the equivalence of the total number of edges and the total coreness k values per time bin, animation movies were finally made using the edge-scaled coreness k index (voxel coreness k values divided by the total edge number per time bin), as well as animated edge-scaled flagplots and edge-scaled coreness k timepoint plots (Suppl. Fig. 5FG).

The qualitative readout of these image-animation sets revealed fluctuating changes in hierarchical intensity (red represents the highest position on the hierarchy, and blue represents the lowest position), and jumping changed from one cluster to another cluster of hierarchically prominent voxels. The hierarchical dominance of clustered voxels seemed to be sustained for a varying short period and then quickly replaced by other clustered voxels. A stacked histogram for the hierarchically highest voxels on $k_{max}$core plots showed the abrupt and sharp transition of the dominant voxel clusters. We called this abrupt transition of $k_{max}$core voxels the ‘state transition’ of the hierarchically highest voxels (Suppl. Fig. 5E). On the animated glass brain images of lateral (sagittal) and dorsal (transaxial) views, we explored the alternating participation of voxels belonging to the same ICs (Suppl. Fig. 5F). Seven well-characterized ICs and their voxels are colored using rainbow colors. Animated $k_{max}$core voxels/IC composition images, either stacked histograms or glass brains, easily revealed the hierarchical dominance of, for example, visual network (VN)-dominant $k_{max}$core voxels or default mode network/central executive network (DMN/CEN)-dominant $k_{max}$core voxels and their transition from the VN to the DMN/CEN from the DMN/CEN to the VN (Suppl. Fig. 5E). Other combinations of $k_{max}$core voxel/IC compositions could also be observed with every possible transition. In our previous report (5), we measured the intraoperator reproducibility of counting the number of state transitions, and the resulting reproducibility was indicated by a Pearson’s correlation coefficient of 0.88 (Suppl. Fig. 6). These surveys were supplemented by animated flagplots per voxel coreness k/IC compositions, which are actually the shuffled voxels displayed in the animation (Suppl. Fig. 5G). This state transition was easily observed in the positive networks of almost all 180 normal adults (3$^{rd}$ to 4$^{th}$ decade in age) but was rare in the negative networks.

Among 180 individuals, the number of state transitions ranged from none to the most frequent in the positive graphs (Fig. 2). State transition was not observed in 17 subjects, as the same state persisted throughout the entire period (Fig. 2AC). Only one state transition was observed with a half-and-half division of states in 10 subjects (Fig. 2E), and prominent, typical state transitions on several occasions were observed in 105 subjects (Fig. 2G). The others exhibited an intermediate pattern; i.e., an intermediate pattern between no transition to one with half-and-half transitions (Suppl. Fig. 7AC), an intermediate pattern between one transition to indistinct/a few transitions, and an intermediate pattern between typical (Suppl. Fig. 8AC) or too-frequent transitions (Suppl. Fig. 9A). Synchronized combinations of state fluctuations were rare but were definitely present in 12 individuals, and the eloquent case is shown in Suppl. Fig. 10A. Asymmetry of module composition of state was easily recognized on edge-scaled coreness k animation images, one of which showed alternating contributions of the CEN (from left to right, then to left, and then to right) (Suppl. Fig. 11A), another of which showed a left frontal lobe reminiscent of Broca's area (Suppl. Fig. 11C), and the other showed ripples in the left cerebellum but not in the right cerebellum (Suppl. Fig. 11F). All of these are shown in the positive graphs. Unlike in positive graphs, in unsigned negative graphs, the state transition was not remarkable but was vague, if any, and the $k_{max}$core and coreness k animation revealed ripples with infrequent unison-like synchronization (Fig. 3). In an exotic case (only one of the 180 individuals, Suppl. Fig. 11C), animated coreness k plots revealed the explicit state of Broca's area twice in negative graphs (Fig. 3B), which was the same combination of the left frontal area of the salience network (SN), dorsal attention network (DAN), CEN, and auditory network (AN) in the positive graph of the preceding time bins of the first half of the image acquisition (Suppl. Fig. 11C). Otherwise, in all the other individuals, the negative networks did not show a characteristic state composition or pattern of $k_{max}$core and were ignorant of the individuation of the individuals on their behalf.

As the k core percolation was performed per time bin independently in the positive or negative graphs produced with the same thresholds in an individual, the total sum of edges varied greatly per time bin in an individual as well as between individuals. In all the individuals, the sum of the coreness k values all over the 5,937 voxels were 1 to 1 matches with the sum of all the edges per graph time bin (Suppl. Fig. 5A). This relationship was without exception. Thus, coreness k timepoint plots were made using edge-scaled coreness k values and are presented as quantitative plots.

Quantitative edge-scaled voxel coreness k values were displayed on the basis of voxels/IC-composition, which enabled us to follow the trajectories of each voxel's coreness k values on the timepoint trajectory plots of voxels per IC. From the DMN to the VN, the seven components and the left/right cerebellum and vermis and their corresponding voxels were displayed using MATLAB or Excel. On both outputs, the 10 ICs showed quasisimilar contours due to the confounding effect of edge numbers if we used edge-nonscaled voxel coreness k for timepoint plots. After edge scaling, transient increases, decreases or ripples remained on the contour surface (Suppl. Fig. 2A, B). The subtle differences in edge-scaled voxel coreness k timepoint plots could not be the source of evaluating the distribution heterogeneity of voxels and their trajectories. In this study, although we used voxel-annotation by ICs derived from group static ICA of 180 individuals' cohorts, voxel trajectories made recognizable modules for each voxel/IC composition. In addition to the collective behavior of the temporal progression of the voxel/IC composition, these timepoint plots enabled us to follow through the voxel behavior of joining by taking turns in the $k_{max}$core. Each voxel has its own characteristic recognizable trajectory or temporal path along the time bins, meaning that 1) heterogeneity was present between voxels taking hierarchical top-tier positions although their associated ICs were the same and 2) despite this heterogeneity, the voxels for each IC collectively constituted a discernible trajectory along the temporal axis according to the ICs to which they belonged. This was the requisite of hierarchical temporal progress of voxels themselves and their given identity according to their characteristic belonging to ICs (Suppl. Fig. 5B).

**Volume entropy, regardless of surplus edges, functions as a global parameter of information flow over resting-state brain functional graphs**

To reduce the computational burden, brain graphs (with a threshold of 0.65 for positive graphs or -0.5 for negative graphs) were downsampled to 1,489 voxels and subjected to volume entropy calculations using the model described in our previous report (38). In short, each graph was transformed to a universal cover made of nodes and edges of the individual original graph (Suppl. Fig. 12). According to the theorem of minimum volume entropy (40), a random walker's journey on the metric surface of the metric ball of the universal cover converges to an asymptotic end as the radius of this metric ball increases to infinity. The topological invariant, volume entropy of a brain graph, h, on the exponential term is now defined on the metric ball of universal cover of the original graph (Suppl. Fig. 12). In the previous study (38) and in the following application study (39), instead of using numerical analysis to find asymptotically converging values of volume entropy, we applied generalized Markov system on edge transition matrix and Eigendecomposition. Volume entropy is a global measure that represents the global sum of the information flow over the edges of functional brain graphs, either positive or negative.

The volume entropy of a graph obviously depends on the numbers of nodes and edges and the graph constitution. We first tried to remove the confounding effects of the number of nodes and then to understand those of the edges. In the positive and negative graphs, we used graphs with the same thresholds of our choice for all the HCP individuals. The number of nodes was greater than 70% of 1,489 voxels (>85% of 5,937 voxels), and we found that the number of nodes had no effect on the volume entropy on the timepoint plots when we compared the volume entropy of time bins per se and the volume entropy divided by the number of nodes (Suppl. Fig. 13A). Thus, though variable, but if above a certain percentage of nodes were included, we could ignore the effect of the number of nodes. The effect of the number of edges was slightly more complex, as expected, regarding how the number of edges and graph structure were related. By varying the thresholds in an individual's graphs and by observing the entire cohort, we obtained the following findings. First, the number of edges per time bin varied greatly (from 5,000 to 200,000) for every individual. Volume entropy decreased proportionally to the number of edges below a certain number of those specific to each brain graph. For example, in an individual (Fig. 4A), an average of 50,000 or fewer edges showed a coarsely proportional decrease in volume entropy. Second, in this individual, with an average of ca. 330 K (threshold of 0.2) to 78 K (threshold of 0.5) edges in positive graphs or ca. 193 K (threshold of 0.2) to 45 K (threshold of 0.4) in negative graphs, the volume entropy was approximately 9 to 12 million in positive graphs or approximately 8.1 to 9 million in negative graphs (Fig. 4A, B). Third, unlike nodes (regularized within an individual, called edge-scaled), the volume entropy divided by the total number of edges waxed and waned in timepoint plots in all the individuals (Suppl. Fig. 13A). Volume entropy with preset thresholds for an individual (with all the time bins) or for the entire cohort of individuals could not be used for comparisons (Suppl. Fig. 14). Nor could the edge-scaled volume entropy per time bin. This observation limited the use of the volume entropy of brain graphs as a global parameter of information flow capacity for comparisons among individuals, but instead, co-produced afferent/efferent node capacity on directed brain graphs could be used. The timepoint plots of the afferent capacity of the positive graphs (Suppl. Fig. 12) revealed module formation and exchange. The same modules with voxel/IC compositions and their changes were found (Figure 4CD). The plateauing relationship between volume entropy and the total number of edges with lower thresholds (e.g., $\leq 0.4$ in positive graphs) is the first observation that the surplus edges did not contribute to the globally cumulative information flow in a graph represented by volume entropy, a topological invariant of a graph (Fig. 4AB).

**Directed functional brain graphs yielded afferent/efferent voxel-node capacities as**

**surrogates for edge transition matrices**

Edge weights of nodes to and from any nodes yielded edge transition matrices, which were asymmetric in our previous study by Lee et al. using 274 node regions (38). By putting in pairwise intervoxel correlation in this study, instead of the previous inter-regions of interest correlation, we could reproduce the asymmetry of directed edge matrices per time bin. The edge matrix was very large, with a rare possibility of direct visualization; thus, the edge metric was converted to a node metric. That is, from the edge matrices, marginal values were calculated to yield afferent (sum of columns representing 'to the node') and efferent (sum of rows representing 'from the node') values. These afferent and efferent node capacities were overlaid on the MRI slices and time-bin merged to make avi files for animation in play. Positive and negative graphs and their afferent and efferent node capacities produced 2 by 2 (total of four) animations per individual (Suppl. Fig. 12).

Animated afferent node capacities of positive graphs were unique in their revelation of voxels' composition of modules and their exchanges along the time-bins axis, but not efferent capacities or afferent/efferent capacities of negative graphs (Fig. 4C). As each time bin's edge capacity and volume entropy calculation were already normalized during calculation (38), comparable modules popped up from the baselines with exponential height, took temporary union in various combinations of ICs, and took turns, which we called 'module exchange' (Fig. 5A). For example, in these positive graphs, DMN voxels frequently coalition with CEN voxels, SN voxels with AN or sensorimotor network (SMN) voxels, or VN voxels with a variety of IC voxels. The transition from prominent DMN/CEN modules to VN modules or vice versa was observed very often during our read-outs. On the animated afferent node capacity of positive graphs, we could see the waving or undulating progress with collective ICs, sometimes almost all the ICs (Suppl. Fig. 15A). In contrast, animated efferent node capacities of positive graphs were almost stationary with small multifocal flickering on MRI-overlay brain animation plots (Fig. 5B). Animated afferent node capacities of negative graphs showed ripples of collective voxels (Fig. 5C). Negative graphs tended to rarely generate modules, which were different from the afferent node capacity of positive graphs but also formed the unison of smaller module collections with much lower maximum heights of modules (1/3 to 1/8 of positive graphs) (Suppl. Fig. 15,16). The animated efferent node capacities of the negative graphs were stationary with multiple small flickers (Fig. 5BD,

Suppl. Fig. 15B).

The quantitatively assessed afferent node capacities were displayed in voxel-run timepoint plots (point line-connection figures) on the basis of the voxels/IC composition, which followed each voxel's spatiotemporal trajectory of afferent node capacity along the time bins for each IC (Fig. 5AC, Suppl. Fig. 15A). Afferent node capacities of the positive graphs showed individually different envelopes of the voxels' trajectories for each IC, module composition with exponential increases/decreases, and on-and-off movement along the temporal time-bins axis. These modules changed from one IC to another and vice versa between any combinations of ICs. This module exchange reminded patterns of state transition of animated stacked histograms of $k_{max}$core voxels.

Interestingly, voxel trajectories were extremely heterogeneous in making modules for every IC. In other words, when we used an Excel chart display for every voxel (n=69−247) belonging to ICs (AN ~ VN) with the capability of each voxel trajectory tracing, a voxel was on the highest swing in one module but on the baseline without any ascent in the next module (Fig. 6). A priori entitlement of voxels could only manifest the correspondence of voxels to a certain IC. This pattern was universal among individuals in the cohort in terms of the voxels' behavior along the temporal axis in that it belonged to an IC macroscopically and groupwise, but the participation was ad hoc and alternated with the colleague voxels (Fig. 6EF). This immediately defied the simplicity of the region of interest (ROI) approach for following up the collective trajectories of voxels in an ROI. The center-of-mass assumption of inter-ROI correlation was computationally friendly but based on an incorrect assumption in the previous ROI that used investigations, including ours (38,39).

**Voxel hierarchy and afferent capacity of functional brain positive graphs represent mental state transition at rest**

We asked questions regarding whether the voxel-node hierarchy of a dynamic functional brain graph can be represented by the afferent or efferent node capacity thereof for each individual, either or both on the positive and the unsigned negative graphs. While voxel coreness k values were derived from the undirected graphs, which were the aggregates of in-and out-degrees of voxel-nodes, afferent and efferent node capacity was derived from the directed graphs produced independently using universal cover/Markov modeling, which separately represented the afferent (incoming) and the efferent (outgoing) edges with their own weights (Suppl. Fig. 12). Both the voxel coreness k value and afferent node capacity timepoint plots and their overlayed MRI animations revealed the module exchanges during the progress of the temporal time-bins, not efferent node capacity plots. In the positive graphs, the animated stacked histogram (or glass brain display) of the $k_{max}$core voxels (Fig. 2ACEG, Suppl. Fig. 7AC, 8AC, 9A, 10A, 11ACE) and the animated afferent node capacity of the voxels/IC composition (Fig. 5AB, Suppl. Fig. 15AB) well disclosed these module exchanges, corroborating each other. Unique to the afferent node capacity of the positive graphs, the timepoint plots of the afferent node capacity well represented the voxel/IC module formation and exchanges and are presented in the figures (Fig. 2BDFH) and supplementary figures (Suppl. Fig. 7BD, 8BD, 9B, 10B, 11BDF). However, this was not the case for the efferent node capacity, even in the positive graphs (Fig. 5B, Suppl. Fig. 15B). In unsigned negative graphs, rare examples of module exchange were present (Suppl. Fig. 16ACEG), even in the afferent capacity and obviously not in the efferent capacity. Thus, positive graphs, but not unsigned negative graphs, were the main source of fMRI evidence of module exchanges. Notably, the voxel coreness k was derived from undirected graphs, and the afferent and efferent voxel capacities were derived from directed graphs.

The next question was whether positive animated stacked histograms of $k_{max}$core and/or positive animated afferent node capacity are complementary or redundant for revealing state transitions by revealing module exchange of the voxels/IC composition. Both methods were found to reveal similar numbers and timing of state transitions, which were sometimes clearer in positive animated stacked histograms of $k_{max}$core voxels and, at other times, recognized more easily in positive animated afferent node capacity brain overlay images. Scrutiny of the time point plots of both the $k_{max}$core and afferent node capacity revealed that both were the same in their ability to find state transition or module exchanges during the resting state. However, these observations in individuals did not overlap entirely, and one did not include another. Eventually, trajectory tracing of single voxels, as an example, enabled us to determine that their trajectories were actually unrelated to each other (Suppl. Fig. 17). Voxel/IC timepoint plots of afferent node capacity and stacked histograms of $k_{max}$core voxels were the best for disclosing the state transitions and module exchanges once they were observed in collective plots along their IC compositions. Both parameters were helpful for revealing the time bins of the start and end of the modules, combination of the modules, and contribution of combination products (simultaneously activating voxels belonging to the ICs at that moment) not exclusive to one or another modality.

Timepoint plots of coreness k voxels per se were found to be strongly dependent on the total number of edges for each time bin; thus, the seemingly self-similar feature per se was an artifact of the confounding effect of varying the total number of edges. In contrast, the $k_{max}$core was free from this confounding effect, at least within certain threshold ranges, and the afferent node capacity on the directed positive graphs was theoretically and datawise free from the edge-number confounding effect. The following enabled the discovery of resting-state fMRI evidence of fluctuating/transiting human mental states: confirmation of the scale freeness of the degree distribution, finding the hierarchical skeleton of brain graphs and the subsequent k core percolation, relabeling the centrality of voxel degrees with hierarchically structured values of coreness k and, finally, an information/graph topological approach producing directed graphs and afferent/efferent node capacity. Both methods independently went on to unravel module switches, reminiscent of mental state transition in their own ways, and corroborated each other. Eventually, the drawing of the timepoint plots of both measures easily showed that they were not adequately proportional to cancel or include one within the other (Suppl. Fig. 17).

The next simple question was whether the voxels' behavior, along the temporal axis of the time bins, occurred on the 2-dimensional plane of the abscissa of the hierarchy, i.e., the coreness k value and the ordinate of the afferent node capacity. From the positive graphs, an exemplary case was selected, and an IC was selected. When the trajectory of a voxel (and others) was drawn along the temporal progress upon this plane, exuberant paths were produced, which need to be modeled in future studies (Fig. 6EF). The heterogeneity of these voxel trajectories of hierarchy and of afferent node capacity was another luxury of these methods of animated display of pairwise voxel-based amplitude correlation studies of resting-state fMRI.

## Discussion

In this investigation, resting-state mental state progression and transition were investigated by two separate methods, k core percolation on the degrees and information flow analysis of constructed directed graphs, both positive and negative. K core percolation overcame the current ignorance of voxels, producing collective trajectories that make emergent modules and state transition at rest. Disregarding the principal components, we attended to the voxels on the hierarchical ladders of functional brain graphs, exploiting their transience of combinatorial actions. Information flow measures of topological invariant searches of volume entropy allowed us to construct the most probable directed weighted graphs and their corollary quantitative afferent and efferent capacities. This latter approach automatically transformed pairwise edge information to node characteristics, which are called afferent or efferent voxel-node capacities. Fortunately, both methods proved the presence of state and state transitions at rest caught by resting-state fMRI and visualized the modules composed of voxels preannotated to ICs with either animation or timepoint plots. Voxel trajectories have now been understood to join popular ICs, i.e., the DMN, CEN, SN, VN or others, and further investigation of trajectory clustering is warranted to understand their within-IC taking-turns in timepoint progress. Of course, the underlying mechanism of state transition/module switching can be elucidated by combining phase coherence and amplitude correlation. The role of negative correlations (i.e., anti-correlations) requires further study, although we considered negative correlations as unsigned correlations in this study and found that their unsigned pseudomeasures did not contribute substantially to the state transition.

Unlike structural brain graph interpretation, functional dynamic brain graph interpretation mandates understanding graph flexibility. When we interpret functional brain graphs, we need to note that functionally, the brain is not a closed system but an open system receiving inherent emergent (41-44) and/or external perturbation stimuli (45). This is also the case for the resting state, although people erroneously assume that the stationary state of the human mind will be revealed in resting-state fMRI studies. Neuron–glia complexes consisting of brain voxels make the voxels connected/independent identifiers of inherent emergent behaviors but with the characteristics of a microcanonical ensemble (5,14,32). This openness of voxels leads to the spatiotemporal expansion of interactions with adjacent and remote voxels, which produces graph structures of positive and negative correlations/coherences. The sliding window (of sufficient duration, e.g., 1 minute) analysis method for brain blood oxygen level-dependent (BOLD) waveforms needed to be chosen to discretize the temporal expansion of the fMRI signals, which made the analysis computationally burdensome. Fortunately, computing resources, either hardware or software, are available and affordable; thus, voxels and their trajectories can now be traced. The observables from this voxelwise volume of the brain graph consist of voxel-sized quanta spatially and 1-min window-sized quanta temporally. Functional interactions between adjacent or remote voxels are endowed with luxurious time for one minute (several rounds of global communication via electrical and chemical interactions between neuron–glia complexes within each voxel), and adjacency or neighborliness in their similarity characteristics of functional correlations are then free from any constraint of structural connections between voxels. The BOLD waveforms of voxels and the intervoxel correlation values thereof now depend only on their observed correlation/coherence. In other words, the behavior of several thousand voxels is more likely to be that of a three-dimensional lattice, the nodes of which can communicate within the observed period of 1 minute regardless of the Euclidean physical or structural distance. Only the observed similarity between voxels dictates further interpretation of brain graphs. We did not integrate phase coherence in this study.

In the Introduction section, the first assumption was announced to be untestable in our investigation (4,24,25); however, on resting-state fMRI, we discovered modular on-and-off phenomena and module exchanges at rest in humans. Until recently, edge weights and simplified adjacency matrices (degrees) have been used for network understanding. Here, we questioned the long-held assumption that every edge was created to be equal and thus could be treated as equal. We also challenged another inaction from the perspective that undirected graphs would suffice to understand the mechanism of functional brain graphs safely using the observed amplitudes and their correlations on fMRI. Correlation supplies only undirected graphs, unlike the already-directed weighted graphs of internet/phone calls/citations networks. We tested the following questions: 1.) What if we obviate the equality of edges and adopt hierarchical repositioning of adjacency matrices (and degrees) into coreness k and 2.)

What if we make the directed graphs using a graph universal cover/Markov model and observe the afferent voxel-node capacity of these directed graphs. This approach overcame the bottlenecks of limited representation and consequent insufficient understanding of the dynamic states of human minds. Post hoc, we assumed that the fMRI state transitions or fMRI module exchanges represented spontaneous mental state transition at rest in humans.

Based on these adjusted assumptions, the following three hypotheses (sliding window for finding state fluctuation/transition, transition by higher-order interactions ~~on~~in pairwise intervoxel studies, and observed intervoxel correlation as the sum of signals and redundancy/noise) were tested for their ability to explain the underlying physiology of mental state fluctuation/transition. In addition to the question whether these assumptions were correct, we were also curious about whether the findings of this study would precede any subsequent explorative pursuit studies; that is, whether they would enable us to generate the next-step research hypotheses. In this vein, the findings and their technical caveats are briefly summarized below, followed by the plausible relationships of this study's results with the shared ideas of the adjacent scientific disciplines.

The second assumption was whether sliding-window time-binning of resting-state fMRI (Suppl. Fig. 1) reveals the waxing and waning behavior of modules or states made by the gathering and/or dissociating of voxels to reveal the continuous changes in the composition of voxels/ICs on timepoint plots (26,32,46,47). Spontaneous fluctuations without any abrupt transition of mental states were clearly observed in the temporal progression of the animated coreness $k/k_{max}$core images of the unsigned negative graphs of all the individuals. In contrast, in the positive graphs of almost all individuals, in addition to the fluctuating voxels/IC composition, clear/clean state transitions were also observed with razor-sharp transitions in most of the cases. The duration of one state varied, and the number of state transitions from one to another varied greatly among individuals (15 minutes in the HCP database (33,34) and within a session (5 minutes in the normal PET/fMRI or Kirby database (48)), as detailed in our preprint in bioRxiv (5). Figures 2, 3, and 5 and Supplementary Figures 7−11, 15, and 16 show the $k_{max}$core and afferent capacity spectra of voxel nodes in individuals. A 15-minute period of resting-state fMRI was sufficient to reveal the prolific diversity of human resting mental states and their transitions. We suspect that we see the luxury of mental states in individuals during 15 minutes of HCP. If we had observed only the first or last 5 minutes, we might have seen only partial results and related our findings with characteristics of the subjects and erroneously categorized them for their traits. In our previous report (5), a person in the Kirby project underwent a resting fMRI study weekly for 3.5 years, which was visualized by our $k_{max}$core voxel/IC composition timepoint plots. The dominant compositions were fairly similar during certain consecutive weeks and months and then changed to another pattern over another certain period. The feasibility of studying the association between dynamic characteristics and individuals' traits is to be questioned. The characteristic cores of DMN/CEN or VN dominance observed with static hierarchy in our previous study (32) should also be reinterpreted for their significance. The question of the possible representativeness of static analysis for time-varying fluctuating/transiting mental states warrants subsequent studies from the points of hierarchy and the afferent capacity of positive and negative graphs.

State transition is one of the holy grails in statistical and quantum physics and is currently being interrogated using complex network theory to determine its underlying mechanisms (49-53). A growing many-body system (29,30) was the basis of the network's hyperbolic geometry, which emerged from the many-body interactions in the simplicial complex (41). By analogy, brain voxels could have been considered many-body interactions such as elements or quanta in physics and fit for hyperbolic embedding with scale freeness and phase transition, which we and others tested previously (32,41,54). Entanglement in quantum science or synchronization in statistical physics are the terms used to describe the emergent coherence of many-body collective configuration or temporal progress during the period of resting-state fMRI. In our previous study, a many-voxel collective configuration showed abrupt transitions in k core percolation (5,32,54) and abrupt (or explosive) changes (5) in the $k_{max}$core voxel timepoint plots. The former was on the configuration space, and the latter was on the temporal axis. The abrupt increase in the afferent node capacity of the directed graphs corroborated these findings in this study. Voxel hierarchy and information flow revealed an impressive analogy with statistical and quantum physical systems. Physical systems address intersite/space correlation and anti-correlation relationships, and similarly, brain fMRI studies have shown intervoxel correlations and anti-correlations. This was the substance of the dynamic criticality of the functional brain connectivity observed in the resting state beyond stability or metastability (55). Specifically, in the positive graphs, a 1-minute quanta with a 3-second shift was enough to produce an easily recognizable state transition as abrupt changes in the voxels/IC composition and an explosive emergent rise and fall of modules made by the afferent capacity, which were not observed in the negative graphs.

The combination of the third and fourth assumptions were whether waveforms of each voxel once observed pairwise (5,52,54), act as the simplest representation of their higher-order interaction (27,28,56,57) and whether the pairwise-observed amplitude correlation as an observable is the sum of signals (functional connectomic dynamics) and redundancy (including inherent and measurement-related error/noise) (58,59). These assumptions were implicit basics for all our analyses of dynamic functional brain graphs, such that the observables, i.e., intervoxel similarity, consisted of signals and redundancy (or noise/error). According to these assumptions, we removed measurement-related errors/noise and inherent redundancy by evaluating log-log plots of the degree distribution of intervoxel pairwise similarity (32,54). Unexpectedly, the afferent node capacity and volume capacity were not affected by the increase in the number of edges by several orders of magnitude (from thousands to millions) when we varied the thresholds from 0.45 to 0.2 and increased the number of edges by several orders of magnitude. Technically, this guaranteed the resilience of the information flow analysis using a topological approach/Markov modeling, and we did not need to remove the redundancy of the weakly connected edges beforehand. Theoretically, this could also mean that the surplus edges do not contribute to the information flow or global information capacity of the graphs. Surplus edges consisted mainly of redundancy and smaller amounts of errors/noises. This would make sense if the redundancy did not confound or help the information flow in the brain graphs with surplus edges. Proving or falsifying this idea requires the input of studies using computational modeling with population codes simulating neuron–glia complexes. Intervoxel similarity of the signals was determined only by the amplitude correlation in this investigation, and we ignored intervoxel phase similarity/differences (60). It is possible that the amplitude-based similarity measures might have constrained the surplus edges to avoid contributing to the calculation of graph volume entropy or afferent/efferent capacity, as the edges were stripped of their phase information.

The thresholding of intervoxel correlations in making brain graphs to expose the skeleton of their hierarchical structures was an essential prerequisite for effective k-core percolation. When we used the unthresholded or underthresholded data of the same individual for k-core percolation, the percolation proceeded but did not yield state transitions on the coreness k animation plots or any abrupt changes in the $k_{max}$core voxels/IC composition timepoint plots. The distribution-free or scale-free characteristics of the degree distribution dictated the lower limit of the threshold. Instead, the nodes (voxels) number criterion, i.e., 85% or more voxels, determined the upper limit of the threshold, which could maintain a sufficient number of nodes to make the giant connected component. Positive correlation-based brain graphs, as well as unsigned negative correlation-based brain graphs, allowed thresholds to range from 0.45−0.7, within which the range of windows, scale freeness, edge number–total coreness k proportionality, and consistency of volume entropy were conserved (Suppl. Fig. 3). We set the thresholds per group of subjects (a cohort) in the HCP or Kirby database (5,48) separately for their positive and negative graphs.

In the 180 individuals in the HCP cohort, with 280 time bins each, there were rare outlier time bins in terms of the number of voxels. Timepoint plots of both total coreness k and volume entropy were not affected by the number of voxels (Suppl. Fig. 13A). The number of edges also did not, but for different reasons, for both measures. The total coreness k was strictly proportional to the total number of edges and was normalized by dividing it by the number of edges. Comparisons could be made between any time bins within or between individuals using edge-scaled plots, animation edge-scaled flagplots, edge-scaled coreness k timepoint plots and coreness k map animation plots. $K_{max}$core plots such as stacked histogram timepoint plots and glass brain voxel/IC animation are free from any edge number effects, as they show top tier voxels irrespective of the remaining edges after thresholding. The volume entropy and afferent node capacity were unaffected by the number of edges. A redundant surplus of edges with full participation of voxels less than a certain threshold (e.g., <0.5 in the positive network in one case) did not influence the calculated values of volume entropy for all 280 time bins (Fig. 4). It was not just the similarity but the sameness. We conclude that the volume entropy of a brain graph (per time bin) reflects the core skeleton of functional brain connectivity graphs and not the surplus of edges. In retrospect, thresholding was not necessary for volume entropy assessment or afferent/efferent capacity quantification. The same thresholds were used in this investigation for both k core percolation and directed graph construction and for determination of the afferent capacity to facilitate side-by-side comparisons. Thresholding effects were evaluated and determined not to confound the quantitation of volume entropy, the production of directed graphs, and the plotting of afferent/efferent capacity.

Unsigned negative graphs were used to make graphs, meaning that the nature of the anti-correlation was intentionally neglected. Before being able to combine signed negative correlation edges with positive edges, we could not but observe the nature of unsigned graphs. As mentioned, the $k_{max}$core (hierarchical structure) and afferent node capacity (summed edge weights of voxels on a directed weighted graph) of positive graphs corroborated each other to reveal spontaneous module switches, suggesting mental state transitions. In unsigned negative graphs, however, the $k_{max}$core and afferent node capacity rarely showed state transition/module switches, as did the efferent node capacity of positive graphs. The quantified timepoint plots of the afferent capacity of the negative graphs showed a much lower capacity than those of the positive graphs. The volume entropy of negative graphs also tended to be lower than that of positive graphs. For the 180 individuals in the HCP cohort, the mean number of average (over 280 time bins) edges per individual in the negative graphs (thresholds: 0.5) was ca. 240 K compared with 490 K in the positive graphs (thresholds: 0.65). Unfortunately, we could not directly compare the measures of positive and negative graphs because these positive and negative graphs are separate entities from the same individuals. In almost all the subjects, negative graphs did not reveal state transitions or module switches. A priori, no switches were expected to emanate ever from sliding window preparation for both positive and negative graphs. Thus, the discovery of state transitions and module switches in positive graphs is remarkable.

In directed functional brain graphs, previously (38,39) and in this study, we observed asymmetry between afferent and efferent flows. In our previous study, asymmetry of the edge matrix was observed on static functional directed brain graphs, but we could not explain this phenomenon. In this investigation, we cropped many (280 time bins per individual) dynamic directed graphs and their edge matrices from every individual. Asymmetry, which cannot be an artifact, prevails with no exception. Animated efferent node capacity was less bright on the brain-overlay dynamic displays. As we calculated afferent and efferent node capacities from the edge capacities on the volume entropy calculation program written in MATLAB in one instance and every instance of calculation was normalized to the volume (38), the brain overlay was normalized to the maximum value of all the voxels of afferent and efferent node capacities per individual. Only the positive afferent node capacity on the animated brain overlay (or on $k_{max}$core stacked histogram timepoint plots) showed definitive modules and module exchanges (states and state transitions) during the session of resting-state fMRI acquisition. Negative afferent nodes showed ripples and small multifocal flickering without any modules or module exchanges. We propose that afferent node capacity is the source of mental state transition in the resting state in humans. This finding reminded us of the earlier report by Logothetis et al. (61) that found that local field potential (LFP) was an immediate source of BOLD signals in a visual stimuli task study using monkeys in a simultaneous EEG/fMRI machine. Local field potential represents the amplitude of afferent (presynaptic) collective inputs at the synaptic buttons in the dendrites of neurons. Postsynaptic potentials were the output from the designated neurons. The difference is that Logthetis et al. (61) used a task paradigm, not a resting-state scheme, and the LFP is not just the equivalent of afferent voxel capacity. We propose this analogy because this finding is new and encourages future computational modeling studies to address this puzzle. We suggest that our approach and its results might be used as input to any modeling scheme in subsequent studies.

Quantitative assessment revealed that the efferent node capacity tended to be much smaller than the afferent node capacity, and animated images showed near stationarity with multifocal small flickering. The state transition was visualized on an animated stacked histogram of $k_{max}$core on a linear scale. Afferent node capacities was were on an exponential scale. We speculate that the state transition might have been affected by the phase similarity (or dissimilarity) of the intervoxel pairwise relationships, which we did not analyze in this study. We suggest the following three implications: 1) The orthogonal phase and amplitude (60) can be used to reconstruct the voxel-pairs relationship authentically if and only if the phase difference or synchrony is integrated into the amplitude correlation. If temporally emergent inherent synchronizing and explosive momentum from both amplitude and phase built within the voxel pairs, the voxels/ICs clusters would show an abrupt pattern of appearance/disappearance of hierarchically structured core voxels with some top tiers, also known as $k_{max}$core; 2) Since realistic higher-order intervoxel interactions have been routinely simplified to the current easily computable pairwise graphs, these higher-order interactions might have promoted and resolved modules in their formation exchanges, the voxels to the top tier in one moment and then the following plunges; 3) Unlike the legitimate analysis of positive graphs, negative graphs were converted to unsigned graphs. If the unsigned negative graphs had been analyzed as signed negative graphs and/or properly integrated into the positive graphs, it might have improved the understanding of the state transition. We are strongly encouraged to combine amplitude correlation and phase synchrony, tame higher-order interactions, and integrate positive and negative intervoxel correlations. In this sense, voxel entities and their pairs of functional brain graphs deserve to be analyzed as hypergraphs or simplicial complexes (27,28,57).

Although state transition per se was a novel discovery by sliding-window hierarchy and information flow analysis, we also observed that the voxel/IC representation enabled us to reveal the heterogeneity of voxel participation at each time bin (5) (Fig. 6. Suppl. Fig. 17). This interesting observation was derived from the selection of the input, not the ROI but the voxels being considered independent entities. As we did not pay attention to the topography of voxels' BOLD signals but rather the pairwise intervoxel correlation edges, we might also assume that voxels were independent during the 1-minute-long 3-second shift time-bin window, even allowing us to disregard spatial adjacency. This is different from the situation of structural connectivity. Recent progress in hardware, CPUs and memory for universal Linux/Windows workstations has led to the routine use of intervoxel pair matrix computation, including eigendecomposition. When we downsampled the original $2 \times 2 \times 2$ mm$^3$ fMRI data to $6 \times 6 \times 6$ or $10 \times 10 \times 10$ mm$^3$ resolution, the propensity of the edges in any individual in any cohort was half positive and half negative. After seeing this inadvertent observation, we could no longer reduce the fMRI data to ROIs such as 274 parcels (38,39,54). Voxel-based calculations seemed justified for every subsequent investigation (32). Independent component analysis (ICA) with a readily available algorithm was only used to annotate every voxel to 7 or more ICs, and the remaining voxels were unclassified (collection of smaller ICs). While positive graphs showed novel findings of state transition, negative graphs showed waving fluctuations in total coreness k per time bin with ripples and infrequent threads or collective waving in the unison of voxels for all ICs.

The trajectories of voxels belonging to the same ICs were surprisingly heterogeneous for both spatiotemporal deployments of coreness k and of afferent capacity. Voxels took turns doing their jobs of emergent construction of modules, rising to the top of the hierarchy during the short periods of 15 minutes (Fig. 6). Edge-scaled coreness k was grossly similar but different in terms of the contours of the collective trajectory maps (Suppl. Fig. 2). The afferent capacity of voxels showed an on-and-off pattern of module formation and switches between ICs, but the efferent capacity did not (Fig. 5, Suppl. Fig. 15). The afferent and efferent capacity of voxels and their module-forming characteristics were initially assumed to form a hierarchical structure of edge composition in a collective to allow the top tier voxels to climb up to reach the $k_{max}$core. This naïve expectation was not true for the role of the efferent capacity , as it was quite homogeneous despite spatiotemporally scattered small flickering (Fig. 5). In a separate analysis of negative graphs in the form of unsigned graphs, we did not observe prominent modules or top tier voxels/ICs with any transitions even in terms of afferent capacity. Thus, we suggest that the afferent node capacity of positive graphs contributes to hierarchical edge characteristics. If so, we needed to survey whether one belongs to another; that is, the two spatiotemporal trajectories of hierarchy or afferent capacity would be inclusive or overlapping if they were plotted together (Suppl. Fig. 17). For any voxel on the surface or 3D plots, no relationship was observed between the two measures, the coreness k and the afferent capacity of the positive graphs. They looked varied on their own upon their axes of abscissa and ordinate. One was not replaceable by the other. We presumed that the underlying mechanism behind their spatiotemporal progress needs to be further scrutinized if we want to understand the nature of state transition on fMRI, which mirrors the mental states of human individuals.

The relationship between the sum of dynamic functional brain graphs and static functional brain graphs was studied using another group of normal PET/MRI studies and yielded crude correlations between the dynamic sum and static brain graphs. Similar results were found with an exemplary case of HCP with the three subsets of an HCP for comparison of divided-3 raw data. If we attend only to the static data, the profuse information contained in dynamic studies is lost and smudged relevantly to a gross sum of findings. This was expected and observed.

Our investigation was distinct from previous studies using various centrality measures in that those studies examined the structure of the networks, but we paid attention to the voxels themselves. Their results revealed that the degree distribution obviously represented global graph characteristics but did not represent the voxel entities in their structural graphs. This created difference in interpretation of their various centrality measures, such as degree centrality, betweenness centrality, Eigenmode centrality, PageRank centrality (62), etc., from ours of coreness k values (32). Interestingly, our volume entropy/afferent node capacity works by taking advantage of the fact that Eigendecomposition is equivalent to the asymptotic measure of a random walker job, regardless of its variations of adaptive signed, return, quantum or hypergraphs (62-64). It would be interesting to determine whether any relationship exists between the afferent/efferent node coreness k measure and other centrality measures on the separated afferent and efferent directed networks we produced in this investigation. Currently, network science or graph research has accumulated sufficient knowledge about network/graph structures with the emphasis of communities hidden in graphs using measure metrics and their geometric meanings (65,66). The communities were determined by the network/graph structures themselves (36,67-70) or separately defined using biological annotations within the networks/graphs (71,72). Among these many, simplicial complexes for higher-order networks or synchronization (73,74), higher-dimensional hyperbolic embedding with popularity similarity (66,75), and generalized k-core percolation (75) led us to the community hidden in the graphs. In our investigation, we annotated each voxel belonging to ICs defined a priori using ICA outputs. It could have been anatomical priors or even trajectory clustering of the current data outputs.

Our findings leave the following questions: The first refers to the nature of state transition, revealed by an observation here using resting-state fMRI. fMRI observables were analyzed simply and easily using 1) a classic statistical physics method (k core percolation) and 2) the construction of directed weighted graphs based on a detection scheme of topological invariant of graphs. Currently, there is likely an underlying explanation for these ‘state transitions’, which were observed universally between individuals and even within sessions in an individual (5). The question of what caused these state transitions has not yet been answered, but clues include 1) nonlinear dynamic interpretations ranging from criticality and self-organizing characteristics of neurons in the literature (76-78), represented as voxels in our investigation; 2) large neuronal theory, recently refined as a neuronal communication system (79-81) and voxels’ communication dynamics in our investigation; 3) information theory perspectives for complex networks (82), von Neumann entropy and its transition interpretation (83,84) or complexity entropy, such as Kolmogorov interpretation (85), a topological invariant named graph volume entropy (40,38) in our investigation to explain the dynamics of intervoxel correlations. However, if we try to exploit these clues, the computational burden immediately intervenes again. We could have moved from the previous schemes that used anatomically or ICA-annotated ~300 ROIs as matrix elements (86,87) to the current 1,500 to 6,000 voxel element studies for dynamic undirected/directed graphs. The graph Laplacian was updated by the Hodge Laplacian and Dirac operator (42,43) to explain the static or dynamic behaviors of graphs. Moreover, quantum explanations of higher-order networks (49) have expanded to adjacent disciplines such as neuroscience. We cannot help using the term 'entanglement' for the collective neurons' behavior connected over the geodesic surface of high-dimensional graphs and their temporal expansions. The spatiotemporal trajectories gained momentum for visualization and quantification in our investigation. Once traced, these outputs can be inserted into refined computational models for dynamic functional brain voxels with higher spatiotemporal resolution along discrete time-bin axes. The possibility of explaining spontaneous or stimulus/task-associated mental status changes in humans, either normal, sleeping, conscious/sedated/semi or comatose, or in the states of disorder/disease during their course or rehabilitation after successful or unsuccessful treatments, is possible.

In conclusion, we have two very distinct methods of producing one spatiotemporal deployment of hierarchical realization of functional brain graphs (coreness k images on animation) and producing another spatiotemporal realization of normalized afferent/efferent node capacity. These two methods were derived from each discipline and model. Exact computing was reproducible in calculations without any interoperator bias over all the thresholds chosen, while there seemed to be a narrow window enabling the observation of state transitions on the $k_{max}$core or a wider window allowing module exchanges on afferent node capacities.

## Methods

*Data preprocessing*

We downloaded resting-state fMRI (rsfMRI) data from the HCP (www.humanconnectomeproject.org). We selected 180 participants aged 22 to 36 years without any significant history of psychiatric disorders or neurological or cardiovascular disease (33,34). We used minimally preprocessed rsfMRI data (35) and performed further preprocessing: smoothing using a 6-mm full-width at half-maximum Gaussian kernel and bandpass filtering (0.01 Hz – 0.1 Hz). Next, we downsampled the data from 91 × 109 × 91 voxels with 2 × 2 × 2 mm voxels to 31 × 37 × 31 voxels with 6 × 6 × 6 mm voxels, which reduced the computational load. We applied a mask to exclude voxels that did not belong to the brain, resulting in 5,937 voxels for k core percolation. For the volume entropy calculation and directed graph composition, another downsampling with a 10 × 10 × 10 $mm^3$ voxel resulted in 1,489 voxels.

*Independent component analysis*

We performed ICA to identify ICs, i.e., resting-state networks, using multivariate exploratory linear decomposition into independent components (MELODIC) (*46*). We obtained spatial maps of ICs and applied a threshold ($Z > 6$) to generate binary masks. In this study, we included 7 ICs, namely, the default mode network (DMN), salience network (SN), dorsal attention network (DAN), central executive network (CEN), sensorimotor network (SMN), auditory network (AN), and visual network (VN).

*Dynamic data analysis: sliding-window analysis*

For the HCP data, we used sliding-window analysis to investigate the nonstationary and time-dependent dynamics of the brain. The window size was set close to 1 minute (84 volumes, 60.48 sec) with a shift of 4 volumes (2.88 sec), resulting in 280 windows. A connectivity matrix of each window was calculated to conduct *k*-core percolation. We implemented k-core percolation for each connectivity matrix after applying the threshold, which meets the scale freeness of the network, to generate a binary matrix. Edges with values greater than 0.65 in positive graphs and 0.5 in negative graphs are assigned values of one; otherwise, they are assigned values of 0.

*k-core percolation*

We conducted *k*-core percolation to investigate an individual's core structure of a functional brain network (32,36). Metaphorically, this method peels the layer of the network as if it were an onion. This procedure first removes nodes of degree 1 ($k$ = 1). As nodes are removed, the degree of the remaining nodes also changes. Some nodes whose degrees were not 1 at first are eventually removed if they meet the removal criteria. The procedure is performed recursively by incrementing k by 1 until no further processing is possible. A subgraph, called the k-core, is obtained when all nodes with degrees less than k are removed. The last surviving core was called the $k_{max}$core. After k-core percolation was performed on each subject's data, we classified $k_{max}$core voxels using IC maps.

*Volume entropy calculation and construction of directed weighted graphs*

During the volume entropy calculation, we modeled the functional brain graphs to have directed-ness using a generalized Markov system on universal covers of the undirected graph matrices. Matrix elements were detected via pairwise correlations of the voxels' waveforms on resting-state fMRI. Universal cover allowed one-directional random walkers to tread all the possible paths from any node to infinity to yield an N-dimensional ball of infinite radius. The topological dynamics representing the capacity of information flow over the graphs were proven to have asymptotic invariant specific for the graph. These intermediary edge matrices were supposed to reveal the in- and out-flow capacity of information via every edge from the standpoint of the information flow along the brain graphs (38). We named the in-flow capacity of certain nodes the afferent node (voxel-node) capacity and the out-flow capacity the efferent node capacity, as in our previous study (39). Thus, edge length (or distance) derived from pairwise intervoxel amplitude correlation was used to define the hidden directed functional brain network.

Once the N-dimensional ball of the universal cover grew over walking to infinity by increasing the radius to infinity, the geodesic sum of the length covered by the random walker was modeled using a generalized Markov system. Eigendecomposition replaced the random walker model and yielded edge capacity matrices. A total of 10 x 10 x 10 $mm^3$ downsampling resulted in a maximum of 1,489 x 1,489 edges, and after thresholding with the same criteria as for the 6 x 6 x 6 $mm^3$ image data, approximately 2 million (incoming and outgoing) edges remained to make 2,000,000 x 2,000,000 matrices for eigendecomposition. The Eigs function of MATLAB® was used to calculate edge matrices via Krylov-Schur decomposition, and then volume entropy was calculated in 2 to 3 hours of computing time per time bin. This led to the completion of the calculation of edge matrices and volume entropy per individual within weeks (4 weeks for HCP data with 280 time bins per individual).

*Data visualization*

Using graphs with 5,937 voxels and 6 × 6 × 6 $mm^3$ brain volume images, k-core percolation was performed. The resulting outputs were visually represented as 1) animation maps of edge-scaled coreness k values overlayed on 36-slice MRI templates, 2) edge-scaled flagplots on the runs of voxels with their IC labels, 3) $k_{max}$core stacked histograms on the runs with their IC labels, 4) edge-scaled coreness k value plots, 5) animation glass brain lateral/transaxial images for 280 time bins of 1 minute duration (84 acquisition bins) with ca. 3 seconds (4 bins equal to 2.88 sec) shifting from 1,200 acquisition bins (0.72 seconds per bin) of the HCP datasets. All these maps are presented as positive and unsigned-negative brain graphs. The total number of edges counted after thresholding was used to normalize (divide) voxel coreness k values, and these edge-scaled coreness k values were used for animation plots, flagplots and voxel trajectory timepoint plots per IC (voxel/IC trajectory timepoint plots).

Using graphs with 1,489 voxels using 10 × 10 × 10 $mm^3$ brain volume images, volume entropy was calculated again for positive and negative graphs while simultaneously producing the directed weighted graphs for each type of graph. The resulting outputs were visually represented as 1) animation maps of afferent or efferent node capacity overlayed on MRI templates and 2) afferent and efferent timepoint plots representing each voxel's trajectory and their collective picture separately according to their belonging to ICs (DMN, SN, DAN, CEN, SMN, AN, VN) and left cerebellum (L_Cbl), right cerebellum (R_Cbl), vermis (V) and the unclassified (Unc).

**Acknowledgments:**

We thank Seonhee Lim for her comments related with volume entropy.

**Funding:** This work was funded by National Research Foundation of Korea (NRF) Grants from the Korean Government (MSIP) (NRF-2017M3C7A1048079, NRF-2020R1A2C2101069, NRF-2022R1A5A6000840, RS-2023-00264160) and network support of KREONET (Korea Research Environment Open NETwork).

**Author contributions:**

DSL: Conceptualization; Methodology; Investigation; Visualization; Funding acquisition; Project administration; Supervision; Writing – original draft; Writing – review & editing.
HK: Methodology; Investigation; Visualization; Writing – original draft.
YH: Methodology; Investigation; Visualization; Writing – original draft.
YK: Methodology; Investigation; Visualization.
WW: Methodology; Investigation; Visualization.
HL: Methodology; Investigation; Visualization.
HK: Conceptualization; Methodology; Investigation; Visualization; Funding acquisition; Project administration; Supervision; Writing – original draft; Writing – review & editing

**Competing interests:** The authors declare that the research was conducted in the absence of any commercial or financial relationships that could be construed as potential conflicts of interest.

**Data and materials availability:** All data are available in the main text or supplementary materials. All data, code, and materials used in the analysis are available through a standard material transfer agreement with POSTECH for academic and nonprofit purposes by contacting the corresponding authors (dsl9322@postech.ac.kr or hkang@snu,ac,kr).

## Figures and Figure Legends

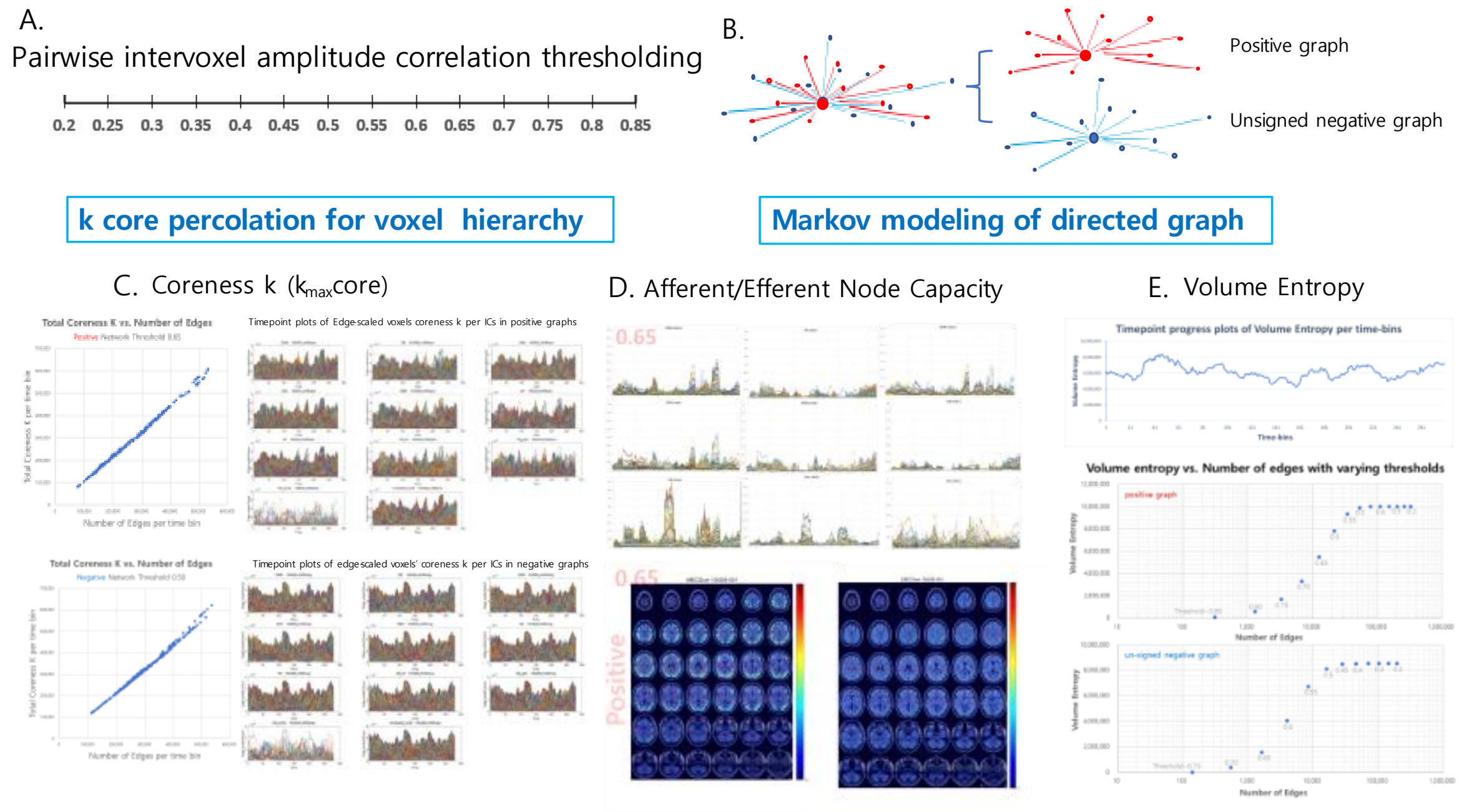

**Figure 1. Pairwise intervoxel amplitude correlation makes positive and negative graphs and their animation, timepoint and other graphic plots following analytical methods.** A. In an exemplary case, the thresholds varied. B. The thresholding yielded positive and unsigned negative graphs that went through k-core percolation to yield voxel hierarchy. The graphs' universal covers were modeled using a generalized Markov system to produce directed graphs to yield the afferent/efferent capacity and volume entropy. C. 1:1 Correlation between total edge number and total coreness k enabled edge-scaling of hierarchy measures: voxel/IC timepoint plots of coreness k (shown here), MRI-overlayed coreness k animation plots, and flagplots. Top tier $k_{max}$core voxel plots on stacked histograms or glass brains need edge number normalization. D. Afferent and efferent node capacity timepoint plots or MRI-overlayed plots of positive and negative graphs revealed voxels/IC modules and their changes, revealing incoming and outgoing flow of information on the directed weighted graphs. E. Each time bin's volume entropy fluctuated on their timepoint plots. Once the thresholds were varied, the edge number varied by several orders of magnitude, but the volume entropy remained the same between graphs with or without surplus edges. A decrease in the number of edges proportionally decreased the volume entropy with increasing thresholds for both the positive and negative graphs.

A.

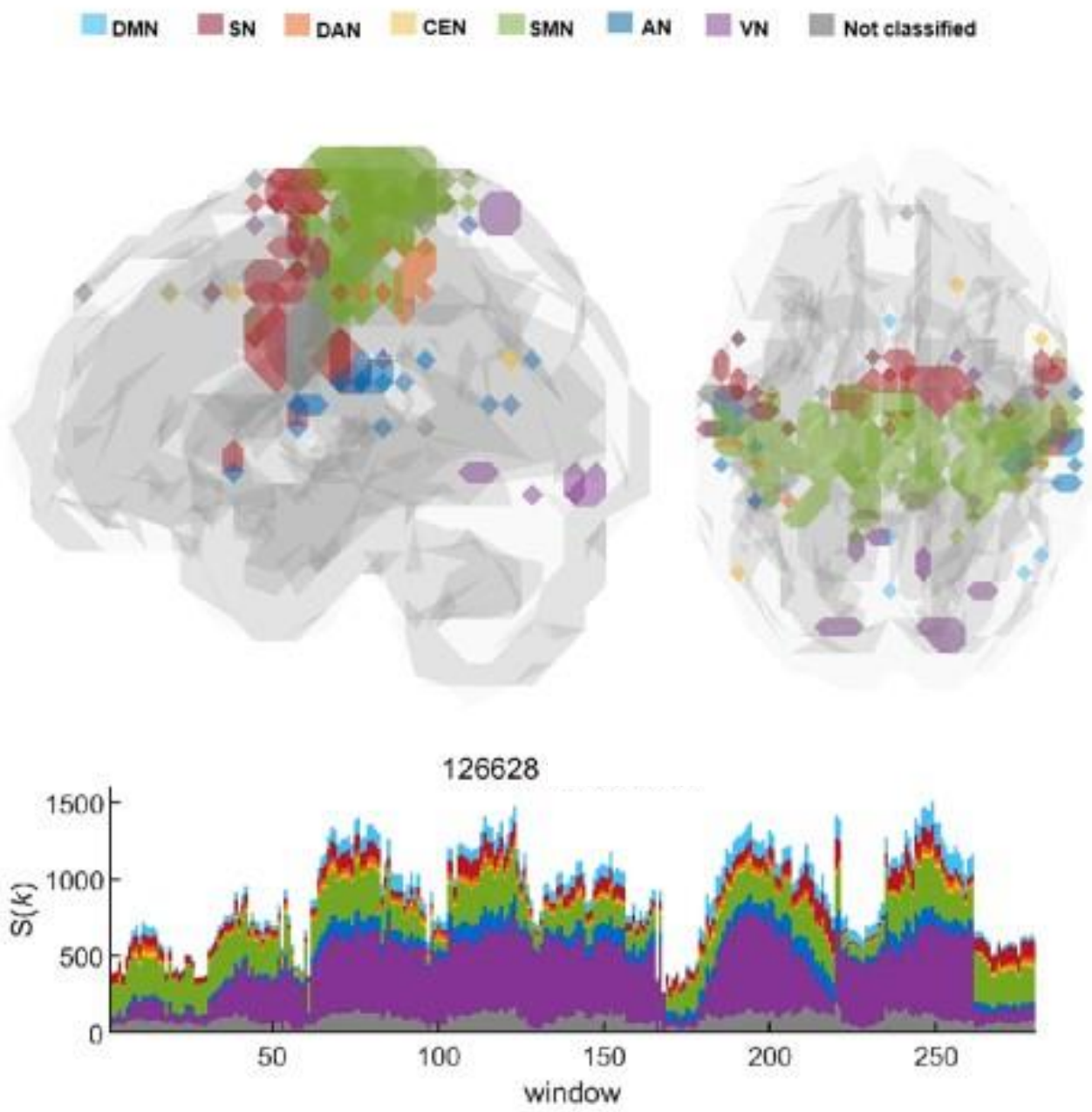

B. **Afferent node capacity**

**Positive Graph**

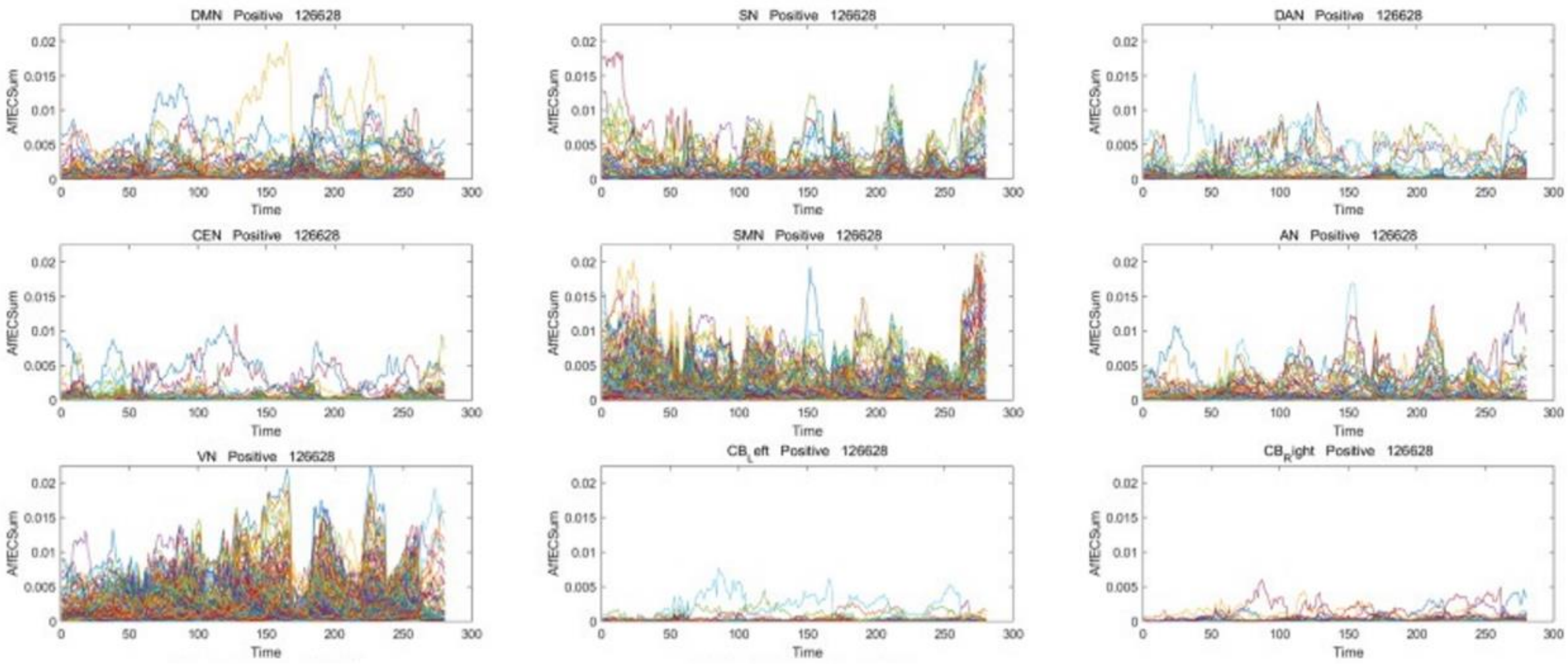

C.

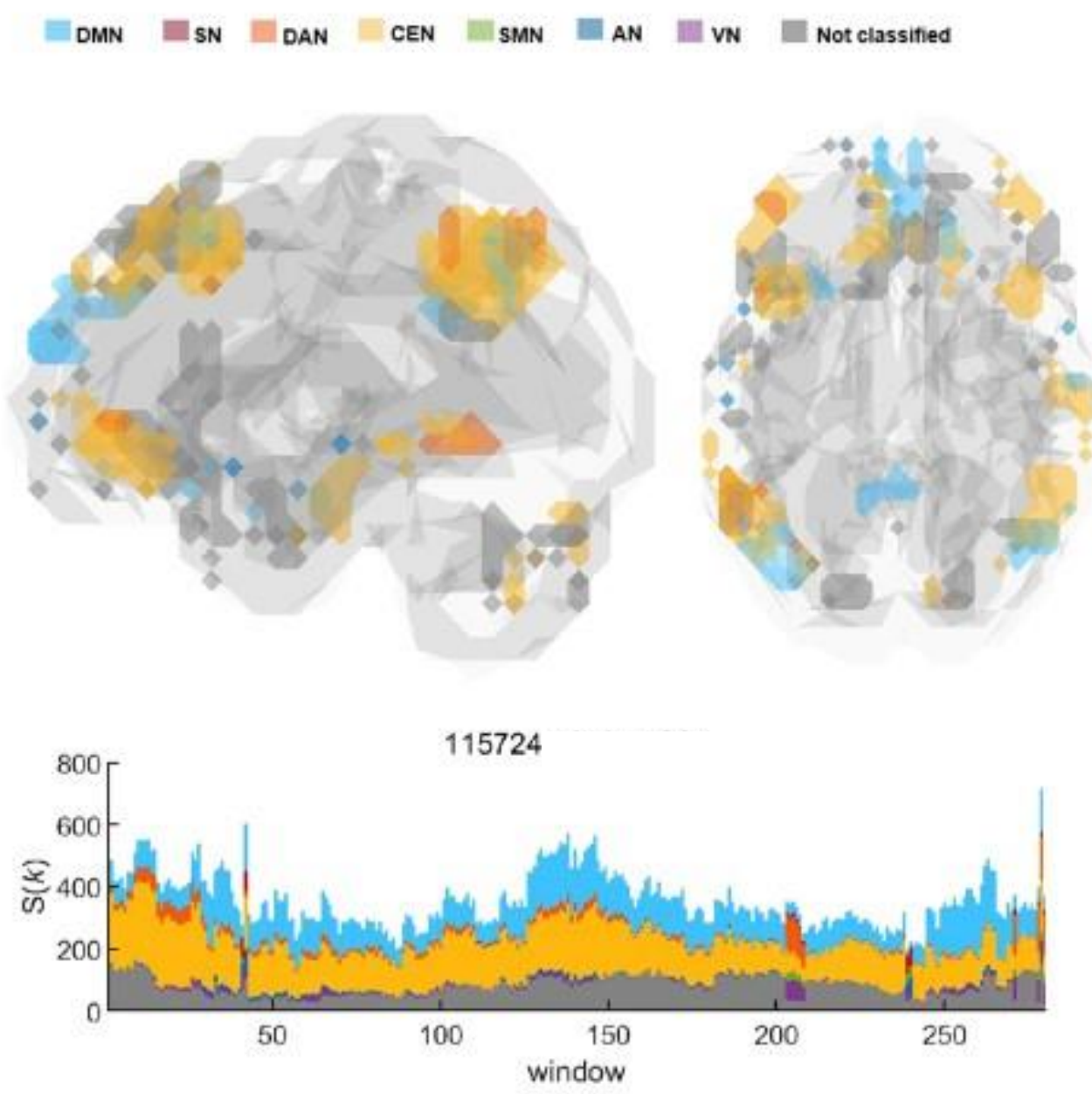

D. **Afferent node capacity** **Positive Graph**

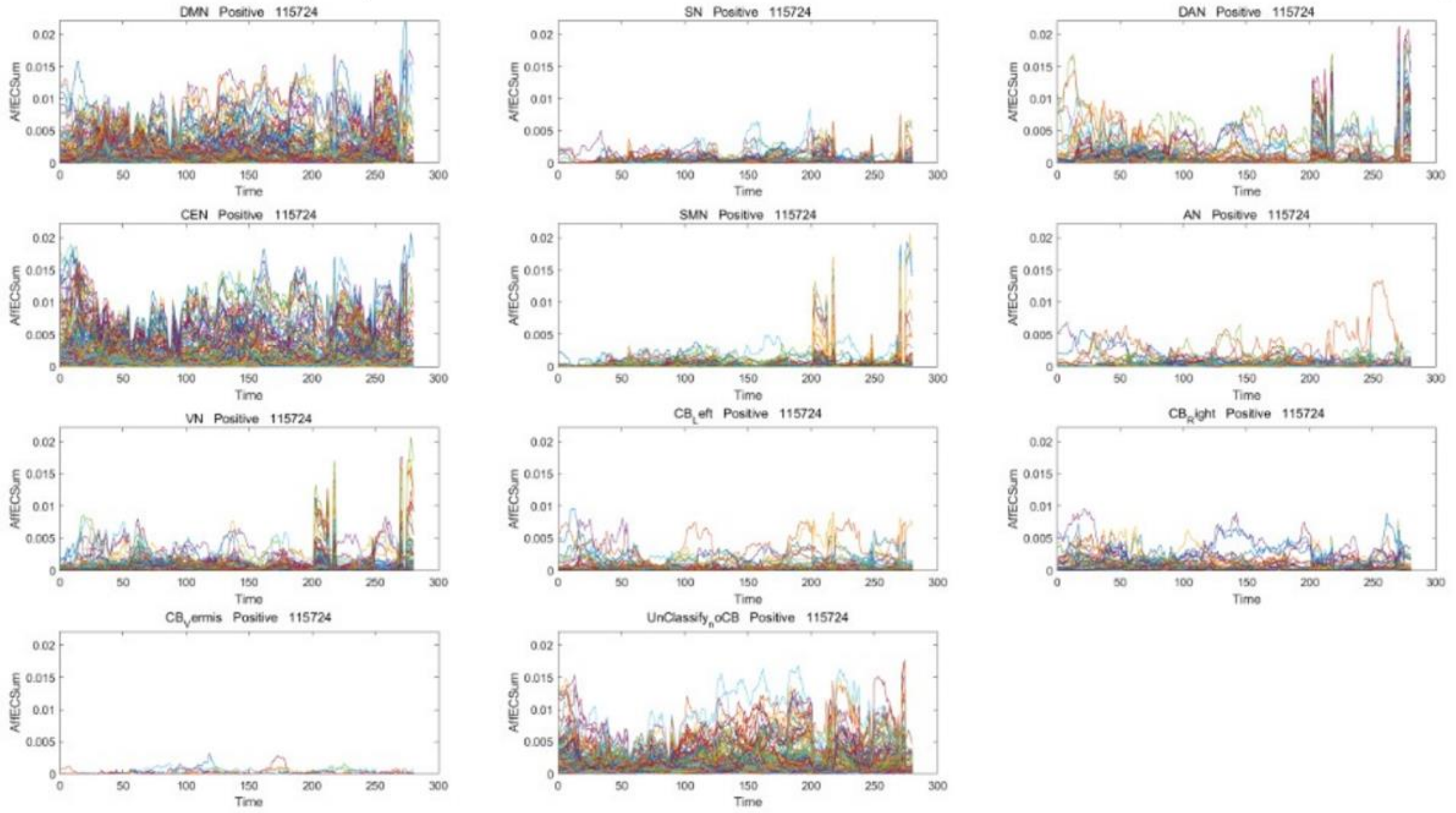

E.

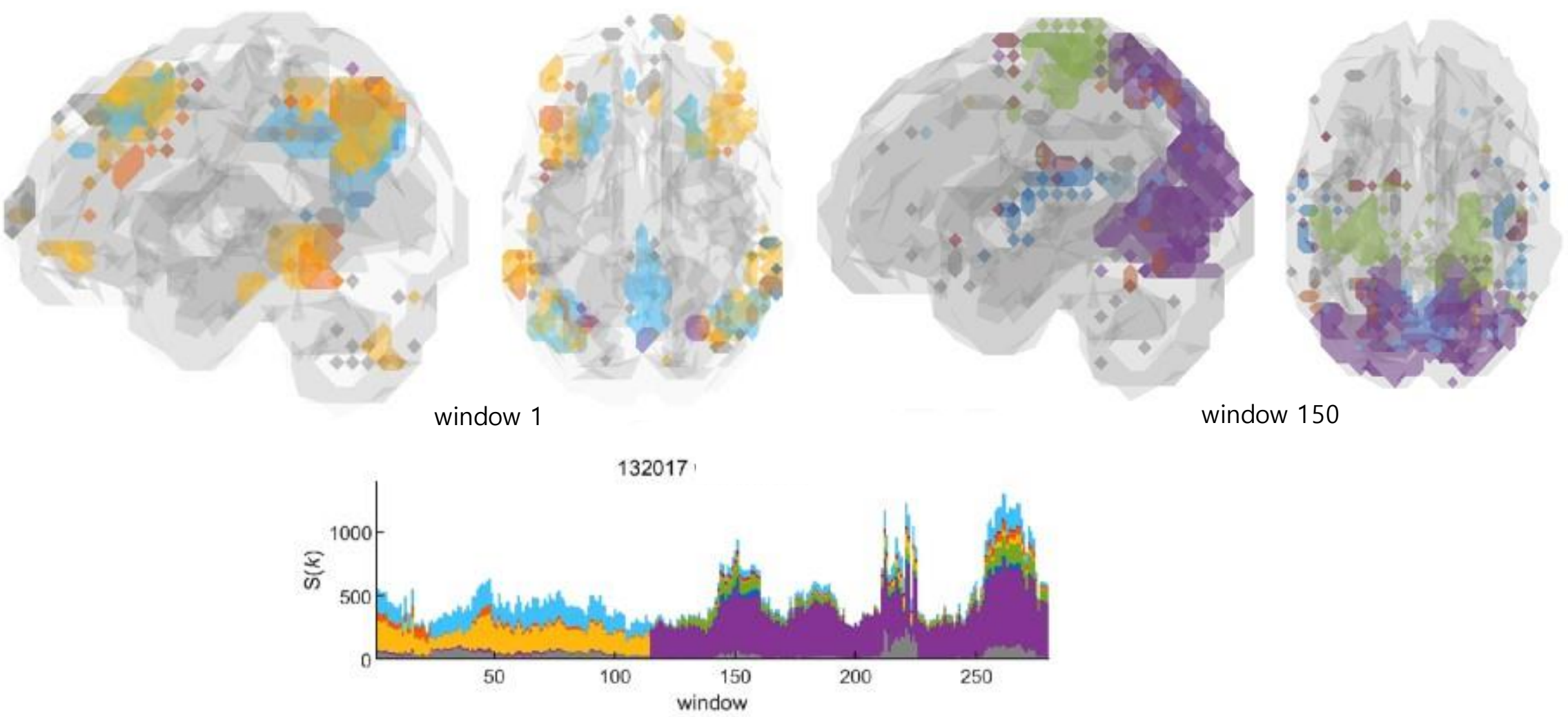

F. **Afferent node capacity** **Positive Graph**

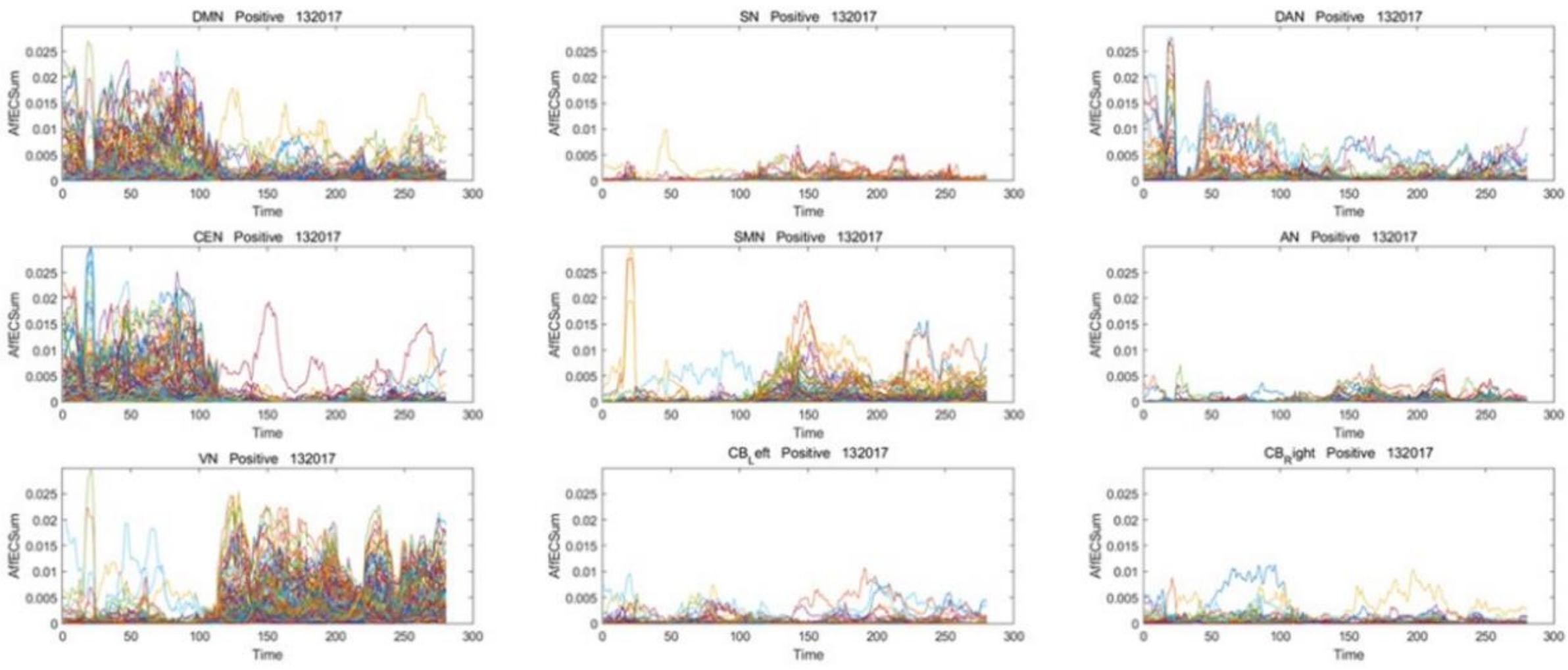

G.

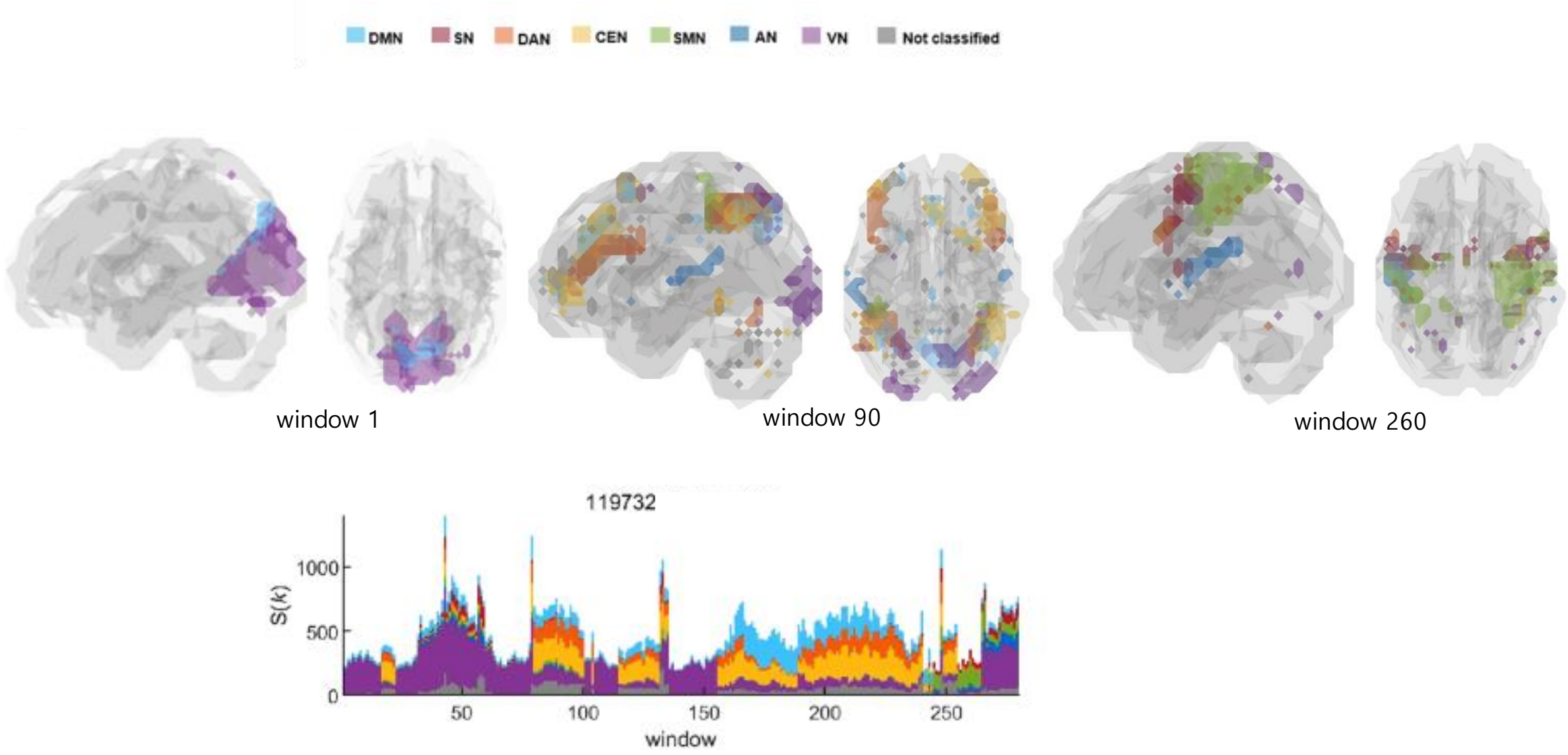

H. Afferent node capacity

Positive Graph

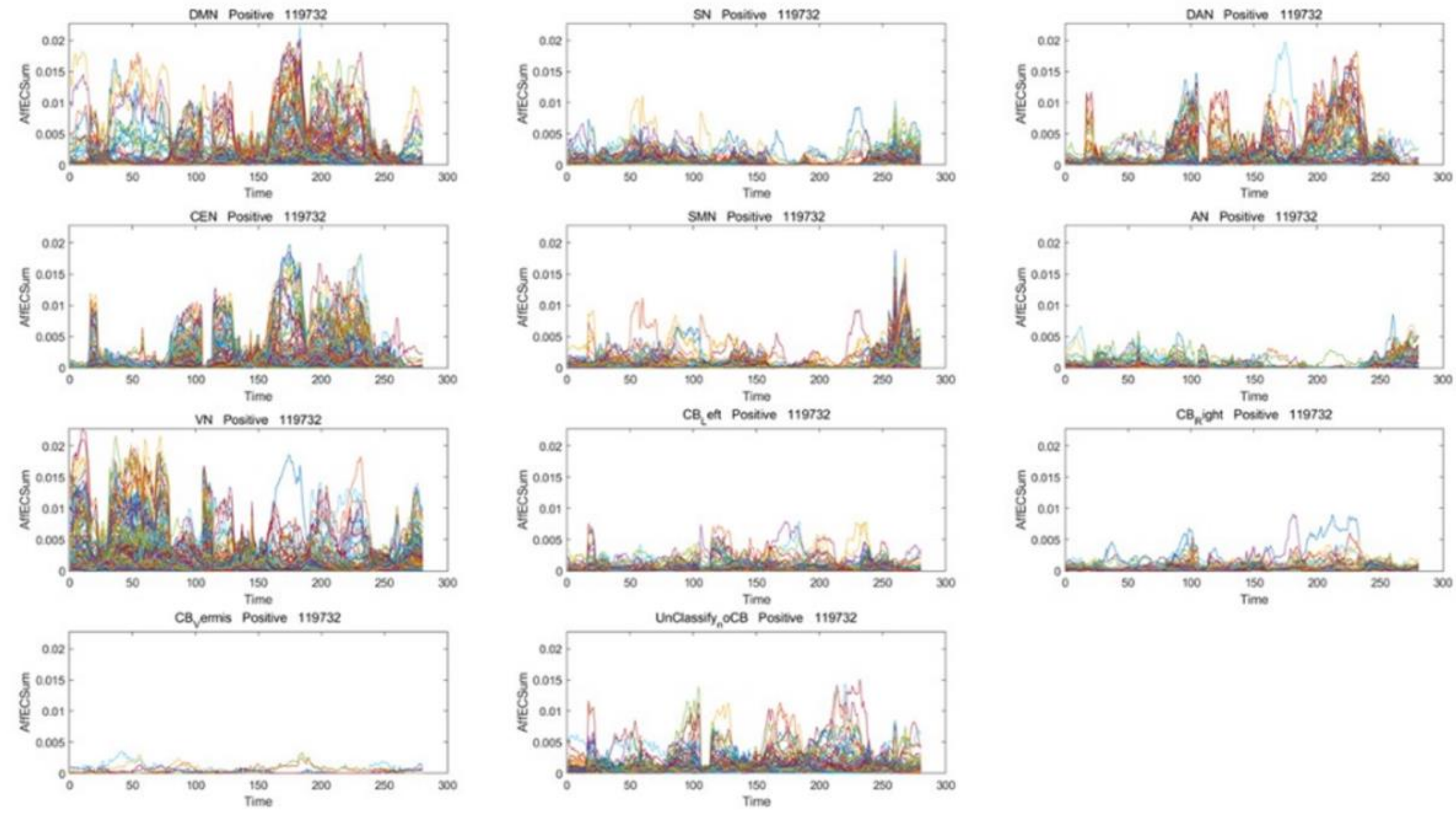

**Figure 2. Various types of states and state transitions on $k_{max}$core voxels/ICs from the DMN to the VN and modules and their exchanges on afferent capacity voxel/IC timepoint plots.** The color strips range from sky blue (DMN), red (SN), orange (DAN), yellow (CEN), green (SMN), blue (AN), violet (VN) and gray (Unclassified).

A. Glass brain image animation and stacked histogram plots along 280 time bins of the

$k_{max}$core. Voxels belonging to VN, AN, SMN, SN and DMN joined the $k_{max}$core all around the time bins. No state transition was observed in this individual, and the total number of $k_{max}$core voxels fluctuated.

B. Afferent node-voxel capacity, labeled AffECSum, was plotted along the time bins separately for ICs annotated for every voxel, such as the DMN, SN, DAN, CEN, SMN, AN, VN, L Cbl, and R Cbl. The density of modules composed of the VN, SMN, SN, AN, etc., varied over time, but there was no drop-out or exchange. A few loose threads were observed in the voxels/IC plots of the DMN, CEN, DAN and AN.

C. Glass brain images and stacked histogram along 280 time bins of the $k_{max}$core. Voxels belonging to DMN, CEN and unclassified group were the dominant and exclusive participants of the $k_{max}$core throughout the time bins. Few state transitions were observed in this individual at approximately the 205$^{th}$ and 245$^{th}$ time points.

D. On time-varying afferent node capacity voxel plots derived from positive brain graphs, voxels belonging to the DMN, CEN and unclassified network dominated with larger afferent node capacities continuously during the entire period. The VN, SMN, SN and AN splashed briefly at the end of the period.

E. In this individual, $k_{max}$core plots revealed initial DMN/CEN dominance and later VN dominance with a small SMN or DMN or others.

F. Afferent node capacity voxel plots corroborated the $k_{max}$core plots, implying that $k_{max}$cores were run by positive afferent node-voxel capacity, the sum of node-linked edge capacities coming in from other voxels, of the DMN/CEN with a small amount of DAN during the first half of the period. At approximately the 115$^{th}$ time point, the dominance of the DMN/CEN voxels abruptly abated, and the dominance of the VN voxels decreased with little help from the SMN and scantly from the DMN and DAN/CEN.

G. Glass brain images and stacked histogram of the $k_{max}$core. State transitions from VN dominance to DMN/CEN/DAN or vice versa are repeatedly shown. At the 80$^{th}$ time point, a sharp transition from sole VN dominance to DMN/DAN/CEN/VN dominance was found. A single petal bin preceded just before the transition. Between the 100$^{th}$ and 150$^{th}$ time bins, VN dominance occurred first, followed by DMN/DAN/CEN/unclassified codominance.

H. Afferent node-voxel capacity, labeled as AffECSum on the ordinate, plots. Previously developed ICA using images of 180 static individuals defined the modules of ICs. Interestingly, the DMN, DAN, and CEN modules gathered together to make module congregates of similar (but with small variations) progress along time bins. The modules were prominent for all ICs, the left and right cerebellums, and for the unclassified. Attention might be given to time bins around the $100^{th}$ one, which showed a razor-sharp transition from the DMN/DAN/CEN comodules to the VN module and back to the DMN comodules. This was a clear representation of the effect of the $k_{max}$core 'state transition' on afferent node capacity.

IC: independent components, DMN: default mode network, SN: salience network, DAN: dorsal attention network, CEN: central executive network, SMN: sensorimotor network, AN: auditory network, VN: visual network, Cbl: cerebellum, L: left, R: right.

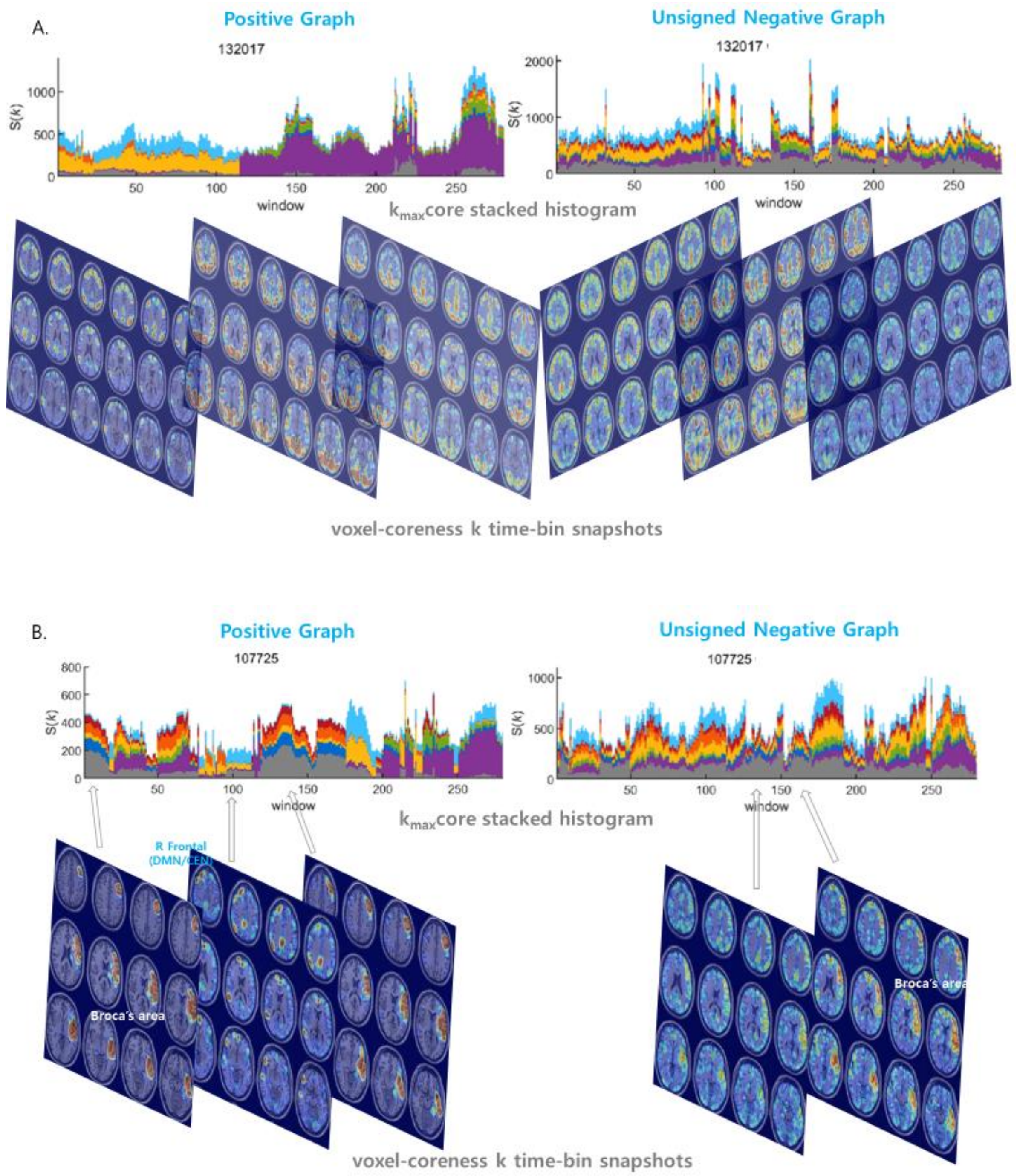

**Figure 3. Hierarchically top tier voxels on stacked histogram timepoint plots of $k_{max}$core voxels or hierarchically colored MRI-overlayed animation plots.** Animation images were captured via snapshots for visualization.

A. In this individual presented in the above Figure 2E, who showed initial DMN/CEN dominance and later VN dominance in positive graph (left side of this picture), in the unsigned negative graph on the right side, a sustained state of voxels belonging to most ICs was observed across all the time bins. This case was representative for the negative graphs of all individuals, in whom the edge number scaled (abb. edge-scaled) coreness k images showed the characterless flickering pattern of ripples.

B. In this individual, the positive graph presented a typical state transition on the stacked histogram timepoint plots of $k_{max}$core voxels but extraordinarily showed the top tier of the left frontal cortex (Broca's area) in the first half of the time-bins' progress, intervened by right frontal lobe leveling up (arrows) on the coreness k animation plots. However, the negative graph did not show any abrupt state transition but only fluctuations in the stacked histogram of $k_{max}$core. At later timepoints (125$^{th}$ to 170$^{th}$), the left frontal prominence of the negative graph accompanied the prominent left frontal area of the positive graph (120$^{th}$ to 180$^{th}$).

IC: independent components, DMN: default mode network, CEN: central executive network, VN: visual network.

A.

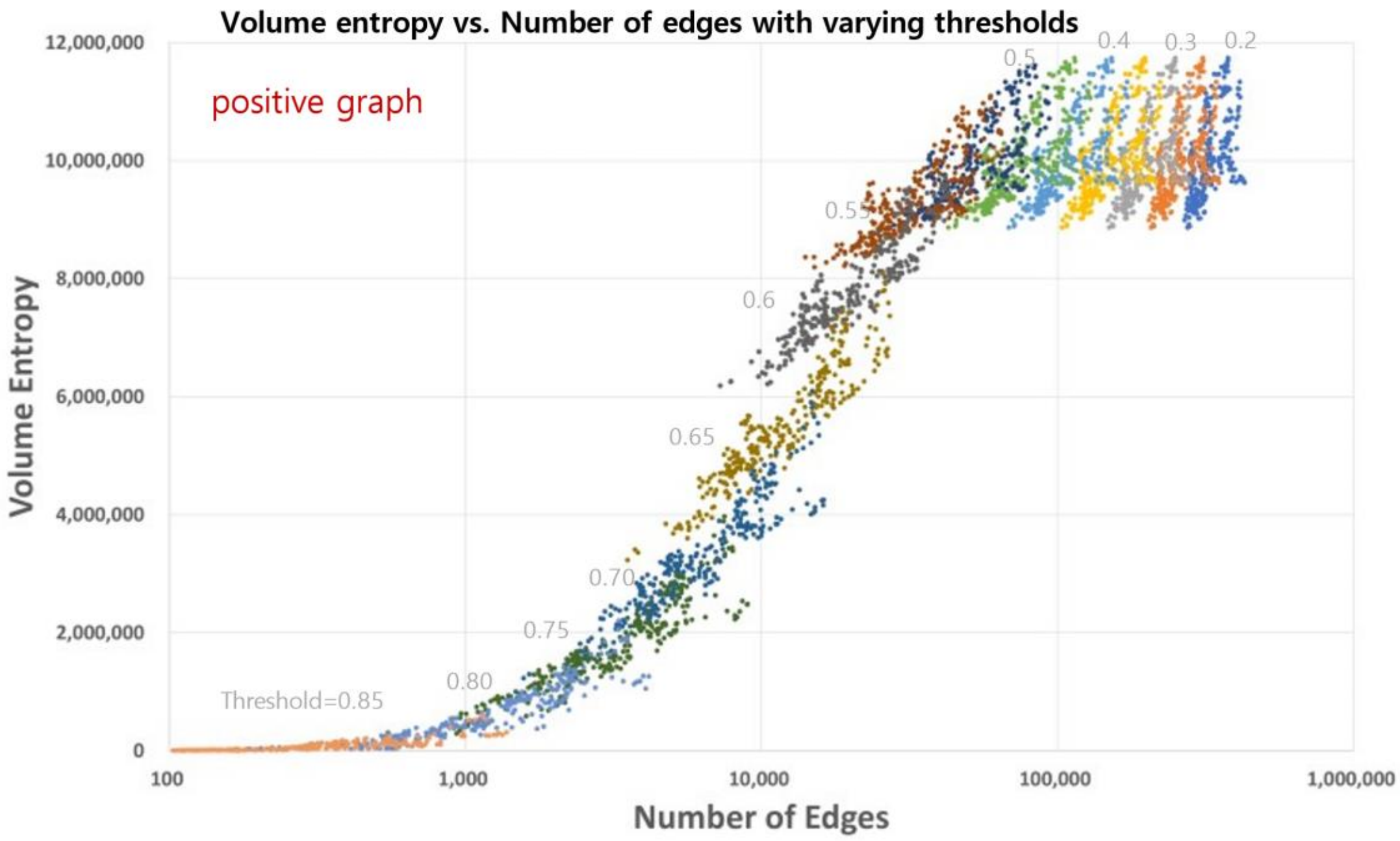

Volume entropy vs. Number of edges with varying thresholds
positive graph
Volume Entropy
Number of Edges
0
2,000,000
4,000,000
6,000,000
8,000,000
10,000,000
12,000,000
100
1,000
10,000
100,000
1,000,000
Threshold=0.85
0.80
0.75
0.70
0.65
0.6
0.55
0.5
0.4
0.3
0.2

B.

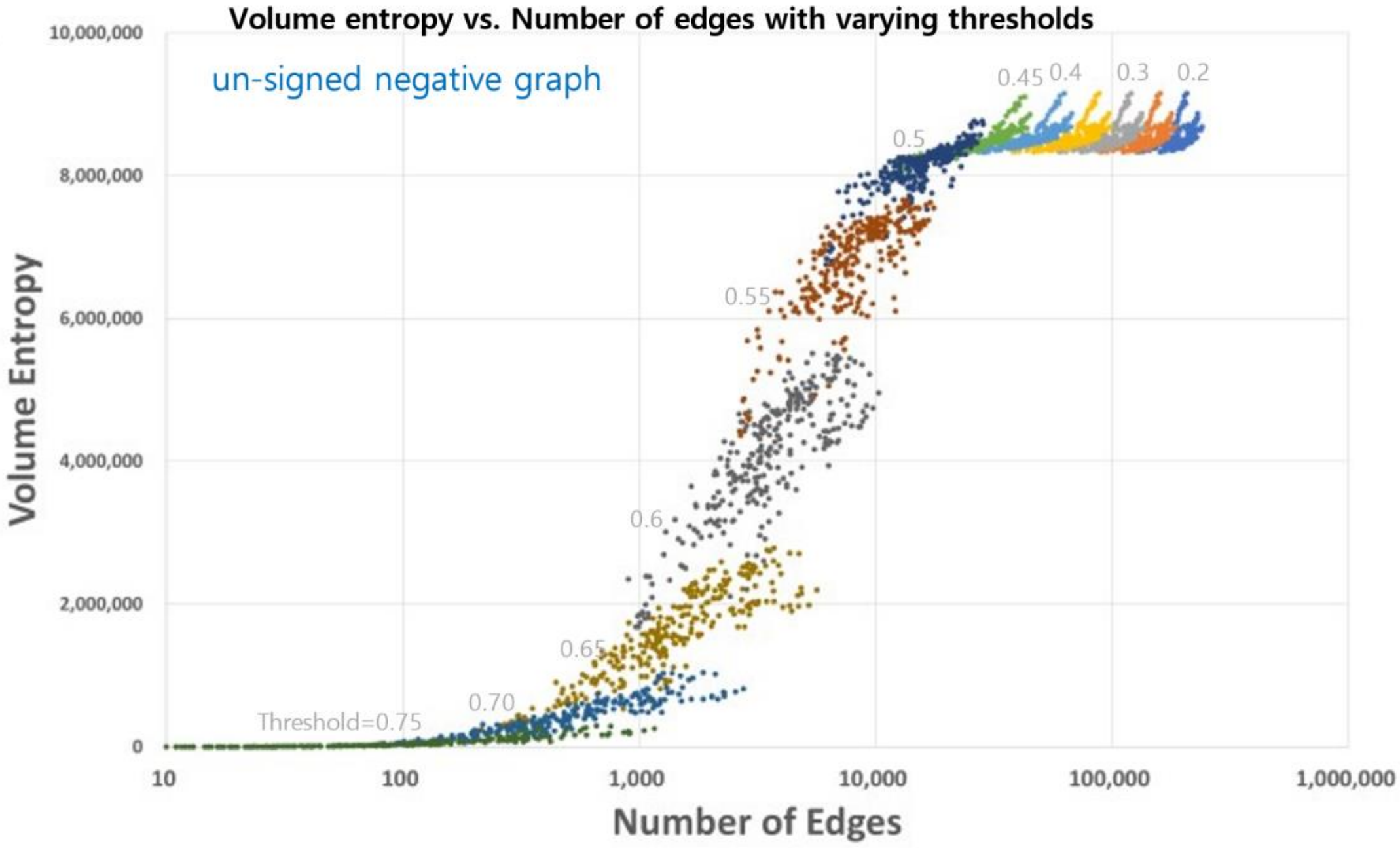

Volume entropy vs. Number of edges with varying thresholds
un-signed negative graph
Volume Entropy
Number of Edges
0
2,000,000
4,000,000
6,000,000
8,000,000
10,000,000
10
100
1,000
10,000
100,000
1,000,000
Threshold=0.75
0.70
0.65
0.6
0.55
0.5
0.45
0.4
0.3
0.2

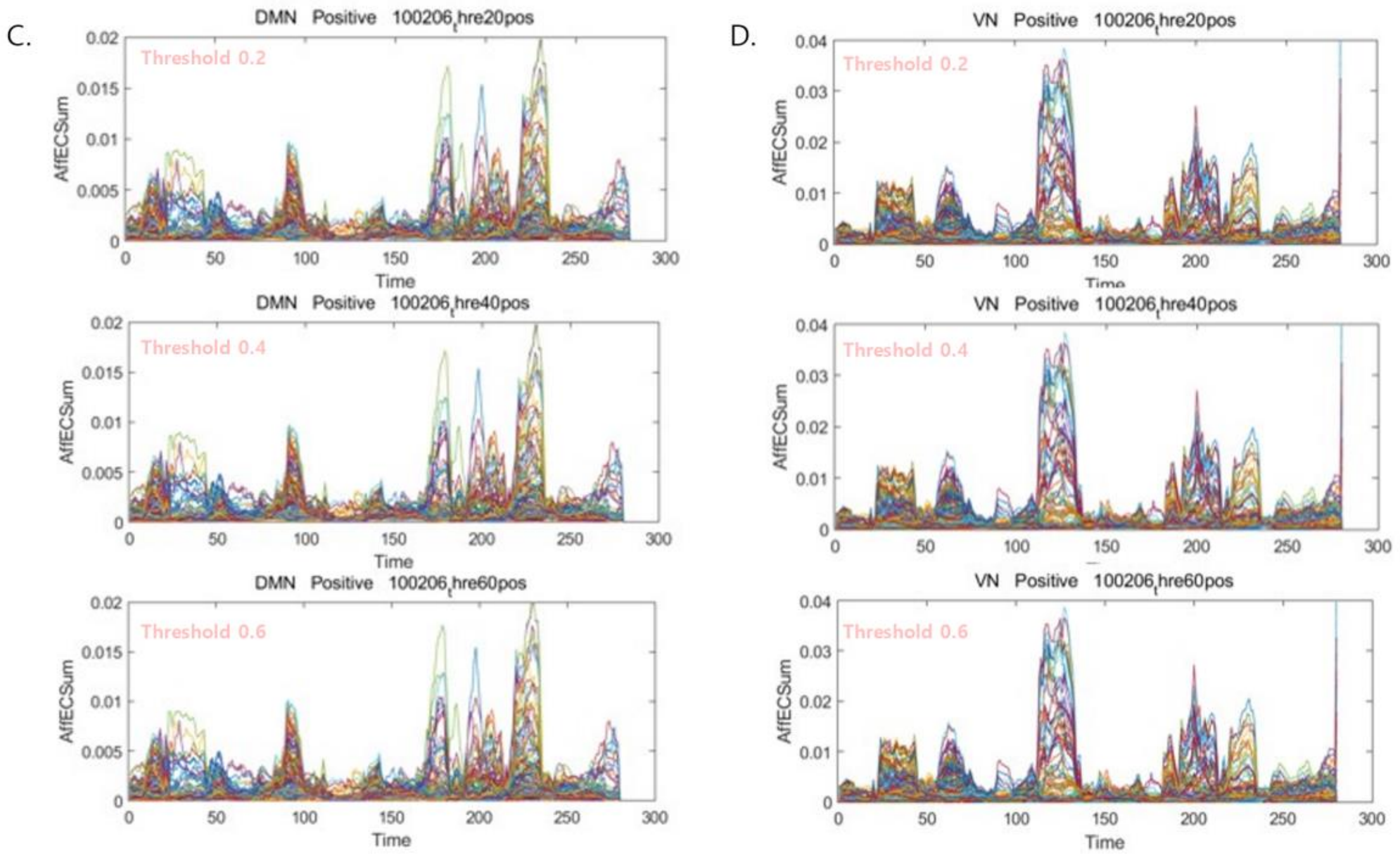

**Figure 4. The volume entropy values and modules and exchange patterns of the voxel/IC composition were exactly the same for the graphs with surplus edges.**

A. In positive graphs with thresholds of 0.2, 0.25, 0.3, 0.35, 0.4, 0.45, 0.5, 0.55, 0.6, 0.65, 0.7, 0.75, 0.8, and 0.85, the volume entropy was calculated separately per threshold (n=14) for each time bin (n=280). The abscissas of the total number of edges were plotted logarithmically because the total number of edges ranged from thousands to millions. In these positive graphs, with thresholds equal to or lower than 0.45, the volume entropy was the same between time bins and between thresholds irrespective of the thresholds and the corresponding numbers of edges. The volume entropy of positive graphs ranged from 9 to 12 million between time bins in graphs with thresholds of 0.45 to 0.2, which was larger than the range of volume entropy of negative graphs with similar thresholds.

B. In the negative graphs, the pattern was similar, and once the number of edges decreased to less than 8 million, the volume entropy decreased dramatically and proportionally with the number of edges. The volume entropy of negative graphs with thresholds of 0.45 to 0.2 ranged from 8.5 to 9 million.

C. Examples of voxels belonging to the DMN or to the VN showing no difference between the positive graphs with varying thresholds from 0.2 to 0.6. Notably, the contour of the formed modules as well as the trajectories inside and the maximum height of the modules were exactly the same. We could be sure that the modules and their exchange did not depend on or vary with the choice of thresholds according to the criteria of Suppl. Fig. 3; at least the number of nodes preserved was greater than the set point (85% in this study).

IC: independent components, DMN: default mode network, VN: visual network.

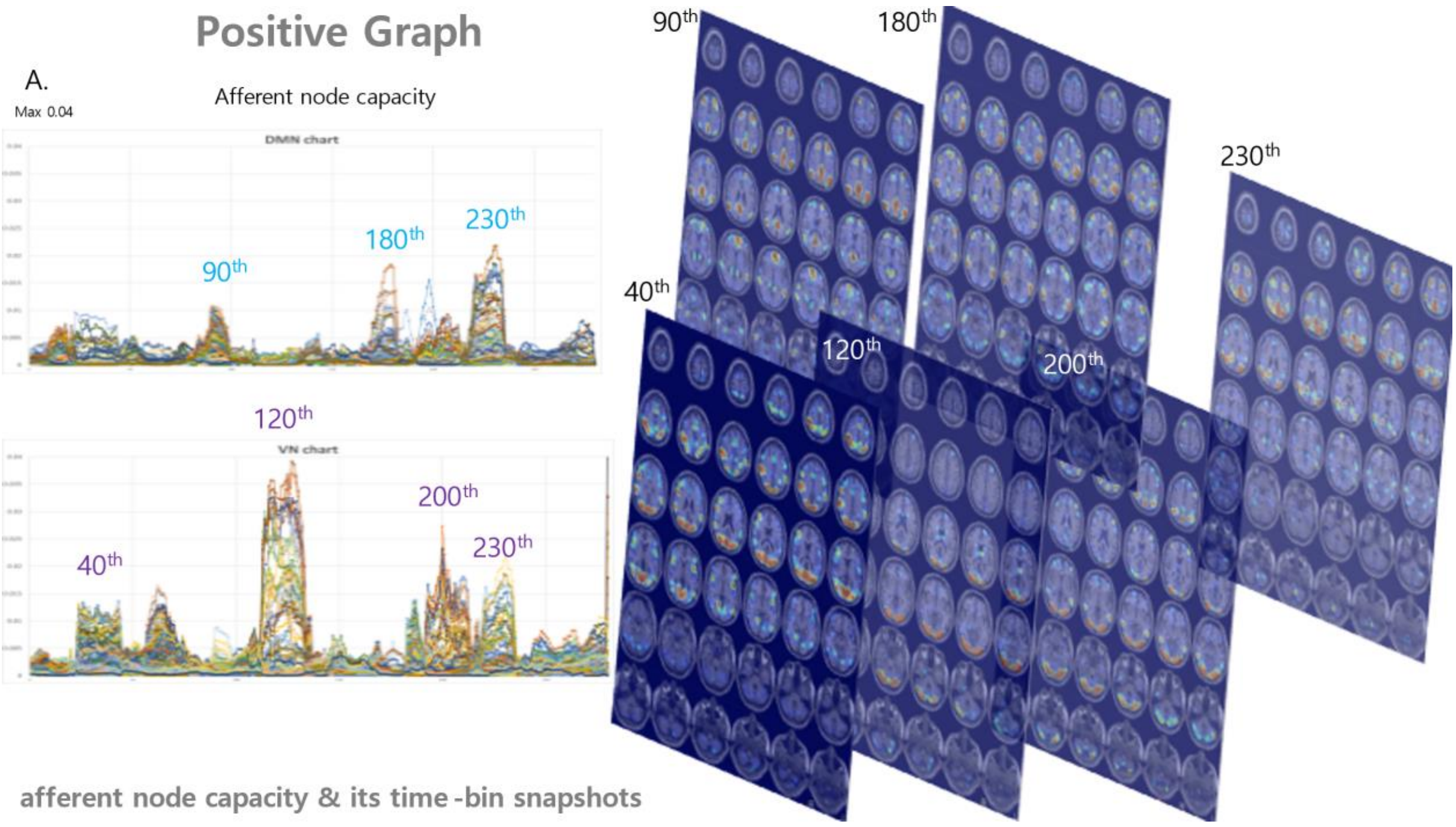
Positive Graph
A.
Max 0.04
Afferent node capacity
DMN chart
90th
180th
230th
VN chart
120th
40th
200th
230th
90th
180th
230th
40th
120th
200th
afferent node capacity & its time -bin snapshots

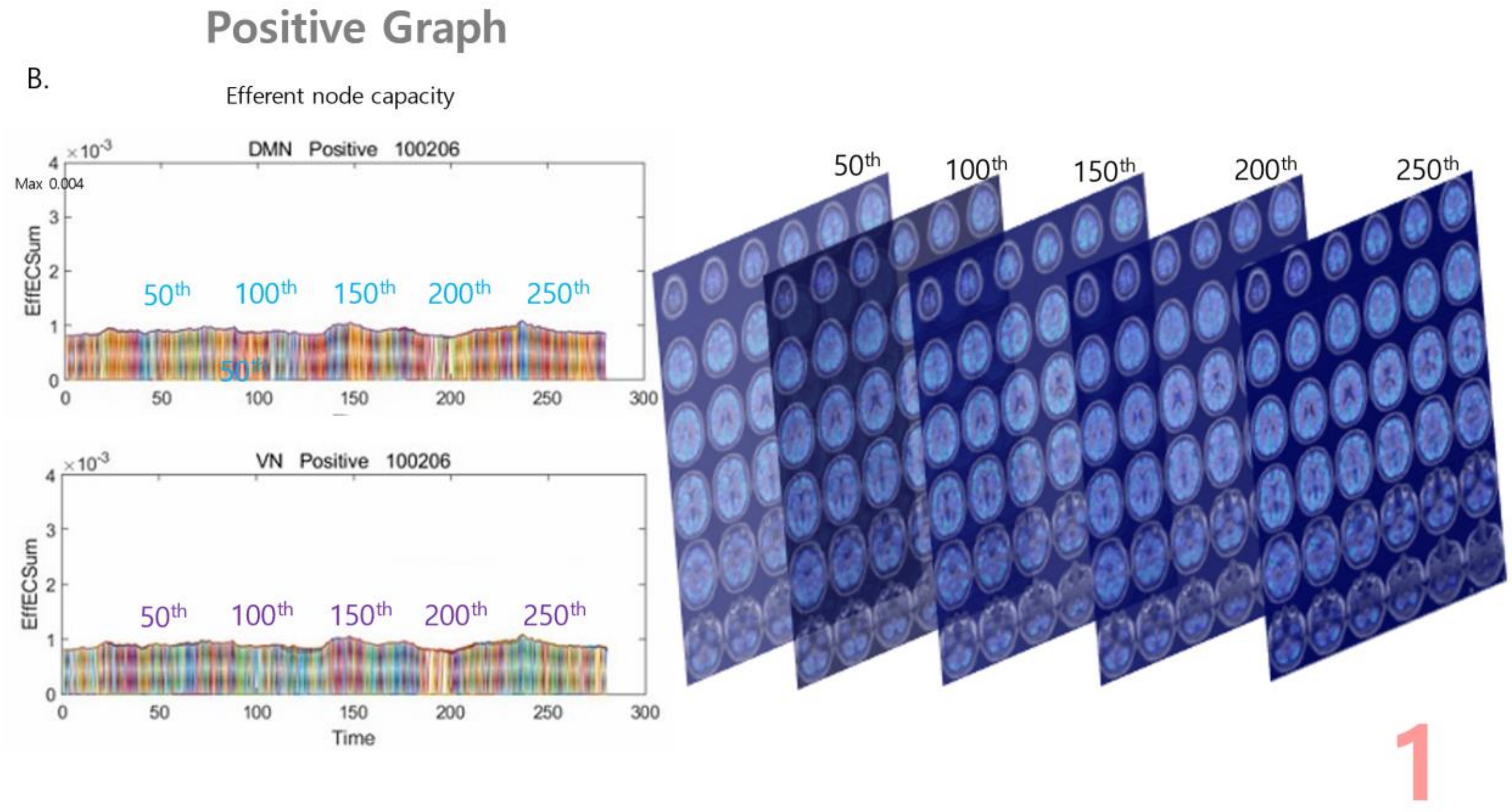

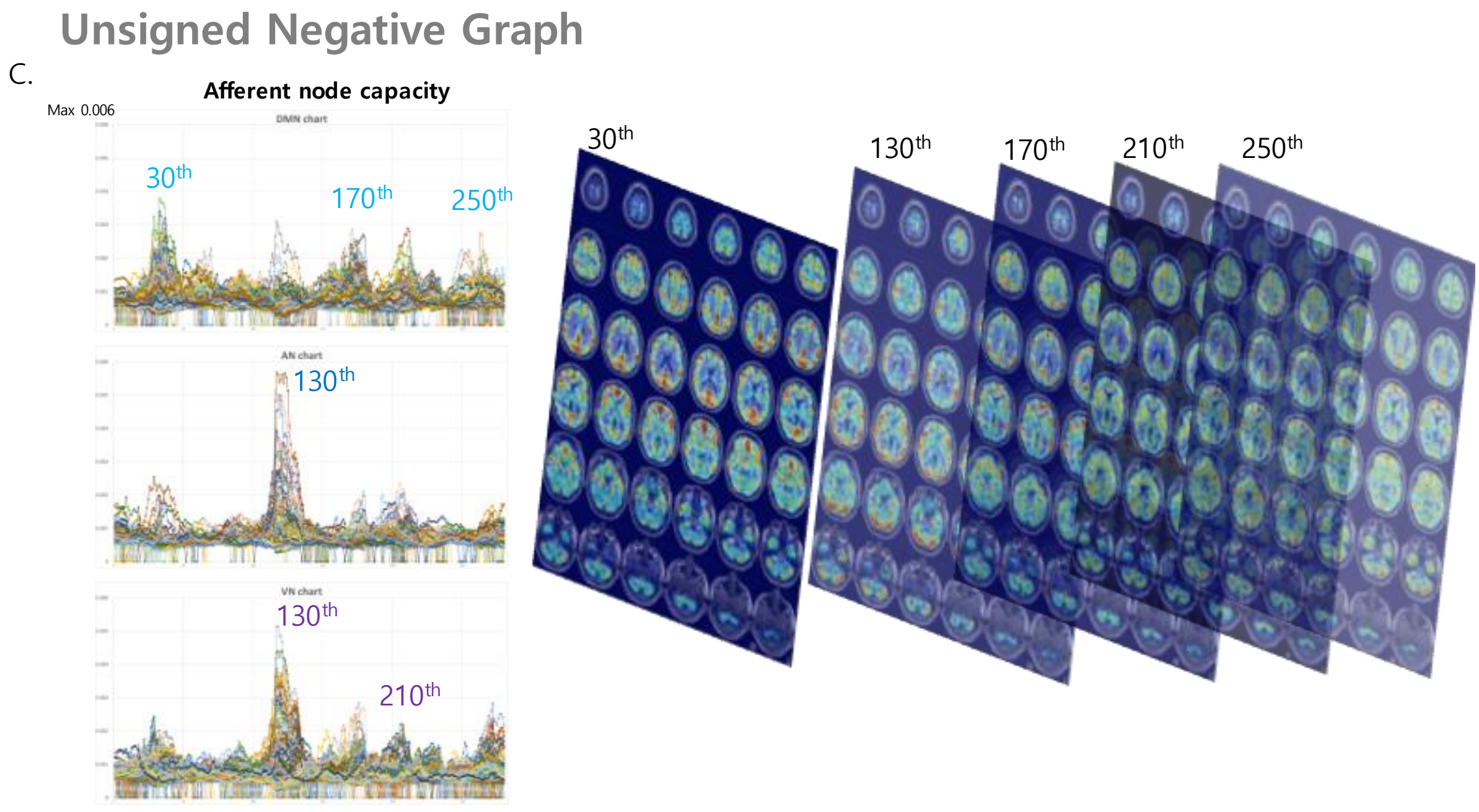

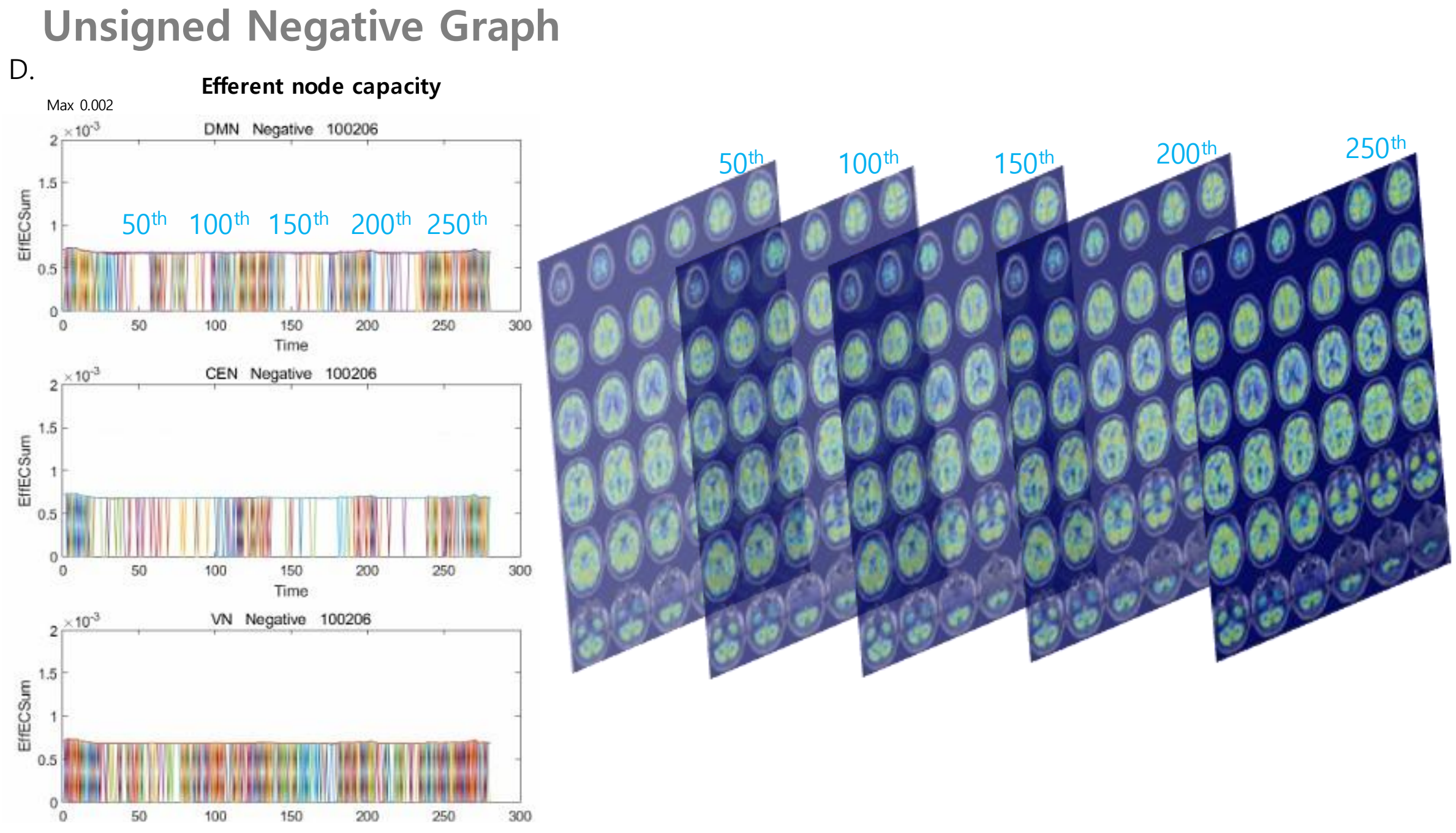

**Figure 5. Differences between positive and negative graphs of afferent and efferent voxel capacity timepoint plots were common on the directed graphs of all the individuals.** Afferent node capacity showed characteristic patterns on both timepoint plots and MRI-overlayed map animations. Afferent capacity was greater than efferent capacity for voxels in general and for voxels/ICs. Afferent capacity of positive graph was greater than that

of the negative graph.

A. Time-bin timepoint plot of voxels/DMN and voxels/VN of afferent node capacity of positive graphs. Module formation and switching were visualized with MRI-overlayed animation plots (shown here with snapshots of the 40$^{th}$ to 230$^{th}$ time bins) for afferent node capacity. The maximum height of the module was 0.04, which was 40 times greater than the 0.001 for the efferent node capacity.

B. Timepoint plots of efferent node capacity were homogeneous, which was also well observed in the animation plots with monotonous snapshots. This characterlessness was common in all the individuals.

C. Timepoint plot for the afferent node capacity of unsigned negative graphs. The contour and trajectory of the voxels/IC (DMN, AN, VN) looked similar to those of the positive graphs. However, the height of the module was only 1/7 that of the positive graphs. Nevertheless, the 0.006 afferent node capacity in the negative graph was almost 9 times the efferent node capacity (0.0007) in the negative graph.

D. Monotonous and smaller efferent node capacities of negative graphs are shown.

DMN: default mode network, CEN: central executive network, AN: auditory network, VN: visual network.

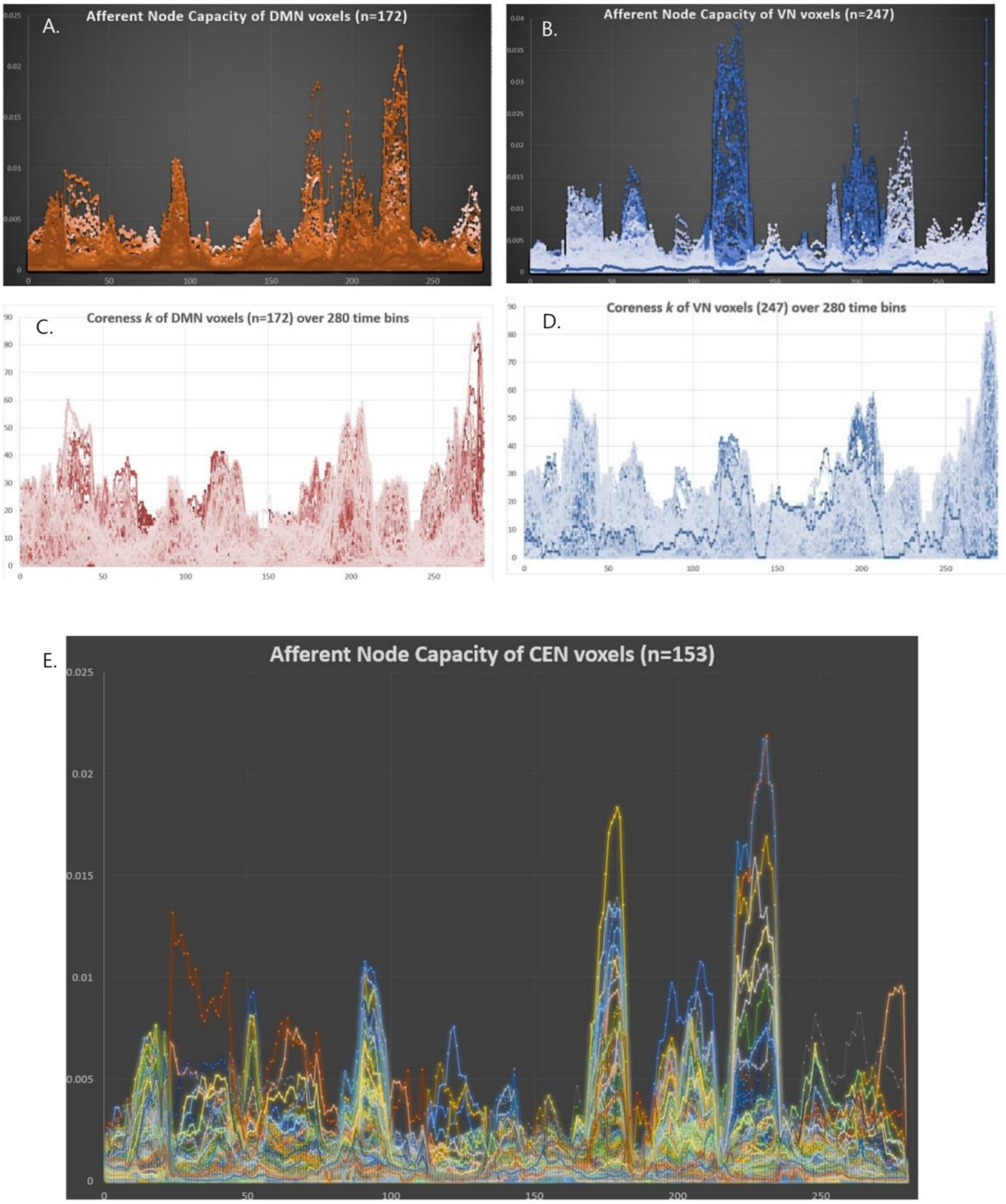
A.
Afferent Node Capacity of DMN voxels (n=172)
B.
Afferent Node Capacity of VN voxels (n=247)
C.
Coreness k of DMN voxels (n=172) over 280 time bins
D.
Coreness k of VN voxels (247) over 280 time bins
E.
Afferent Node Capacity of CEN voxels (n=153)
0.025
0.02
0.015
0.01
0.005
0
0
50
100
150
200
250

F.

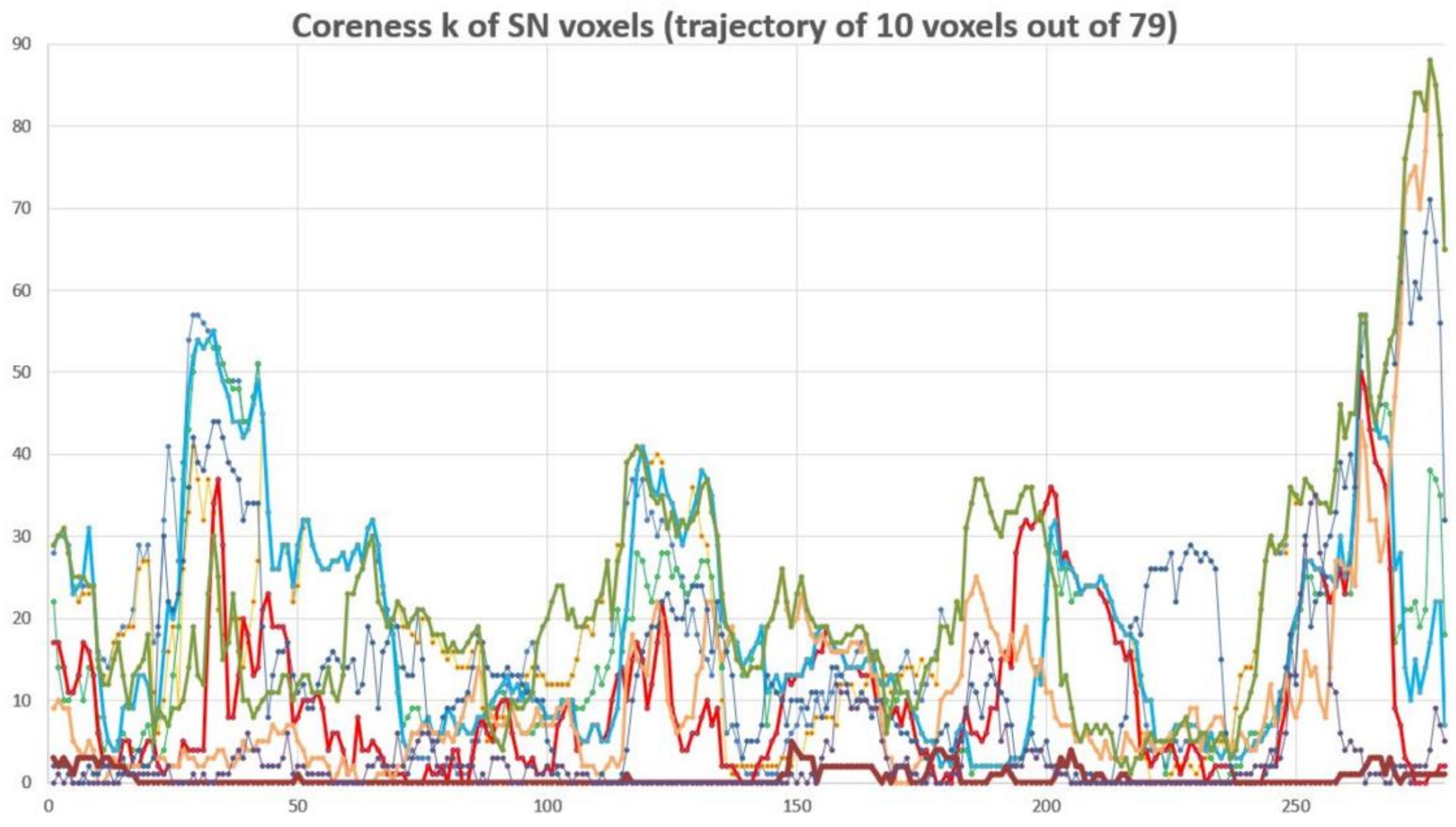

**Figure 6. The voxel trajectories showed extreme heterogeneity in terms of the afferent node capacity and edge-scaled coreness k timepoint plots.**

A. A total of 172 voxels of the DMN were traced along the 280 time bins. The plot revealed that each voxel took its own trajectory, meaning that it contributed once to form a module and then at the next module, either idled or contributed, and then at the third module, unexpectedly behaved, etc. I also revealed that despite these behaviors of individual voxels, an IC in this case, the DMN made a plausible module to be named a characteristic IC with the support of the voxels' unpredictable trajectories.

B. The behavior of the 247 trajectories of VN voxels was similar to those of the DMN voxels. VN voxels seemed to consist of two clusters to make modules; i.e., if we labeled 1 to 5 for medium- to large-sized modules from the start, whitish modules of 1 and 5, bluish modules 3 and 4, and mixture, the two clusters might have been dissociated as separate ICs beforehand.

C and D. Coreness k (Edge-nonscaled) showed the same contour for the DMN and VN. This was because the total number of edges per time bin was the major determinant of total coreness k. The behavior of the trajectories of each voxel, either the DMN or VN, followed its own fate to take any or no route of contribution precariously to form ICs but collectively made the rise and fall of each peak.

E. A total of 153 CEN voxels were followed for their trajectories of afferent node capacity. One can easily follow the luxury of heterogeneity of each voxel.

F. Trajectory images with entire voxels/IC would have made resolving impossible; thus, we chose 10 random trajectories of the SN (79 voxels). Timepoint plots of these 10 voxels showed the heterogeneity of trajectories, making it unbelievable that they belonged to the same IC or SN.

IC: independent components, DMN: default mode network, VN: visual network, CEN: central executive network, SN: salience network.

**Supplementary Materials**

List of Supplementary Figures

Supplementary Figure 11. Asymmetry of module composition of states in three patients showing frontal alternating (A, B), recurrently appearing in left frontal area (C, D), and left cerebellar asymmetry (E, F) patterns.

Supplementary Figure 12. Scheme to estimate volume entropy and afferent/efferent node capacity by making directed weighted graphs from the observed pairwise intervoxel undirected amplitude correlations after thresholding separately on positive and unsigned negative graphs.

Supplementary Figure 13. Volume entropy (A) and afferent node capacity (B) with timepoint plots and their corresponding coreness k (C) and $k_{max}$core timepoint plots (D).

Supplementary Figure 14. The relationships between the volume entropy and total coreness k of the time bins of the positive and negative graphs of the individuals are presented in Suppl. Fig. 13.

Supplementary Figure 15. Example of voxels/IC composition timepoint plots and their afferent and efferent node capacity animation maps of positive graphs in an individual (#128632).

Supplementary Figure 16. Back-to-back representation of afferent node capacity of voxels/IC timepoint plots of unsigned negative graphs and their corresponding positive graphs in representative individuals, followed by their matching $k_{max}$core stacked histogram timepoint plots.

Supplementary Figure 17. Trajectory tracing of a voxel along the time-bin progress of its own coreness k values and afferent node capacity.

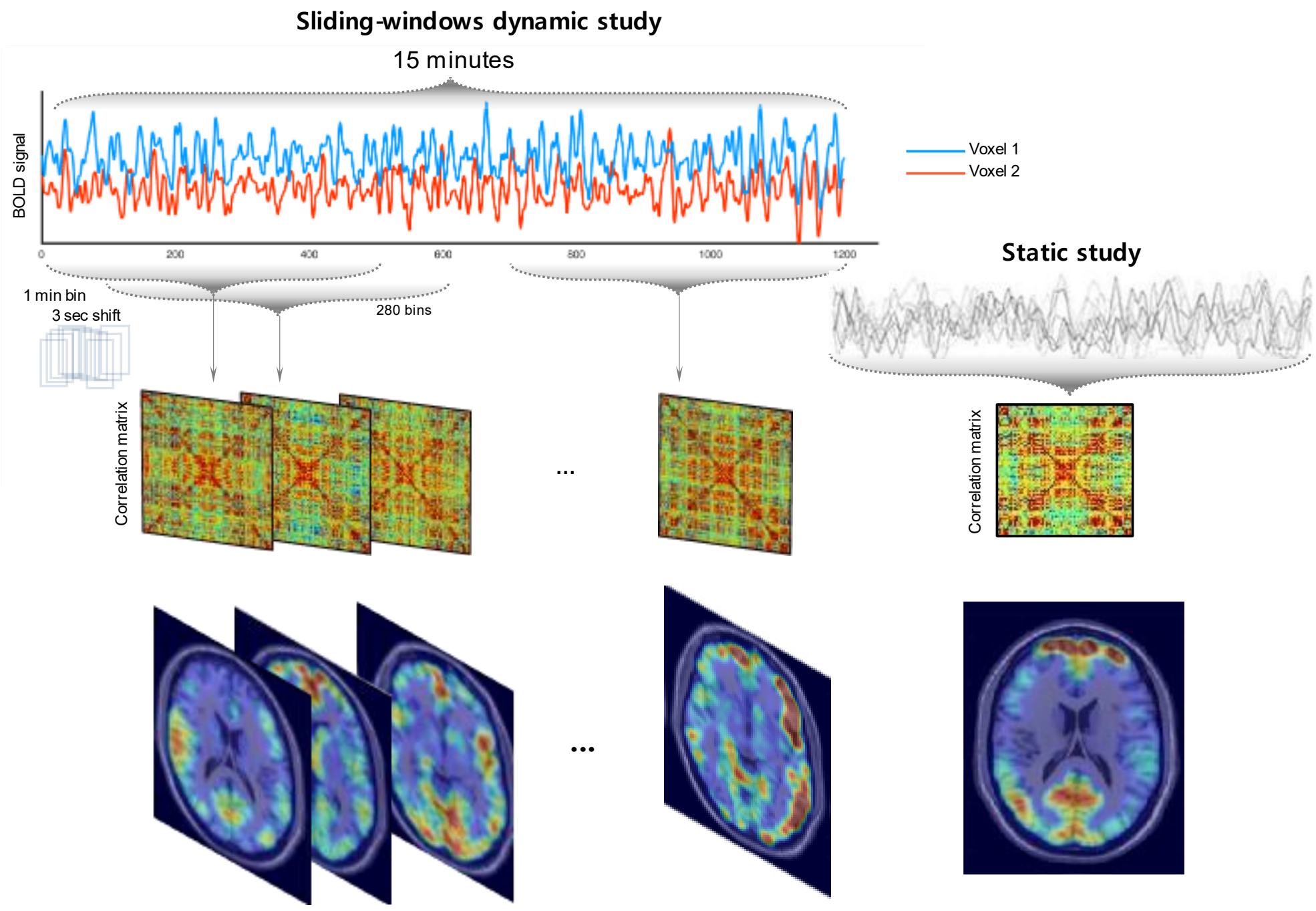

**Supplementary Figure 1. Sliding window representation of resting-state fMRI and its spatiotemporal visualization in matrix and MRI-overlayed animation plots.** Resting-state fMRI acquired 16 minutes, 0.72 seconds for each frame were converted to 1 minute time bin, 3 seconds of time-bin shifts and then 280 time-bins of timepoints were obtained for HCP cohort. For 180 individuals from Human Connectome Project, from 1,200 frames for 2 x 2 x 2 $mm^3$ fMRI images, 280 time-bins were derived and the down-sampled total number of voxels were 5,937 voxels for 6 x 6 x 6 $mm^3$ and 1,489 voxels for 10 x 10 x 10 $mm^3$. The former was used for k core percolation, and the latter was used for directed graph construction. Brain graphs consisted to 280 time-bin images of 5,937 x 5,937 (or 1,489 x 1,489) matrices or 280 time-bin images of 36-slices brain MRI-overlayed coreness k map plots or afferent/efferent node capacity map plots. For both images, final output in the form of MRI-overlayed brain images were in avi files and best viewed with animation play software of any kind.

A.

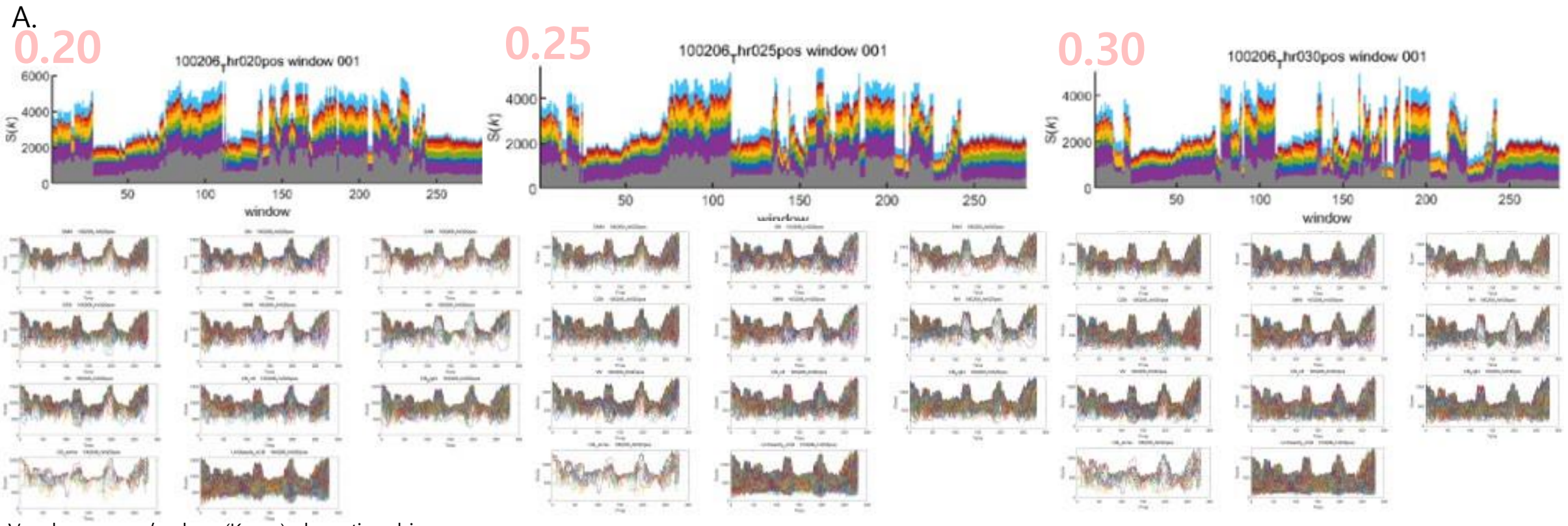

Voxel coreness *k* values (Kcore) along time-bin progress

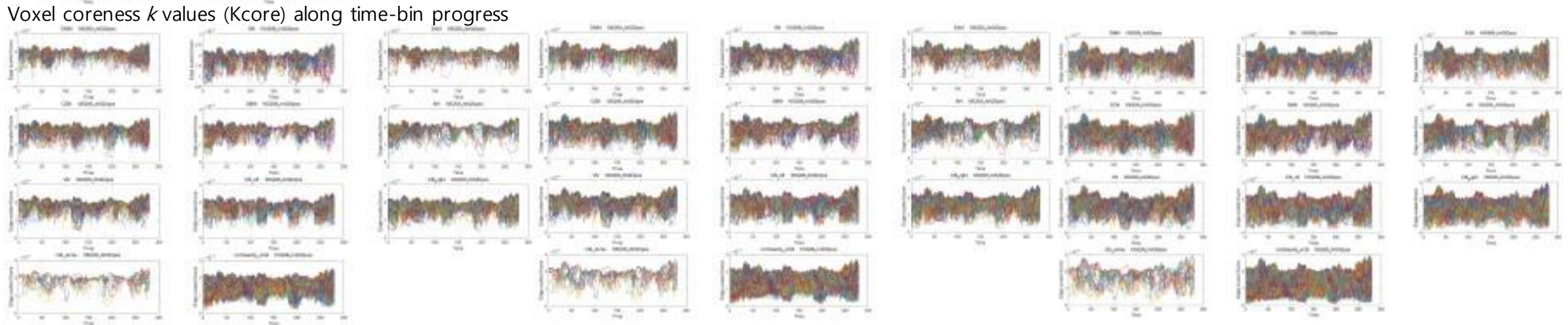

Time-bin graph-total number of edges- (Edge-) scaled voxel *K*core

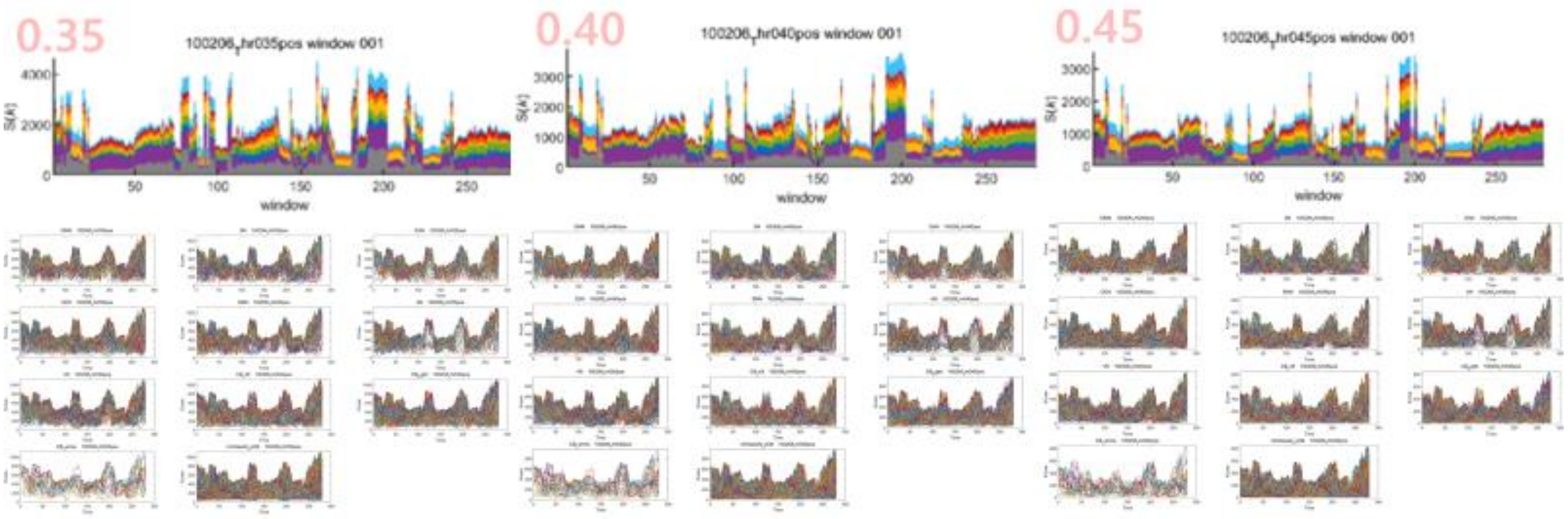

Voxel coreness *k* values (Kcore) along time-bin progress

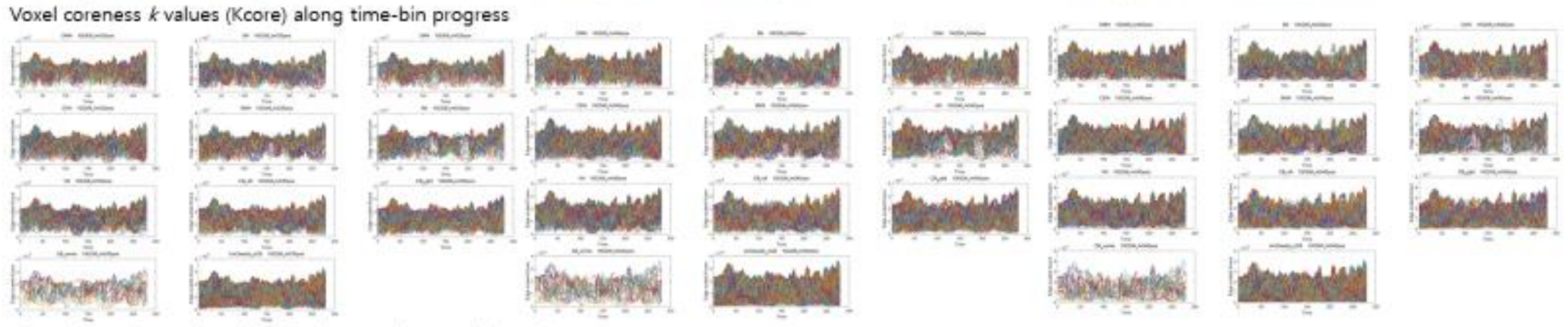

Time-bin graph-total number of edges- (Edge-) scaled voxel *K*core

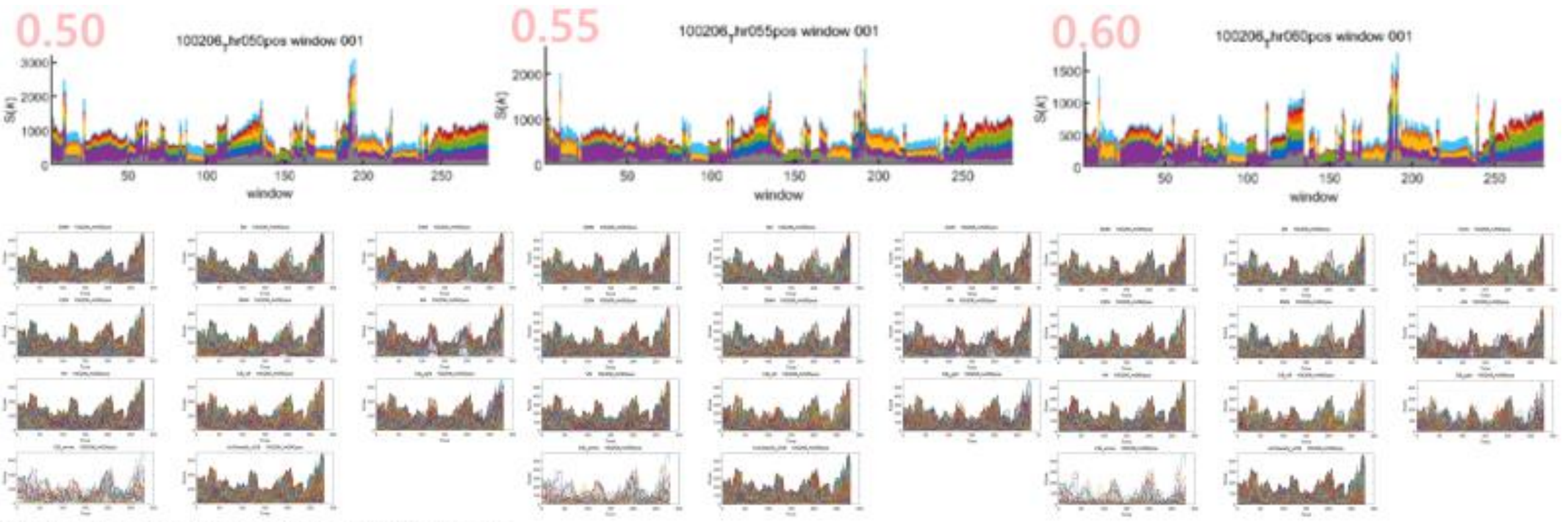

Voxel coreness $k$ values (Kcore) along time-bin progress

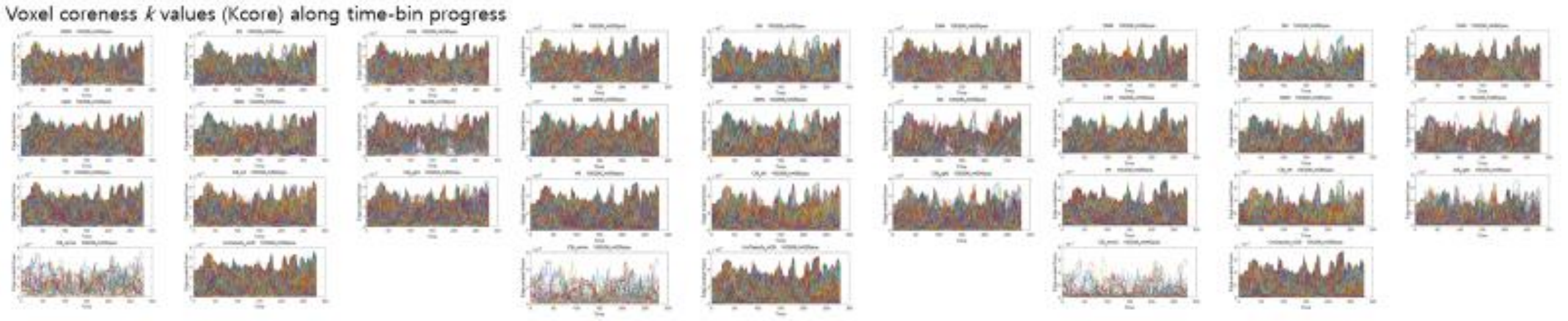

Time-bin graph-total number of edges- (Edge-) scaled voxel *K*core

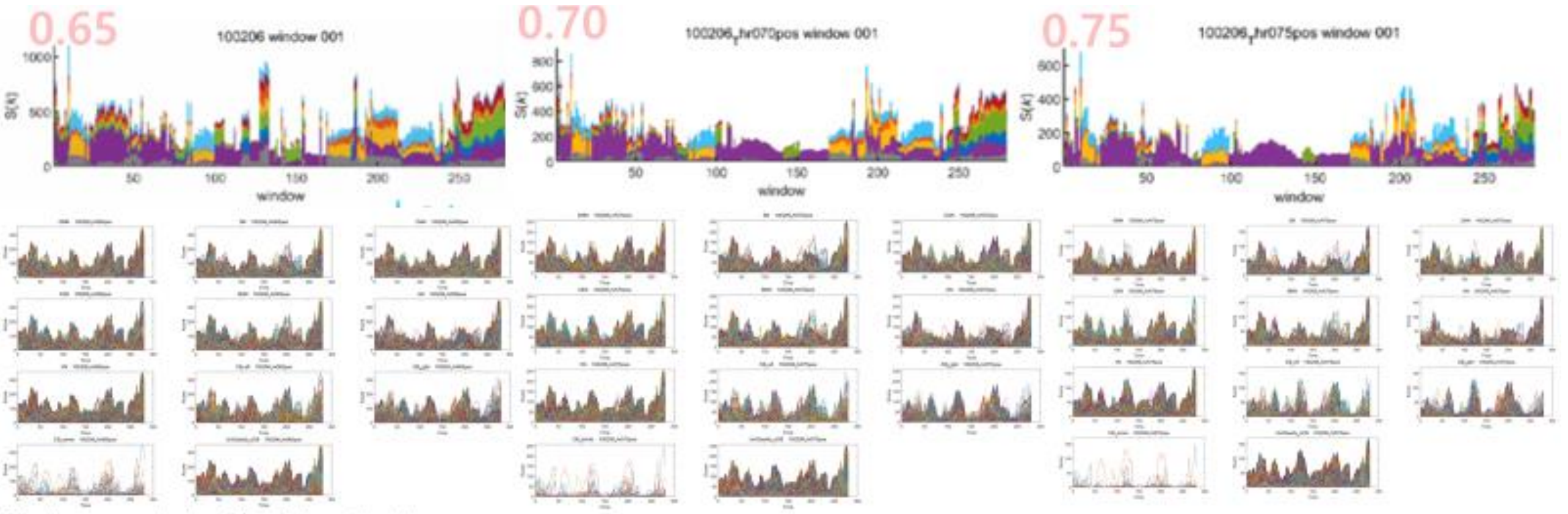

Voxel coreness $k$ values (Kcore) along time-bin progress

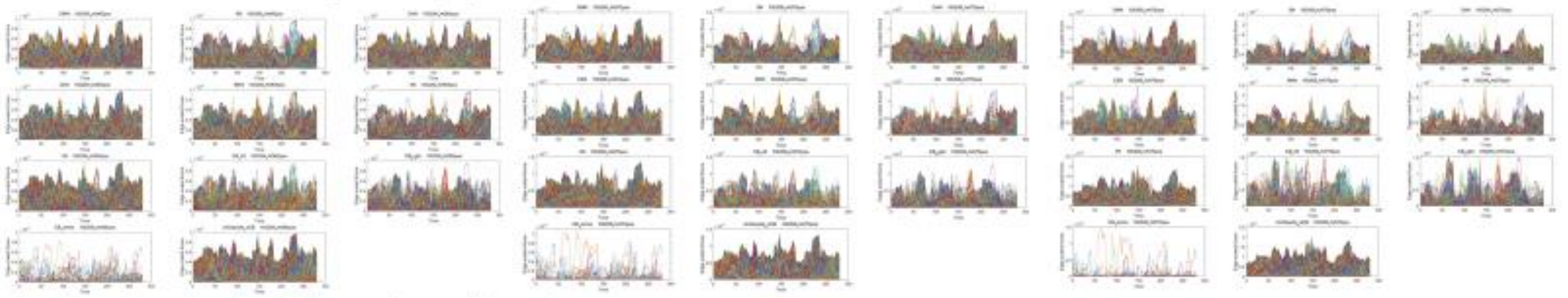

Time-bin graph-total number of edges- (Edge-) scaled voxel *K*core

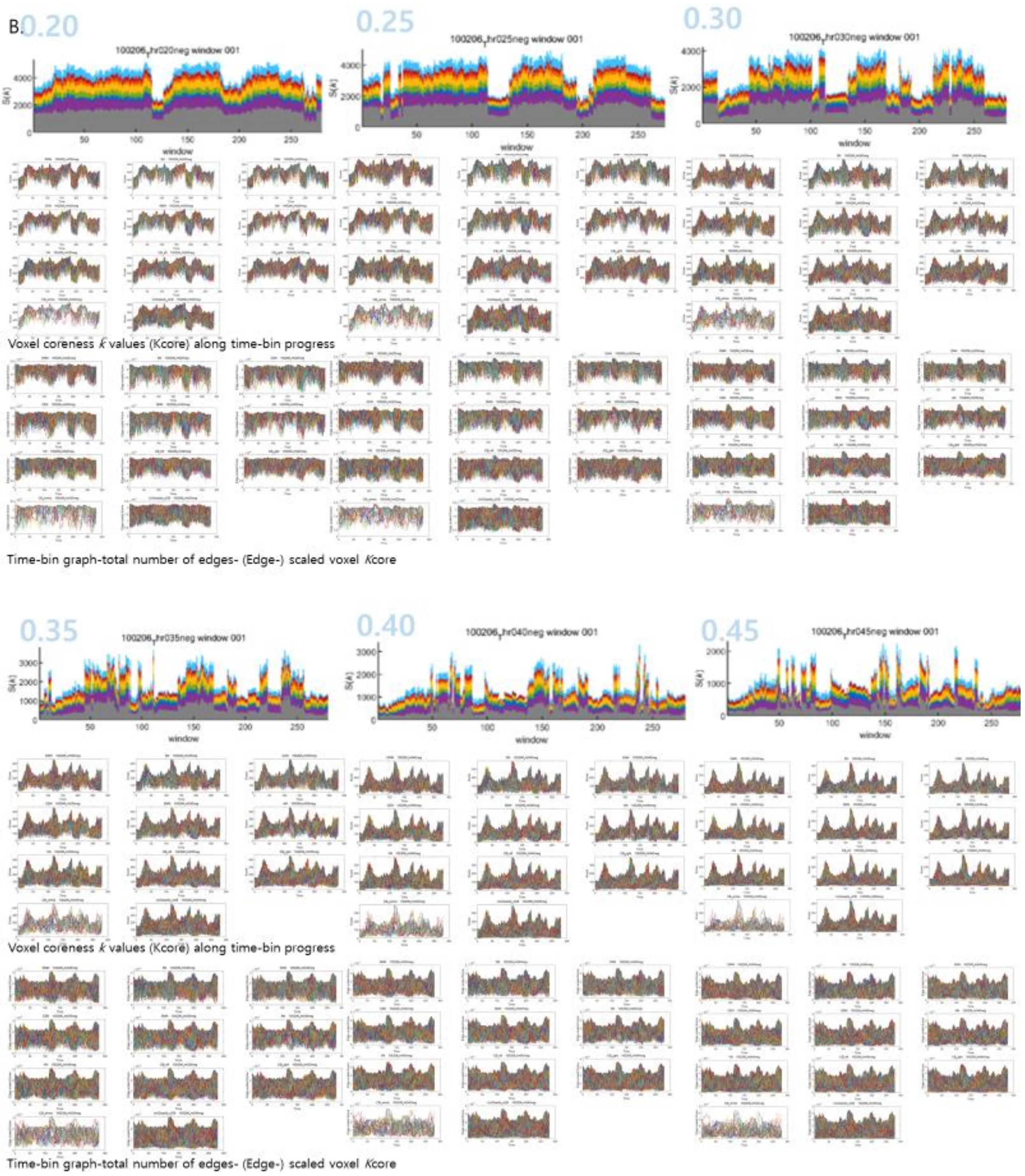
B.
0.20
100206_hr020neg window 001
0.25
100206_hr025neg window 001
0.30
100206_hr030neg window 001
S(k)
window
Voxel coreness k values (Kcore) along time-bin progress
Time-bin graph-total number of edges- (Edge-) scaled voxel Kcore
0.35
100206_hr035neg window 001
0.40
100206_hr040neg window 001
0.45
100206_hr045neg window 001
S(k)
window
Voxel coreness k values (Kcore) along time-bin progress
Time-bin graph-total number of edges- (Edge-) scaled voxel Kcore

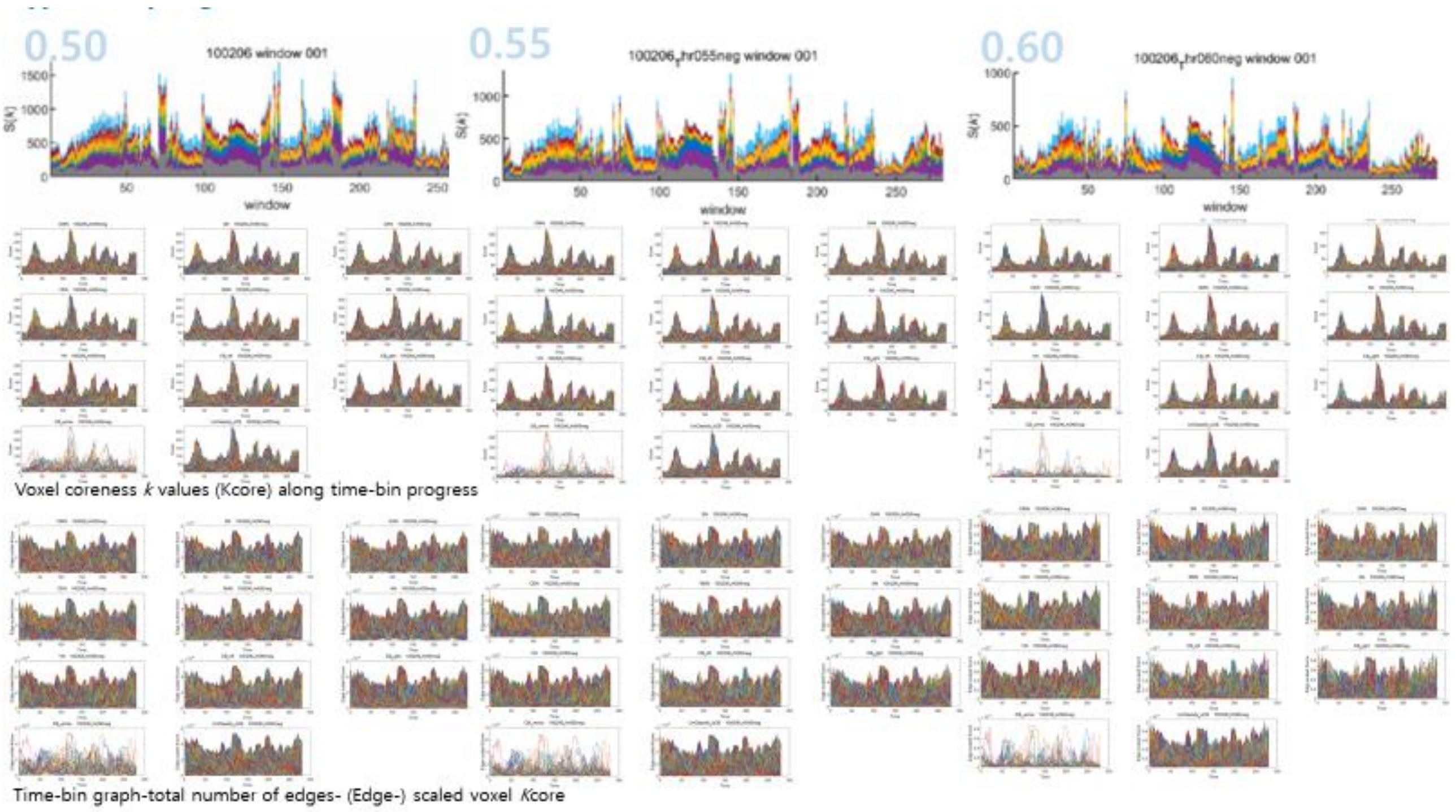
0.50
100206 window 001
S(k)
window
0.55
100206_hr055neg window 001
0.60
100206_hr060neg window 001
Voxel coreness k values (Kcore) along time-bin progress
Time-bin graph-total number of edges- (Edge-) scaled voxel Kcore

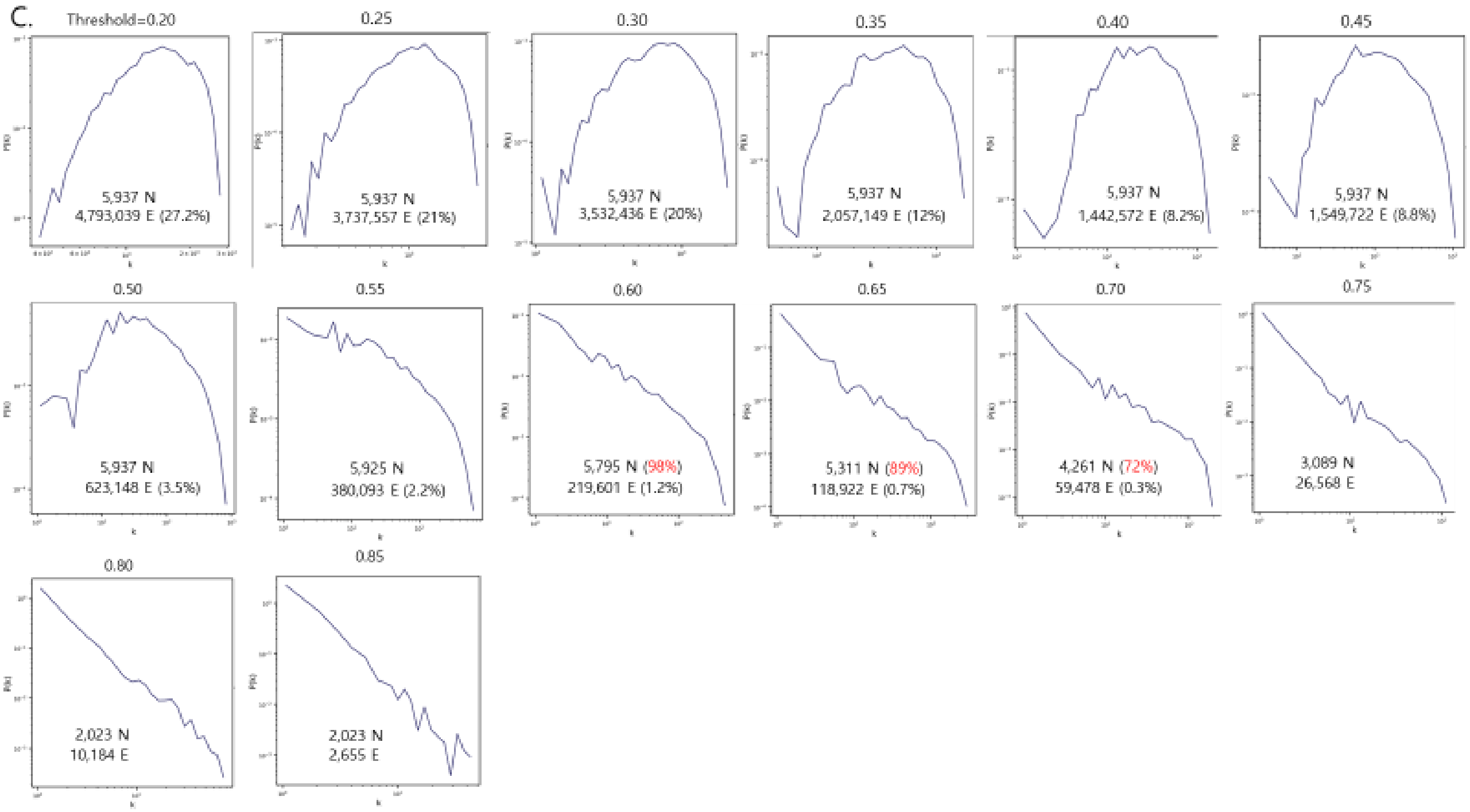

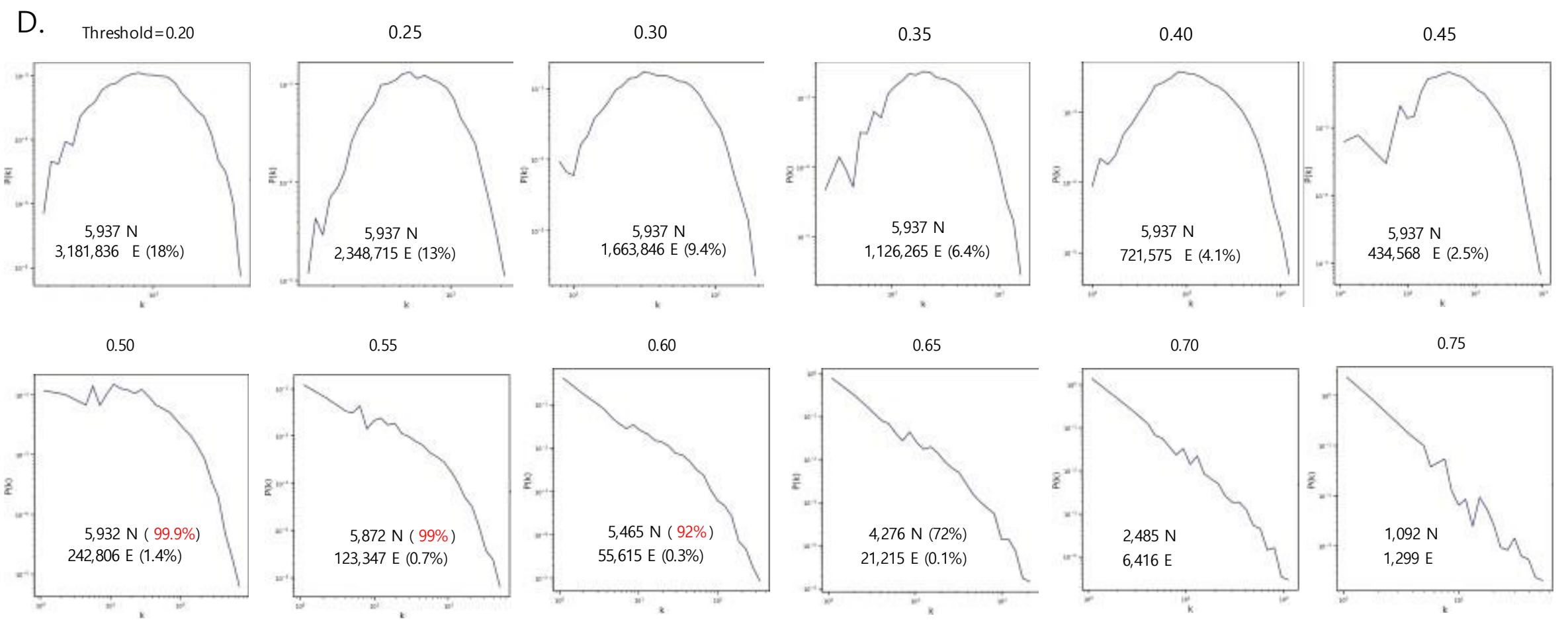

**Supplementary Figure 2. The ability to discover the hierarchical top tier voxels according to the varying thresholds, which are the same for all the time bins per individual, to make the preprocessed input data of pairwise intervoxel amplitude correlations**.

Individuals' intervoxel correlations of 5,937 voxels were assumed to contain the edges showing the characteristic scale-free-ness of their degree distribution and the surplus edges making the distribution have median/mode/mean structures. Varying the thresholds disclosed the degree-distributions of thresholded intervoxel edge weights (correlations) and these correlation matrices and their adjacency matrices were used as input for k core percolation. Thresholds were varied from 0.20 to 0.85 (total 14 thresholds) and k core percolation yielded timepoint plots, MRI-overlayed coreness k animation map plots, and stacked histogram plots of $k_{max}$core.

A. Stacked histograms and their IC/voxels timepoint plots of coreness k values in positive graphs of the case identification #100206. Each IC contained 732 voxels for DMN, 351 for SN, 363 for DAN, 682 for CEN, 483 for SMN, 289 for AN, 1,104 for VN and 2,691 for the unclassified. On the stacked histogram plots of $k_{max}$core with the thresholds from 0.2 to 0.35, state transition was not found because the voxels from all the ICs participated evenly in the top tier voxels. Beneath the first raw of the stacked histogram plots, coreness k trajectory timepoint plots showing the coreness k values (Kcore) were plotted first as the second group raw. Beneath this raw, voxels' Kcore were divided (edge-scaled) by the total number of edges per time bins. The same threshold used for all the time bins let the edges vary in number of several orders (several thousands to millions) and thus the edge-scaled voxel Kcores showed less variation than the non-scaled ones. Stacked histograms of the thresholds from 0.45 to 0.75, state transitions were clearly shown and the timepoint of state transition were almost at the same points and the same duration.

B. Stacked histograms and their IC/voxels time-point plots of coreness k values in unsigned negative graphs of the same case. The influence of the threshold of negative graphs were similar to that of the positive graphs but more dramatic. With the thresholds between 0.2 to 0.3, the numbers of $k_{max}$core voxels alone changed without the composition. With the thresholds of 0.35 to higher, the total $k_{max}$core voxel numbers decreased but the voxels/IC composition were still homogeneous. Apparent state transitions might as well show up but with less confidence. Change after edge-scaling, the contour of the timepoint plots came to be further more homogeneous.

C. In positive graphs, degree distribution showed typical changes according to the varying thresholds. For each threshold, number of voxels remaining were from 5,937 to 2,023 and number of edges from 4.9 million to two thousands. Percent indicators in the boxes of the thresholds 0.60, 0.65, and 0.70 in red showed the threshold of 0.70 are not allowed as the number of nodes was less than 85% (set threshold in this study). Scale-freeness were noted from the threshold of 0.55 to the higher ones. Individual variations between 180 subjects were noted but in a few and we could set the universal threshold of 0.65 for the positive graph studies in this HCP cohort.

D. In unsigned negative graphs, degree distribution changes were the same as the positive graphs. Red percent indicators say that we needed to use the threshold equal or lower than 0.60. Edge numbers ranged from 3.1 million (threshold 0.2) to 1,229 (threshold 0.75). Scale freeness could be found at the threshold 0.5 or higher. Changing pattern according to the thresholds was similar to that of positive graphs. After looking at the changing pattern of scale-freeness and the remaining number, we chose the threshold 0.50 for the negative graphs in this HCP cohort.

HCP: human connectome project, IC: independent component, DMN: default mode network, SN: salience network, DAN: dorsal attention network, CEN: central executive network, SMN: sensorimotor network, AN: auditory network, VN: visual network

A.

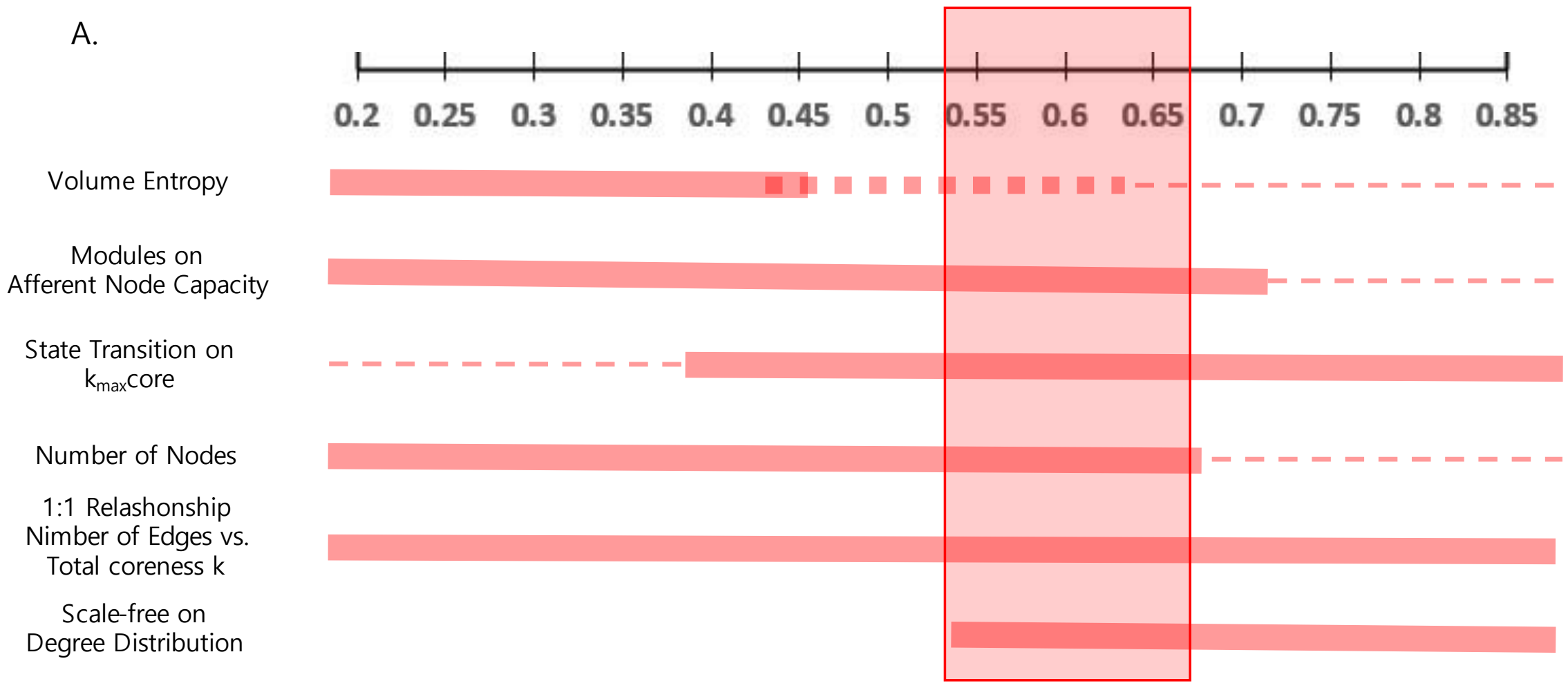

B.

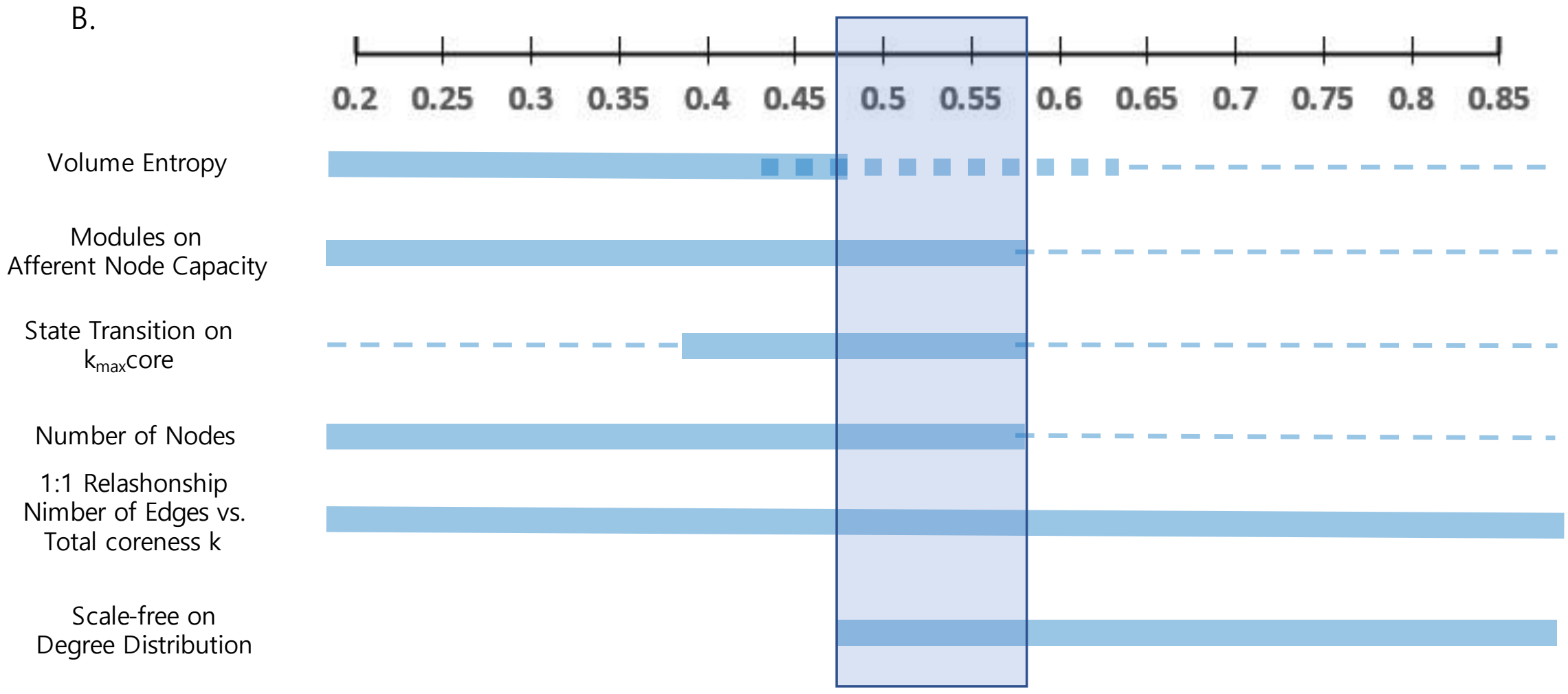

**Supplementary Figure 3. The criteria for acceptable brain graphs, positive and unsigned negative, were set as follows: 1) prescreening with the number of nodes and scale-freeness of the degree distribution according to the varying thresholds and 2) the consequences of thresholding on volume entropy, modules on afferent node capacity, state transitions and the relationship between the number of edges and total coreness k.**

A. In an example of positive graph of an individual (#100206), number of nodes more than 85% of 5,937 voxels were guaranteed until the threshold was 0.65 and degree distribution was scale-free (linear descent on log-log plot of degree-distribution plots) for all the time-bins of brain graphs whose thresholds were equal or higher than 0.55. Post-hoc observation of volume entropy showed the same values for the graphs with the thresholds ranging from 0.2 to 0.45 and with higher thresholds, slowly decreasing till 0.6 and then linearly decreased according to the total edge numbers of each time-bin graphs. Directed graph and its afferent node capacity revealed exactly the same modules (voxels/IC composition) and module exchanges until the threshold of 0.7. K core percolation results, especially $k_{max}$core stacked histogram timepoint plots yielded the evidence of state transitions, the same from the threshold 0.4 to the last one of 0.85. One-to-one relationships between the total number of edges and the total coreness k of the brain graphs per time bins maintained all over the thresholds.

B. In unsigned negative graph of this person, number of nodes larger than 85% were until the threshold of 0.55. Scale-freeness was observed when the threshold was equal to or higher than 0.5. Volume entropy were the same until the thresholds were less than 0.5, and modules were found until the threshold increased to 0.55. Implicit evidences of state transition were barely observed when the thresholds were between 0.4 and 0.55. 1:1 relationship between number of edges and total coreness k maintained all over the thresholds.

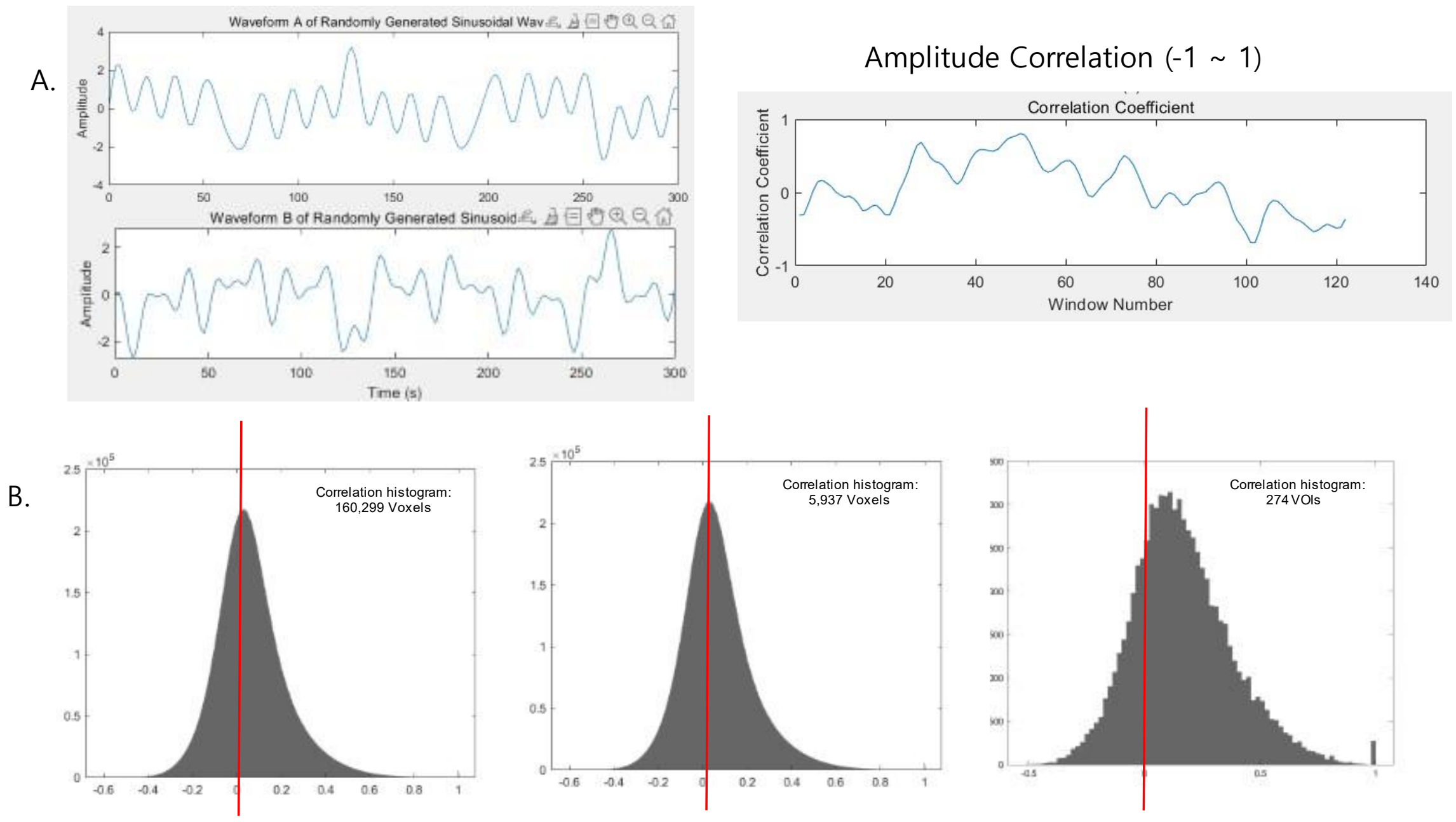

**Supplementary Figure 4. Observed pairwise intervoxel correlations of brain graphs of the Human Connectome Project using the initially reconstructed 2 × 2 × 2 mm$^3$ resolution (160,299 voxels), 6 × 6 × 6 mm$^3$ resolution (5,937 voxels) and 274 anatomical regions of interest.**

A. Two waveforms of 300 seconds of duration were randomly generated with Matlab program and their intervoxel correlations were plotted using sliding window methods (1 minute duration with 2 seconds shifts and thus 120 time-bins). Correlations ranging from -1 to 1 were produced with their time-point plots on the right side.

B. Propensity of intervoxel correlations derived from the 160,999 voxels (2 x 2 x 2 mm$^3$) with 12.9 billion undirected edges, 5,937 voxels (6 x 6 x 6 mm$^3$) with 17.6 million undirected edges and 274 anatomical volumes-of-interest (VOIs) with 37K edges. Positive shift (less amounts of negative-valued edges than positive-valued edges) was already there but in small fraction in the initial 2 x 2 x 2 mm$^3$ or 6 x 6 x 6 mm$^3$ resolution brain graph, however, in brain graphs with 274 VOIs, negative correlations ranges from -0.3 to 0 and the area was just quarter of positive correlations.

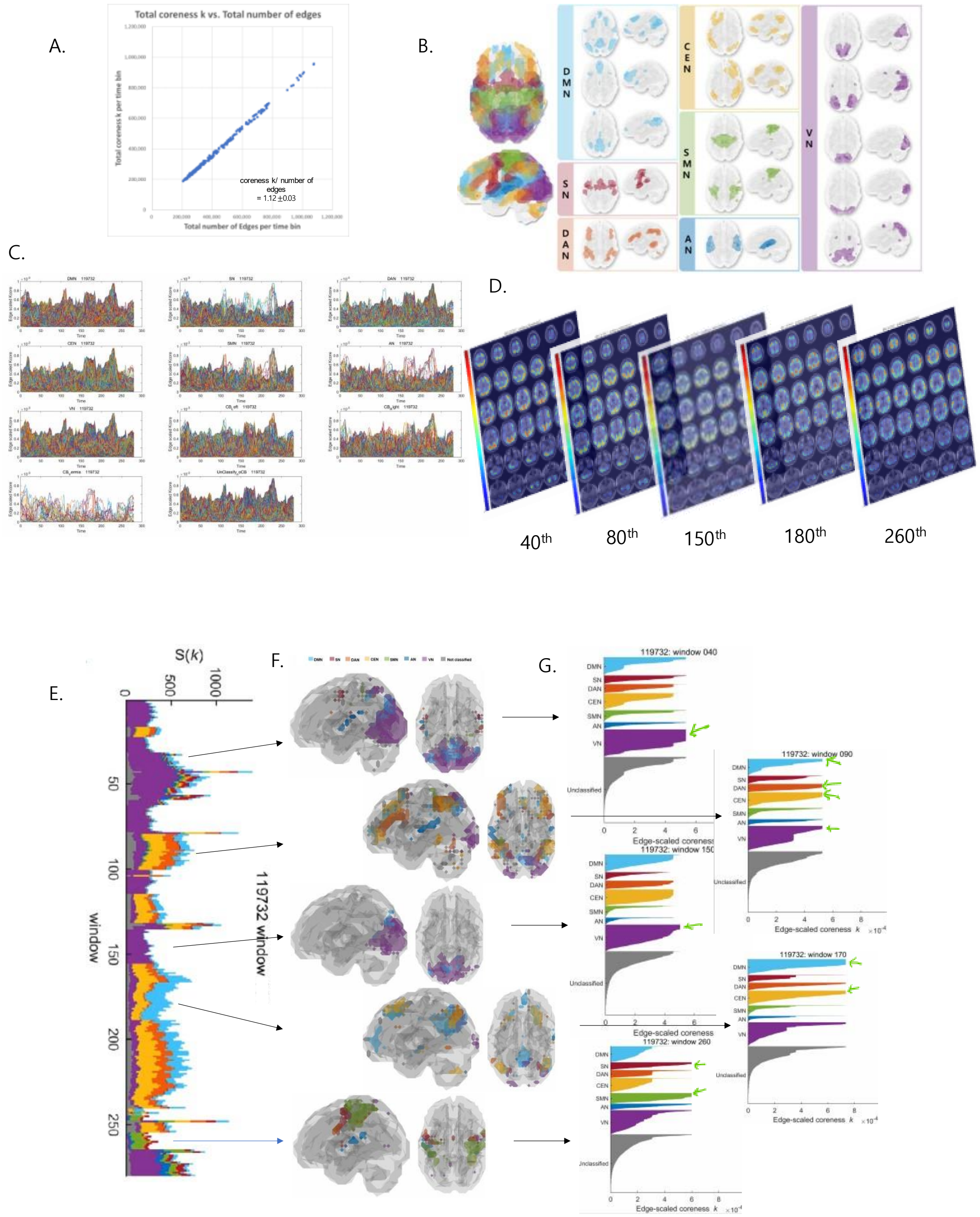

**Supplementary Figure 5. Characteristic measures on the graphs and their animations and timepoint plots of a representative individual for functional brain graphs of positive (and unsigned negative) intervoxel correlations.**

A. After k core percolation, each node obtained its own coreness k number. As expected and confirmed with data in this study, the total number of edges of a time-bin graph and the summed amount of coreness k values over all the voxels of that graph revealed 1:1 relationships in every individual, regardless of thresholds, preset or varied, and also regardless of positive or negative graphs.

B. Voxels were annotated for their belongings to ICs, which were, a priori, determined by independent component analysis of 180 static intervoxel correlations derived from the cohort of HCP consisting of 180 individuals. ICs were designated to each voxel of total 5,937 of 6 x 6 x 6 $mm^3$ and also of total 1,489 of 10 x10 x10 $mm^3$. Around 10% of voxels were annotated for two ICs or in a very few three ICs.

C. Using the relationship of total edge number and total coreness k of each graph, all the voxel coreness k values were normalized by dividing them using their corresponding total edge numbers. The resulting edge-scaled coreness k was plotted for the 289 (AN) to 2,691 (unclassified) voxel trajectories and visualized as timepoint plots per each IC, left cerebellum, right cerebellum, vermis and the unclassified IC. Contour was similar but the trajectories inside each IC was heterogeneous beyond description in any simple terms. 'Ripples and a few threads' were the description of the timepoint plots of coreness k voxels/IC trajectories.

D. Coreness k values of all the voxels were divided by total number of edges to yield edge-scaled coreness k values. These edge-scaled coreness k values were overlayed on the 36 slices MRI in radiologic convention and were displayed as avi file on animation. The 40$^{th}$ to 260$^{th}$ time bin snapshot images were displayed as an example.

E. Timepoint plots of stacked histogram of an example case showed the states of clustered time bins with similar composition of $k_{max}$core voxels/IC compositions, which were colored to visualize clearly the abrupt changes of the voxels/IC compositions at certain points. We called this time-bin switching of voxels/IC composition as 'state transition. In this case, at least 11 state transitions were recognized.

F. Glass brain images of left lateral and superior views were matched with the time-bin points of $k_{max}$core stacked histogram. The uppermost show VN dominant with scanty companion ICs state, the next one DMN/DAN/CEN major and VN/AN minor, the third VN dominant, the fourth DMN/CEN main, and the lowest show SMN with SN.

G. On the animated flagplots (snapshots here for explanation), small green hand-drawn arrows indicate the $k_{max}$core voxels (and their ICs) on the top-tier (rightmost tier).

HCP: human connectome project, IC: independent component, DMN: default mode network, SN: salience network, DAN: dorsal attention network, CEN: central executive network, SMN: sensorimotor network, AN: auditory network, VN: visual network

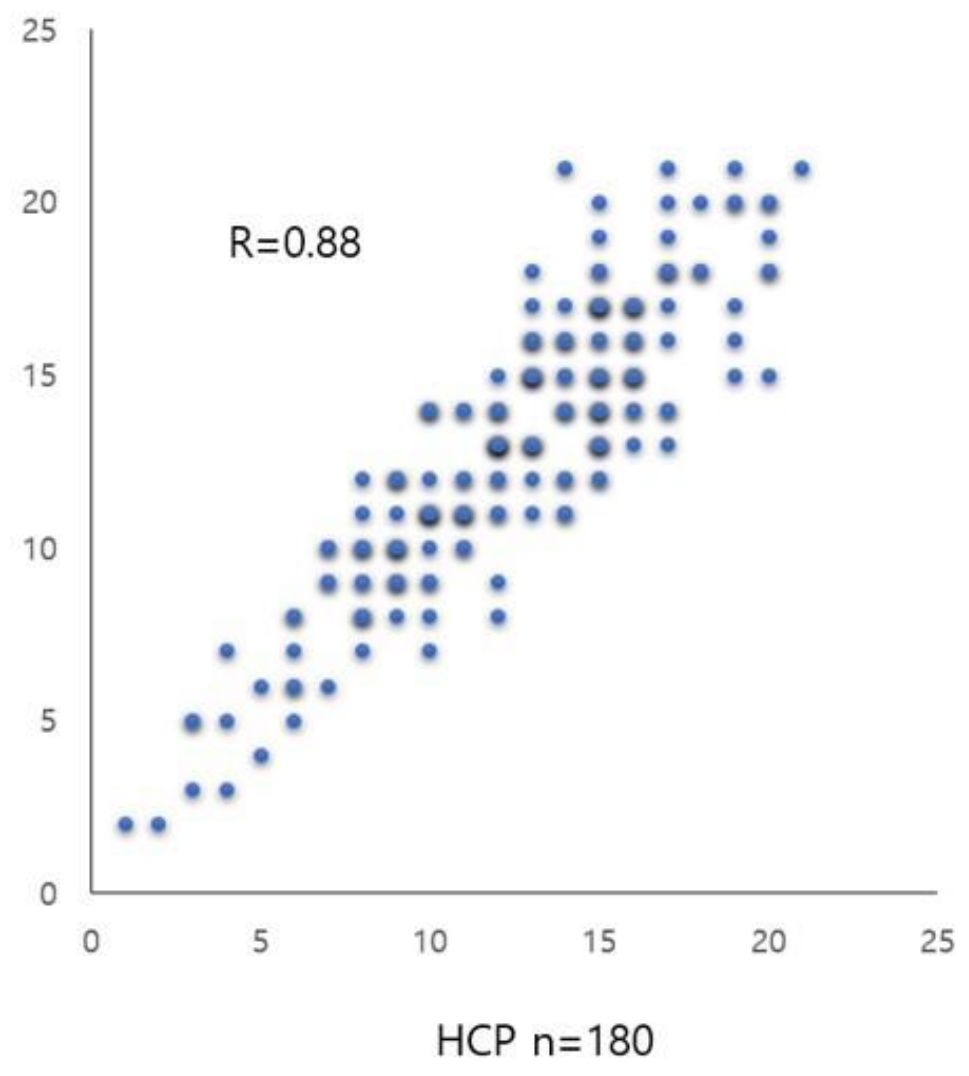

**Supplementary Figure 6. Reproducibility of designating 'state transition' by counting the numbers by one operator among the authors using the positive graphs.**

Even taking into consideration of the operator's learning curve and no explicit definition of state transition on stacked histogram of $k_{max}$core voxels/IC composition a priori, intra-operator reproducibility was acceptable (Spearman's rho of 0.88). Later on, the state transition were defined as 1) abrupt change of $k_{max}$core voxels/IC composition within one time bin, 2) state should be at least one or more time-bin duration, 3) conservatively, voxels' fraction changes of IC composition did not constitute a state, meaning that different states are defined more importantly by IC composition changes with or without fraction changes, and, nevertheless, 4) in ambiguous occasions, heuristic decision whether this or that abrupt change deserves 'state transition' was allowed. Based on the third and the fourth oracles, negative graphs lost many of the state transitions because the subtle changes were not so abrupt nor so clear as those of positive graphs. Interestingly, the stacked histogram of $k_{max}$core voxels/IC composition, once mixed blindly of positive and negative graphs, we could easily dissociate the positive and negative groups.

A.

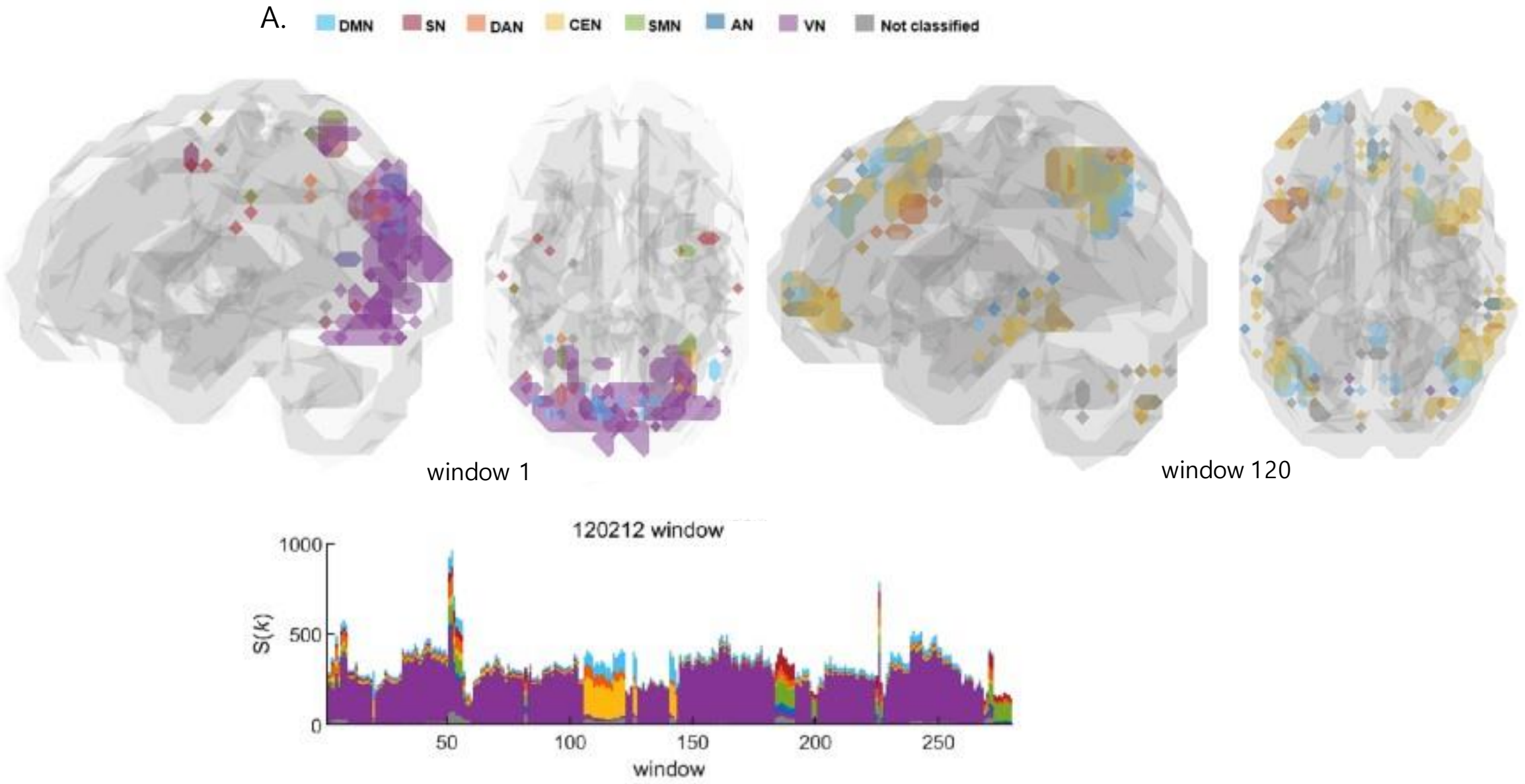

Positive Graph

B. **Afferent node capacity**

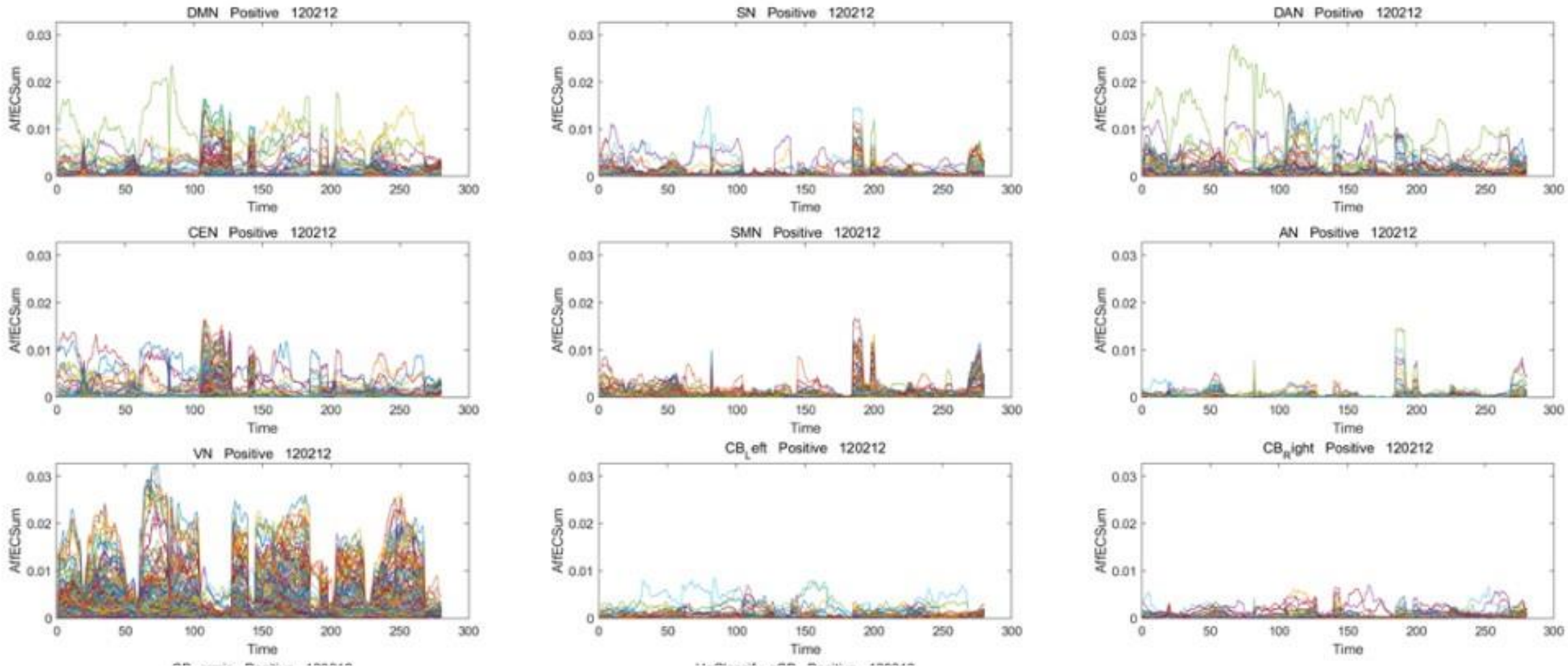

C.

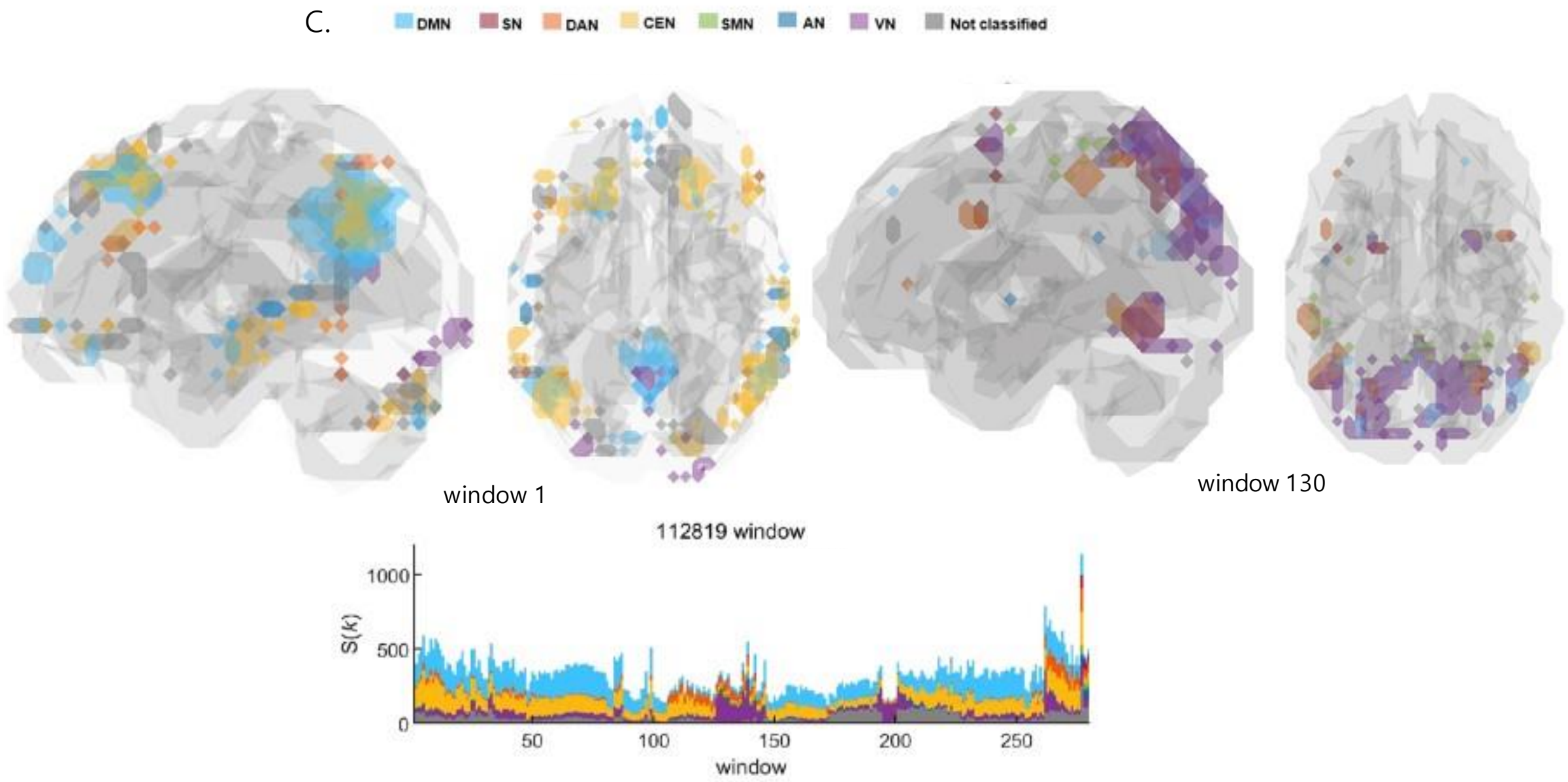

D.

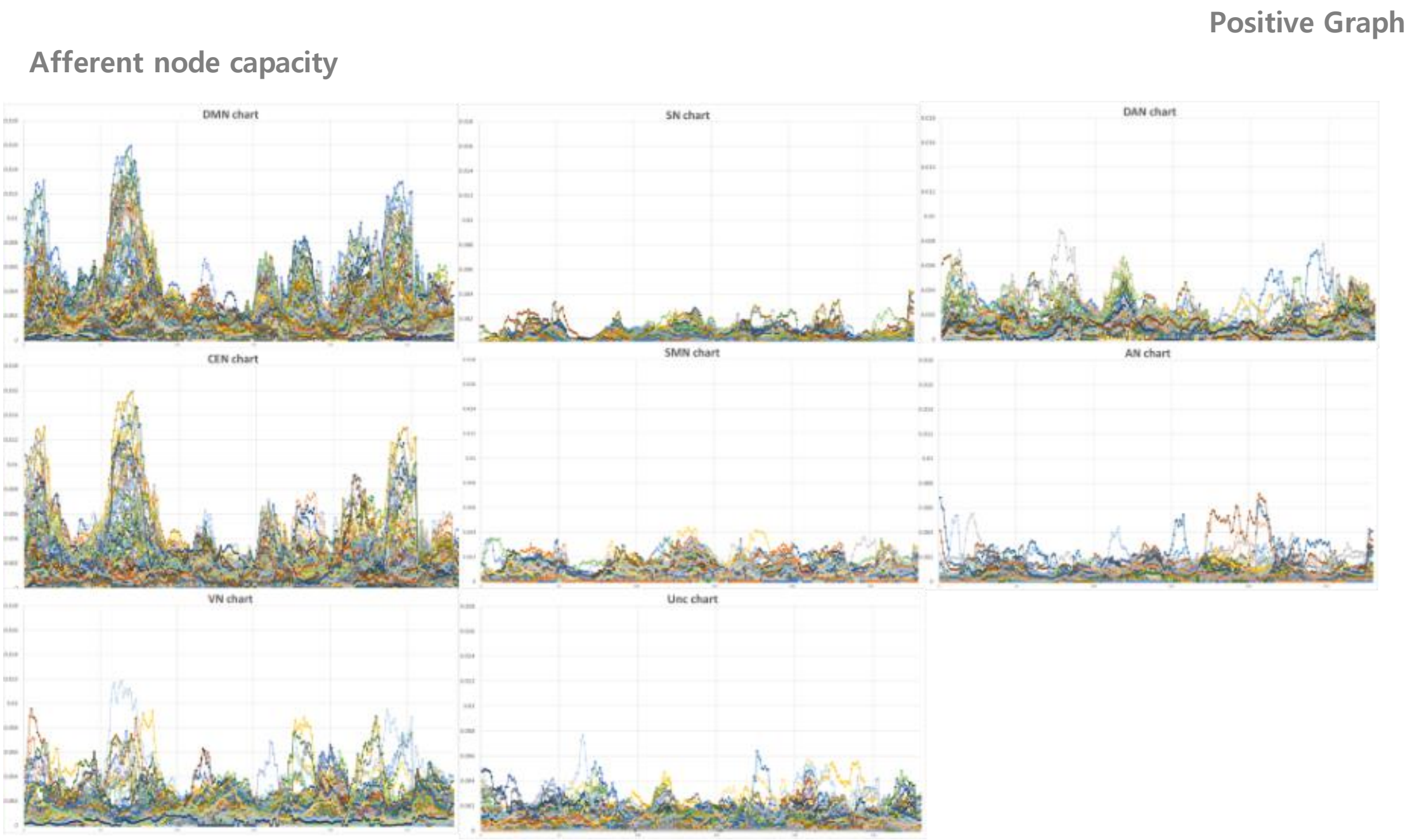

**Supplementary Figure 7. Glass brain animation plots and stacked histogram timepoint plots on undirected positive graphs and afferent capacity timepoint plots showing module formation and switches on directed positive graphs: intermediate pattern between no transition and ‘the state transition before and after half the period’.**

A. In this individual, VN voxels dominated throughout all the time-bins of the $k_{max}$core plots except at the middle of the period. After $100^{th}$ time-bin, DMN/DAN/CEN took over the top tier on the hierarchy for some while and SN/SMN/AN did just before 200th time bin for a short while. Single IC (VN) dominated then for another several minutes while yielding to its combinatorial competitors for moments and took the hierarchical supremacy again.

B. On afferent node capacity plots of this individual, VN voxels dominated every time bin except time bins after 100th and before 200th. After 100th time bin, DMN, CEN and DAN voxels filled the gap between the VN dominances, and just before 200th time bin, SN, SMN and AN joined to fill another hierarchical supremacy.

C. In another individual, DMN/CEN voxels dominated almost all the time-bins of the $k_{max}$core plots. However, in the middle of the period (between100th and 150th time-bins), DAN/CEN first and VN plus other smalls took charge in the top tier of the hierarchy. Around 200th time bin VN alone occupied the supremacy.

D. In afferent node-voxel capacity plots of this individual, DMN and CEN voxels showed almost the same pattern of temporal progress while DAN joined intermittently in part. Grossly similar to the findings on the $k_{max}$core plots, however, the prominent state transition to DAN or VN between 100th and 150th of were not clear on the afferent capacity plots.

DMN: default mode network, SN: salience network, DAN: dorsal attention network, CEN: central executive network, SMN: sensorimotor network, AN: auditory network, VN: visual network

A.

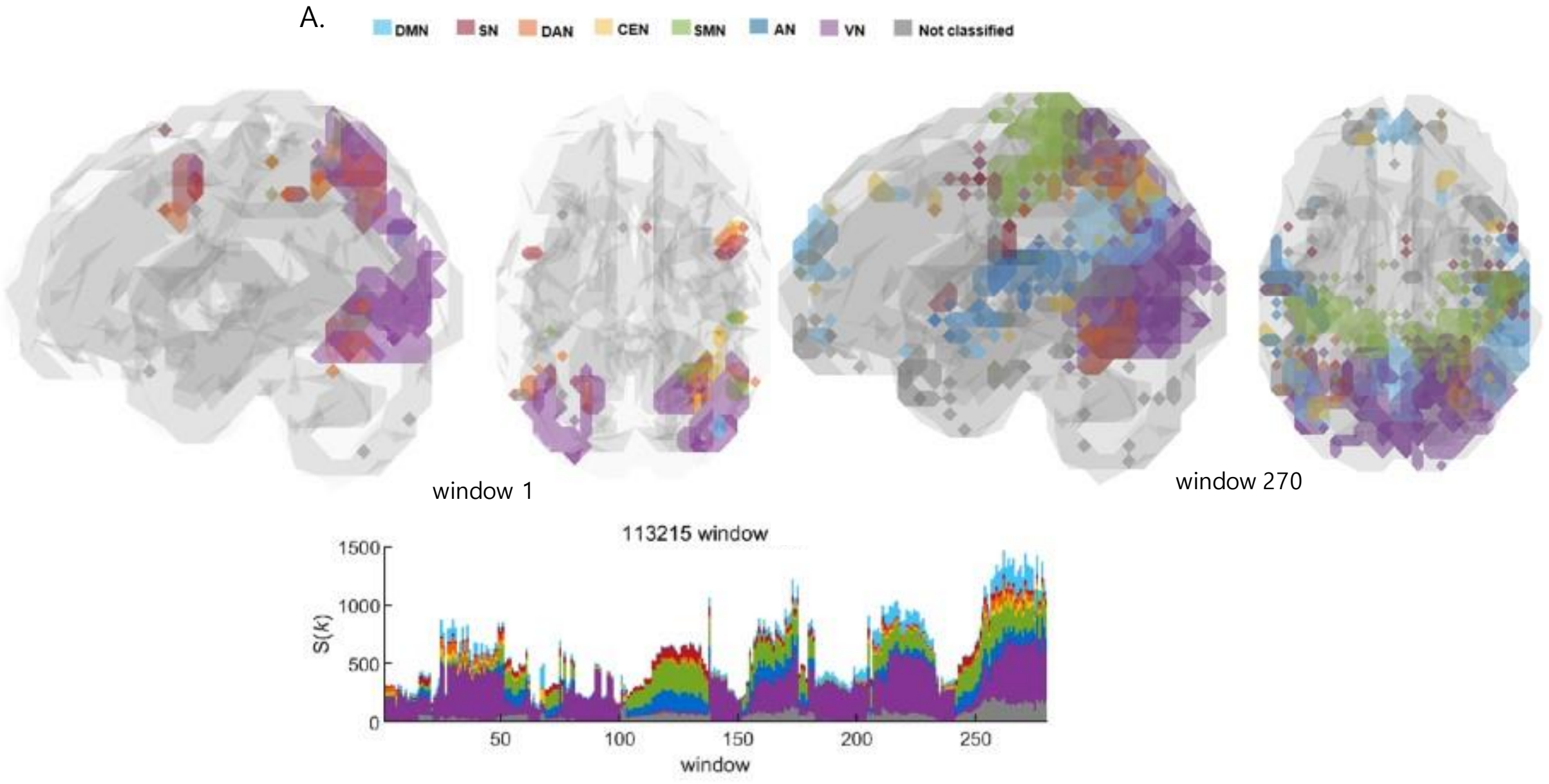

Positive Graph

B. **Afferent node capacity**

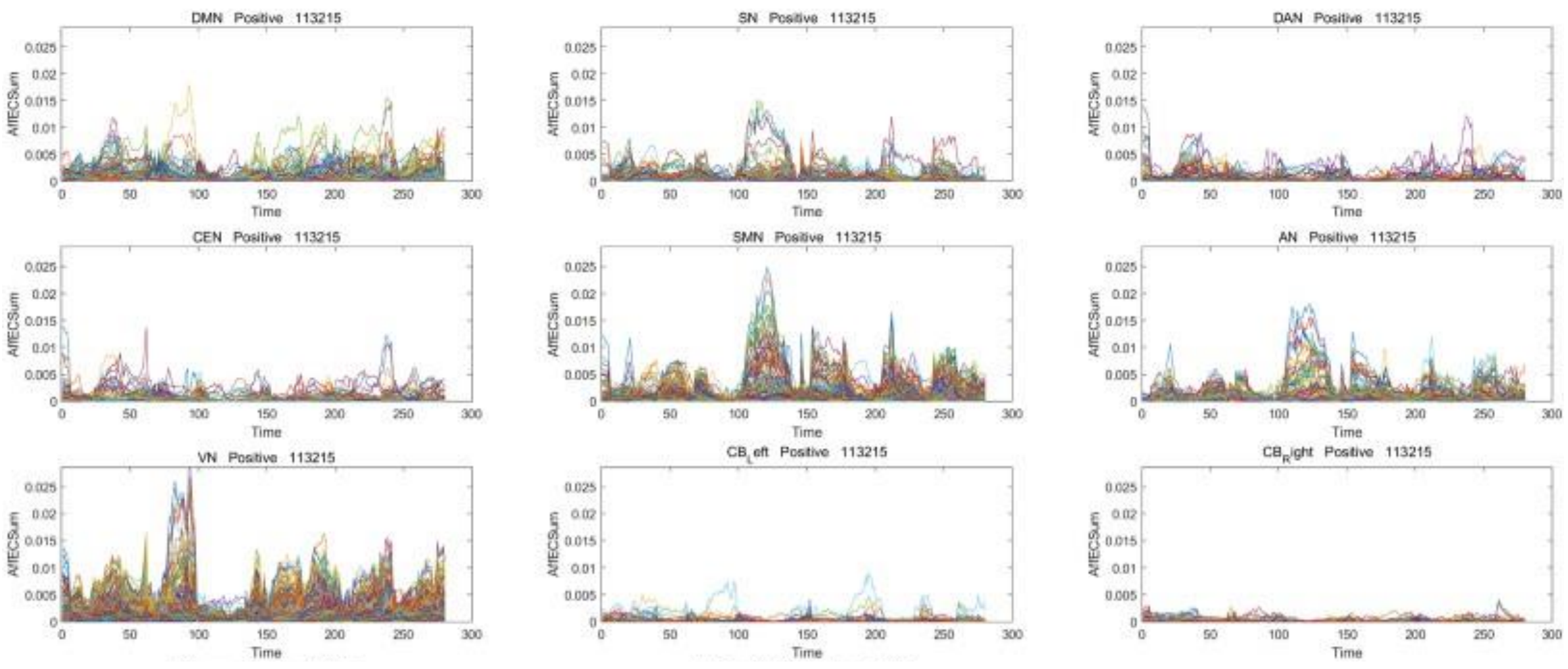

C.

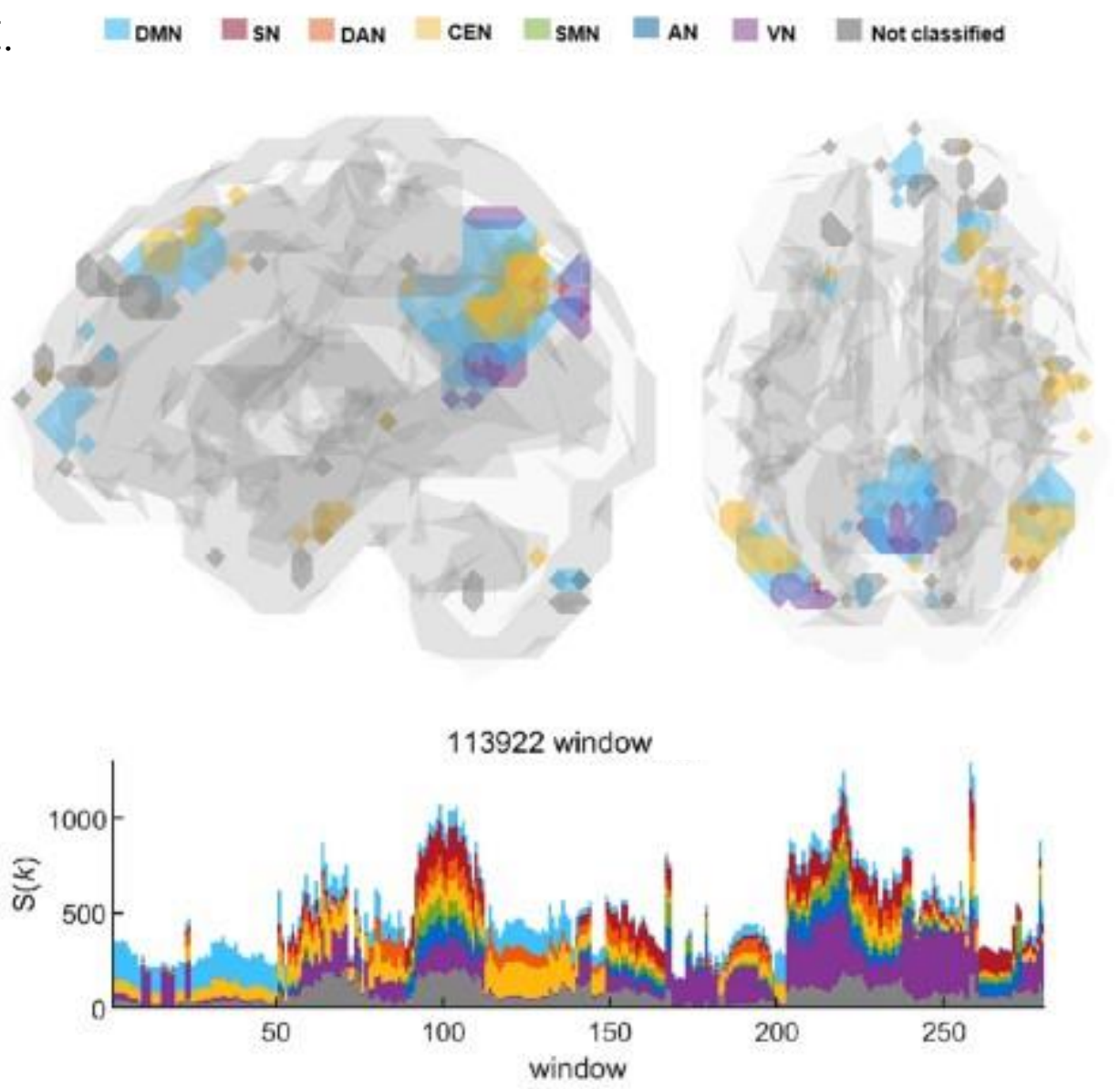

Positive Graph

D. Afferent node capacity

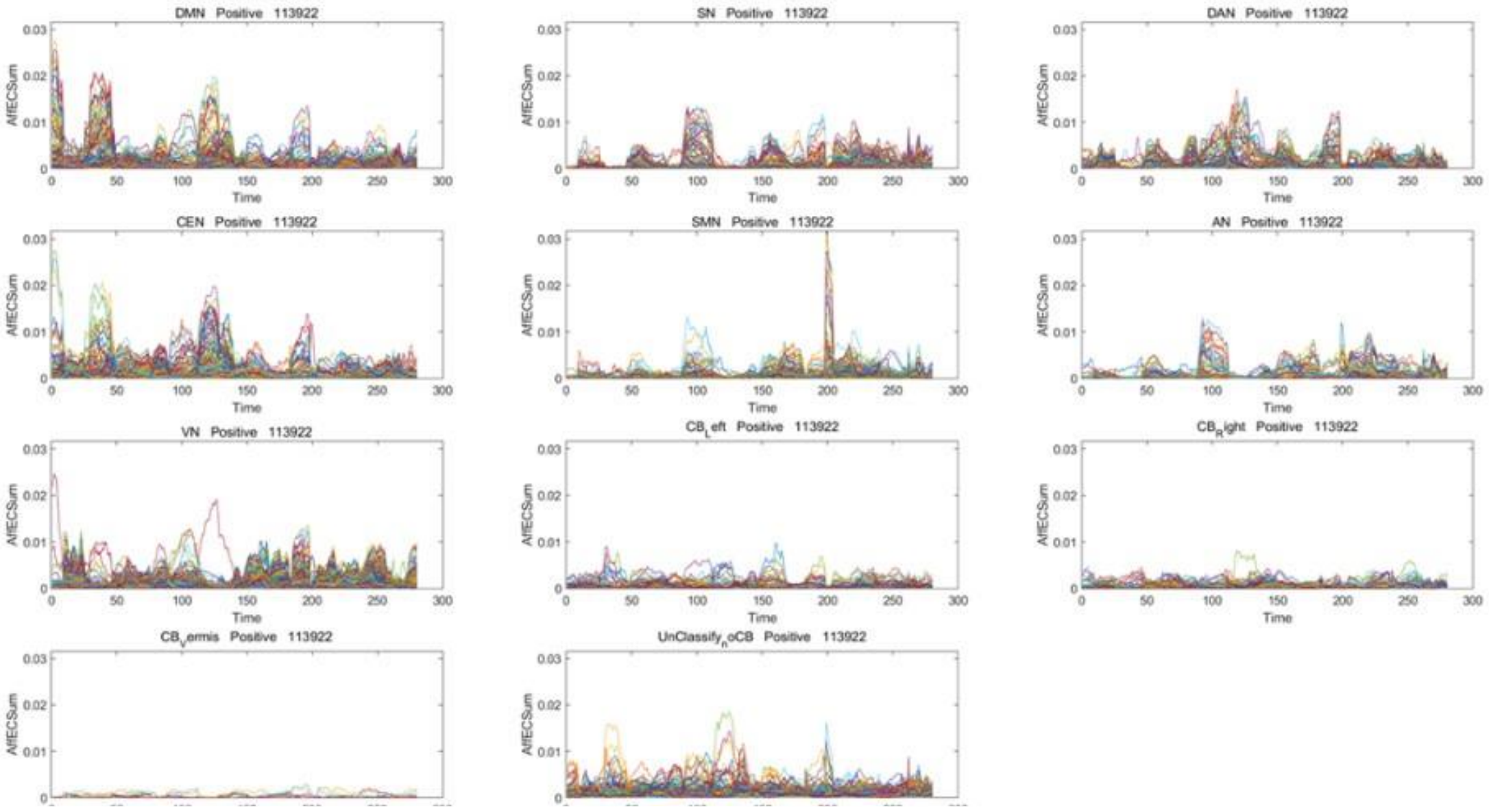

**Supplementary Figure 8. Glass brain animation plots and stacked histogram timepoint plots on undirected positive graphs and afferent capacity timepoint plots showing module formation and switches on directed positive graphs: examples of typical state transitions.**

A. In this individual, state transition was mostly from VN and its allies to SMN/AN/SN and back to VN dominance. Unlike frequent transitions from DMN/CEN to VN and from VN to DMN/CEN, this type of transition from VN to SMN was less frequent. During the period between 80$^{th}$ and 150$^{th}$ time bins, SMN/AN/SN co-dominance was surrounded by preceding

VN and following VN dominances, and sharp transition in-between was easily recognized.

B. On afferent node-voxel capacity plots of this individual, characteristic absence of VN module and replacing SMN/AN/SN co-modules was noted during the period between 100$^{th}$ and 140$^{th}$ time bins.

C. In another individual, state transition was shown from DMN/CEN to VN and back at the earlier part. Around 50$^{th}$ time bin, DMN/CEN dominant state was replaced by all-modules participation, called distributed (ICs) state, which reached 90$^{th}$ time bin, constituting rainbow-type collective modules. This rainbow/distributed dominance was followed by DMN/DAN/CEN at 110$^{th}$ time bin. Several state transitions followed from the distributed to VN, VN to DMN/DAN/CEN/VN, to very short DMN/CEN, and then rainbow/distributed states. Many state transitions could be recognized like this in almost all the 180 individuals. This was the source of counting the numbers of state transitions on $k_{max}$-core plots (5) (Suppl. Fig. 6).

D. On afferent node capacity plots of this individual, between 25$^{th}$ and 50$^{th}$ time bins, and only DMN/CEN were conspicuous with the vacancy of VN and others. Around 100$^{th}$ time bin, reminding distributed module of $k_{max}$core plots, all the modules including the unclassified showed up. Rainbow-type distributed dominance on $k_{max}$core plots were called 'unison' not orchestration or symphony.

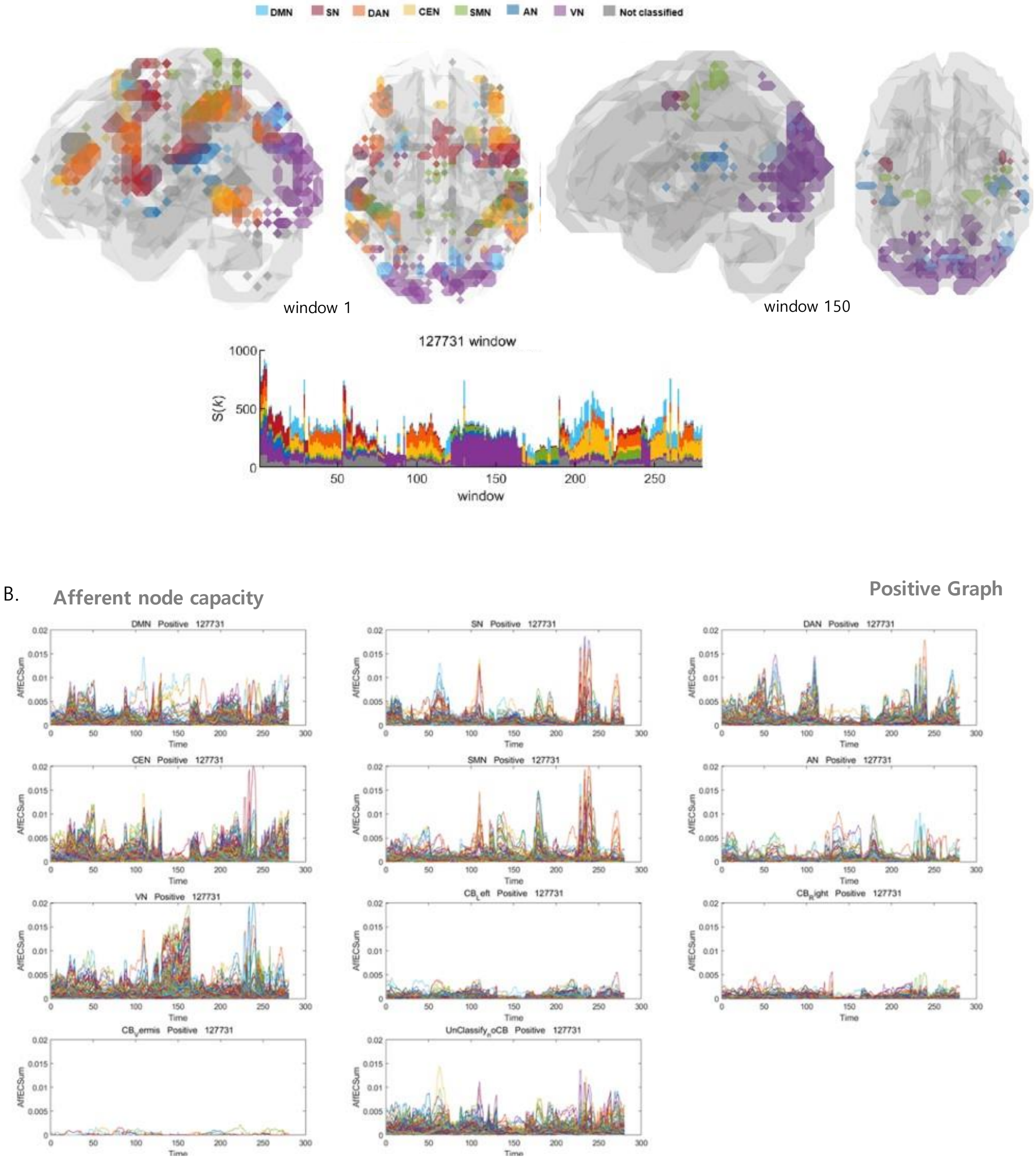

**Supplementary Figure 9. Glass brain animation plots and stacked histogram timepoint plots on the undirected positive graphs and afferent capacity timepoint plots showing module formation and switches on the directed positive graphs: an individual showing too-frequent transitions.**

A. In usual real cases like this individual, $k_{max}$core plots were more flamboyant. Stationary state sustained for variable duration even from one (3 seconds) to several tens (e.g. 50 seconds) of time bins. In this casen, state transitions could have been defined in many ways as arbitrarily as the investigators wanted. However, based on final heuristic decision, considering that stacked histogram plot consisted of the numbers of $k_{max}$core voxels colored according to IC compositions, this individual was found to have 17 state transitions. Flamboyance in state progress and transition and the definition of state transition still remained a challenge.

B. Afferent node-voxel capacity of this individual showed slim, sharp, rapidly changing, and differently gathering and dismantling of voxels on every instance of observation of time bins. Nevertheless, during the period between $125^{th}$ and $170^{th}$ time bins when VN dominance was prominent on the $k_{max}$core plots, VN/SMN/AN modules were built temporarily. Then, interestingly VN module disappeared suddenly and DMN/CEN filled the vacancy and replaced VN.

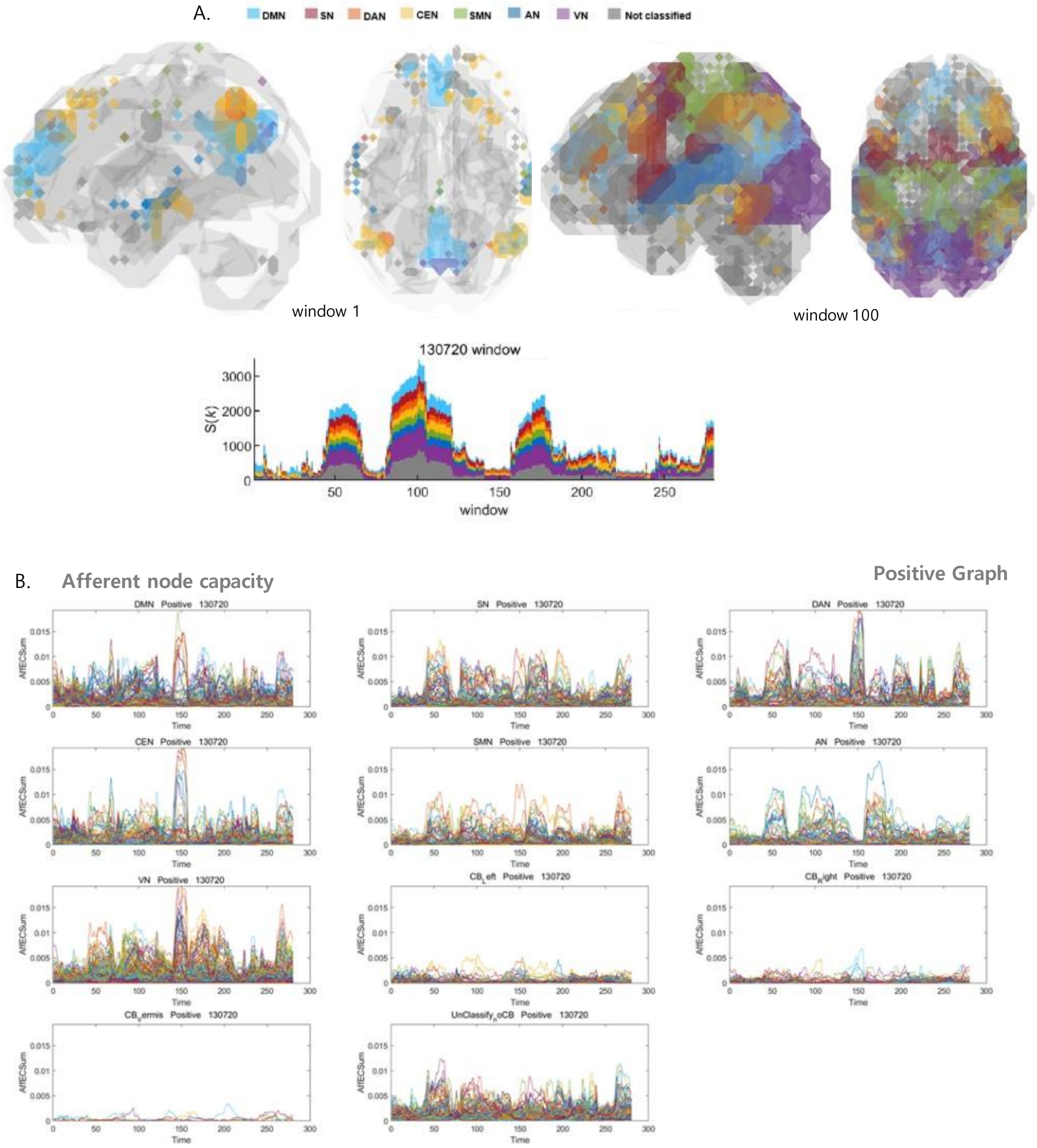

**Supplementary Figure 10. State fluctuation of synchronized voxels on glass brain animation plots and stacked histogram timepoint plots and on afferent capacity timepoint plots.**

A. In this individual, $k_{max}$core timepoint plots showed characteristic rainbow-type distributed modules interspersed by smaller co-modules. Unison of participation of ICs of clusters were similar to each other in the entire period, though replaced by a few ICs taking the temporary top tier intermittently.

B. Afferent node-voxel capacity showed two similar co-modules, i.e., one DMN/CEN/VN and another SMN/SN/AN/Unclassified. Interestingly, DAN mostly mimicked SMN/SN/AN except for a period around 150$^{th}$ time bin, when it adopted the feature of DMN/CEN/VN at that time bin.

A.

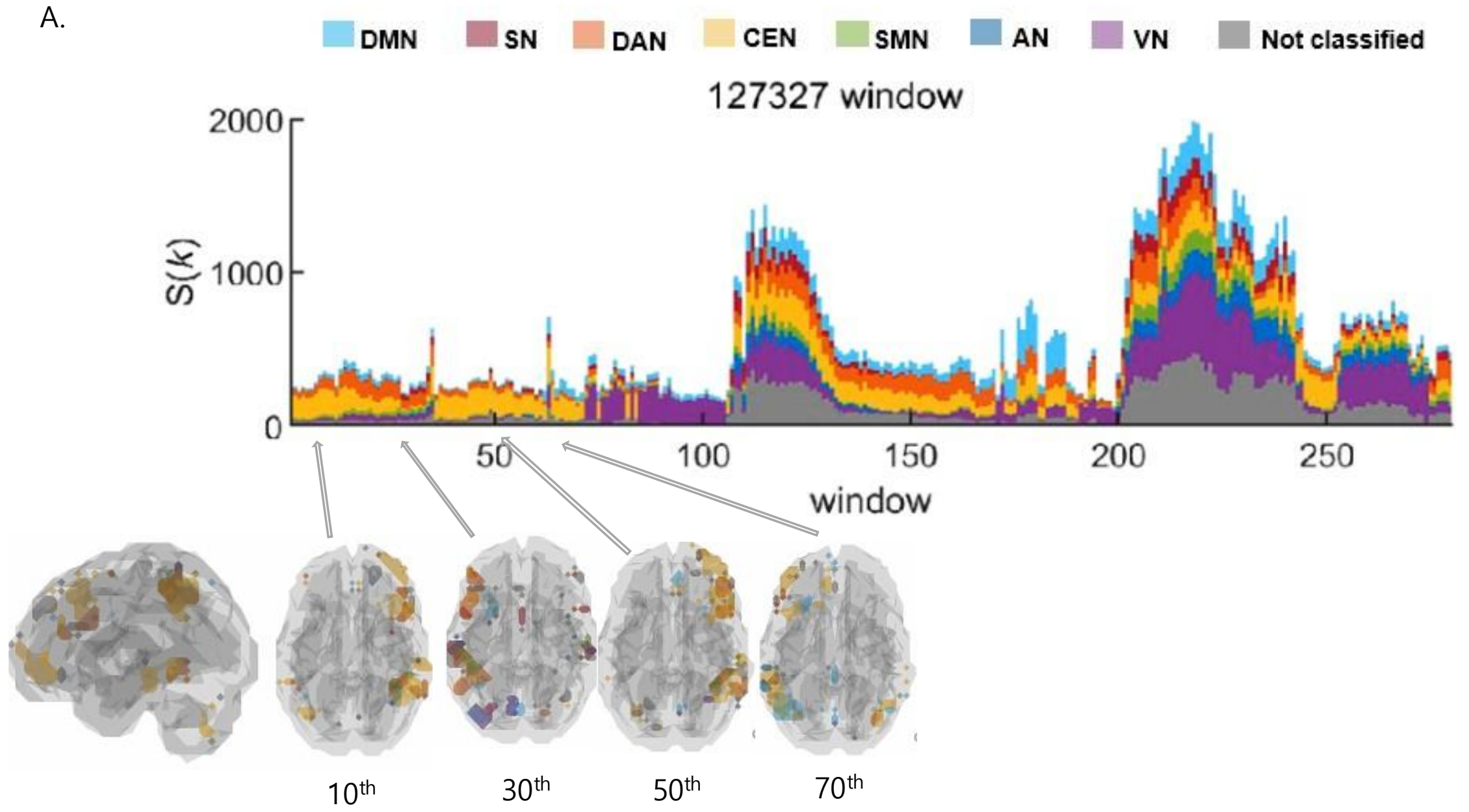

Positive Graph

B. **Afferent node capacity**

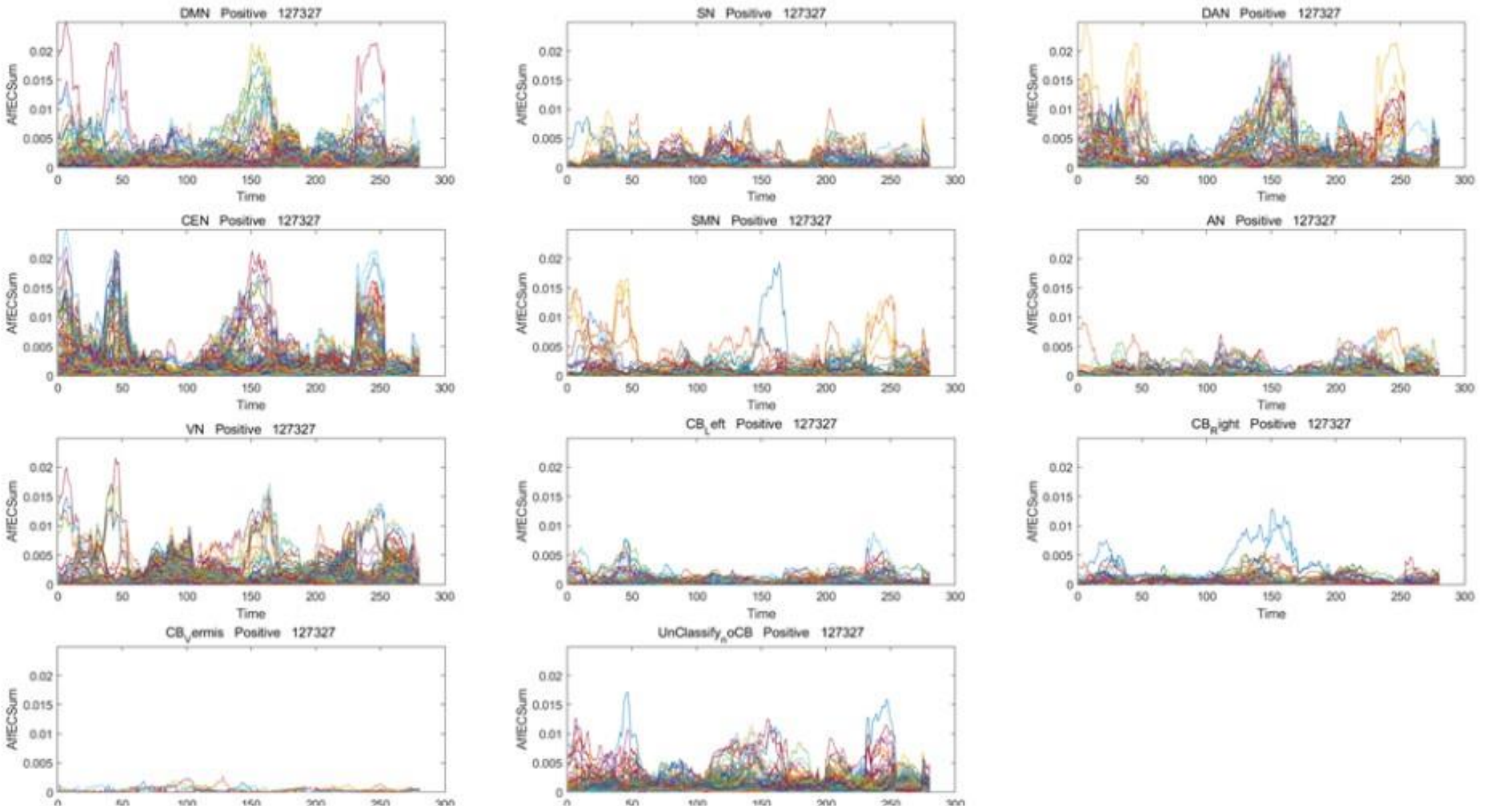

C.

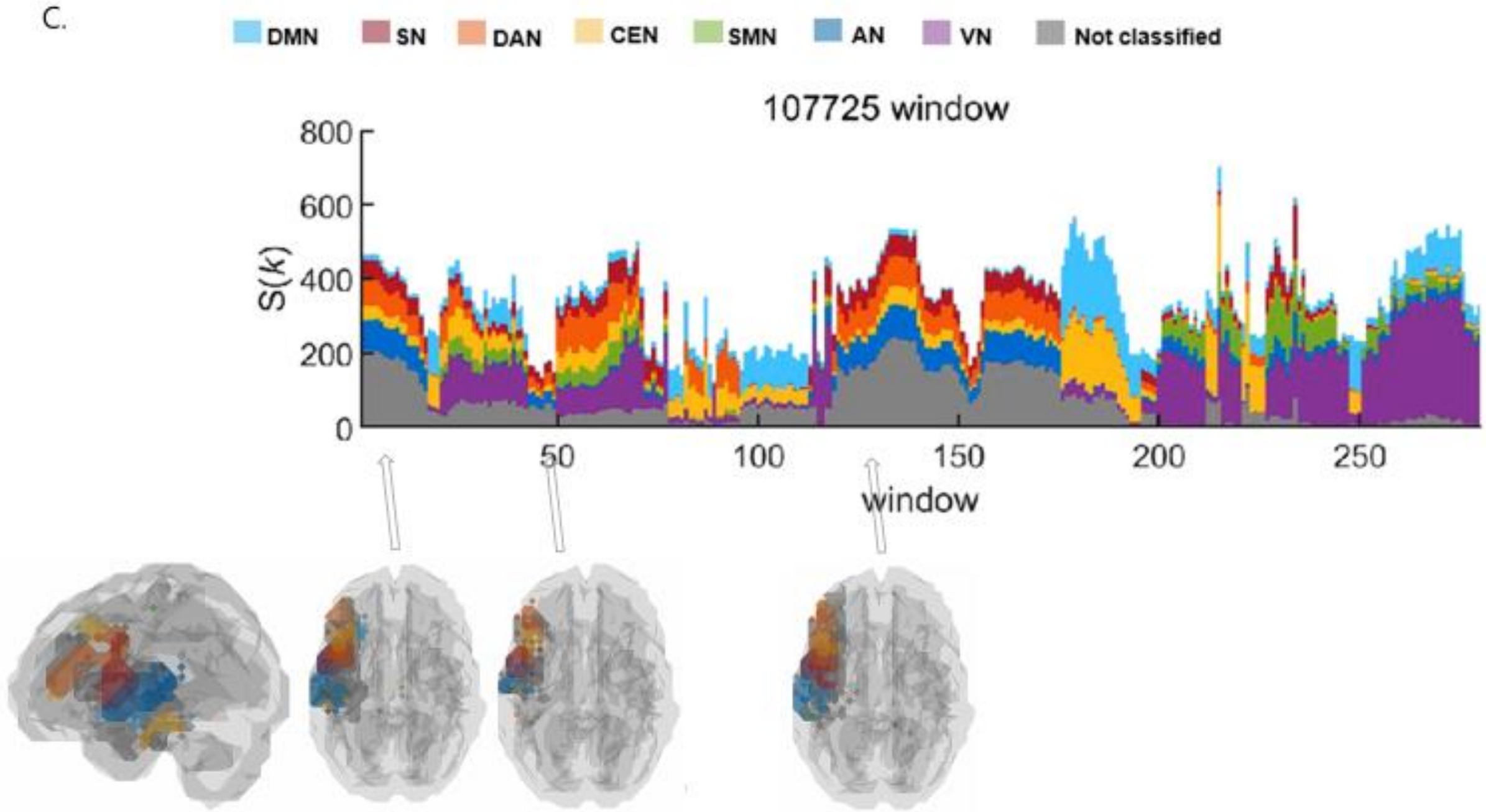

D. Afferent node capacity

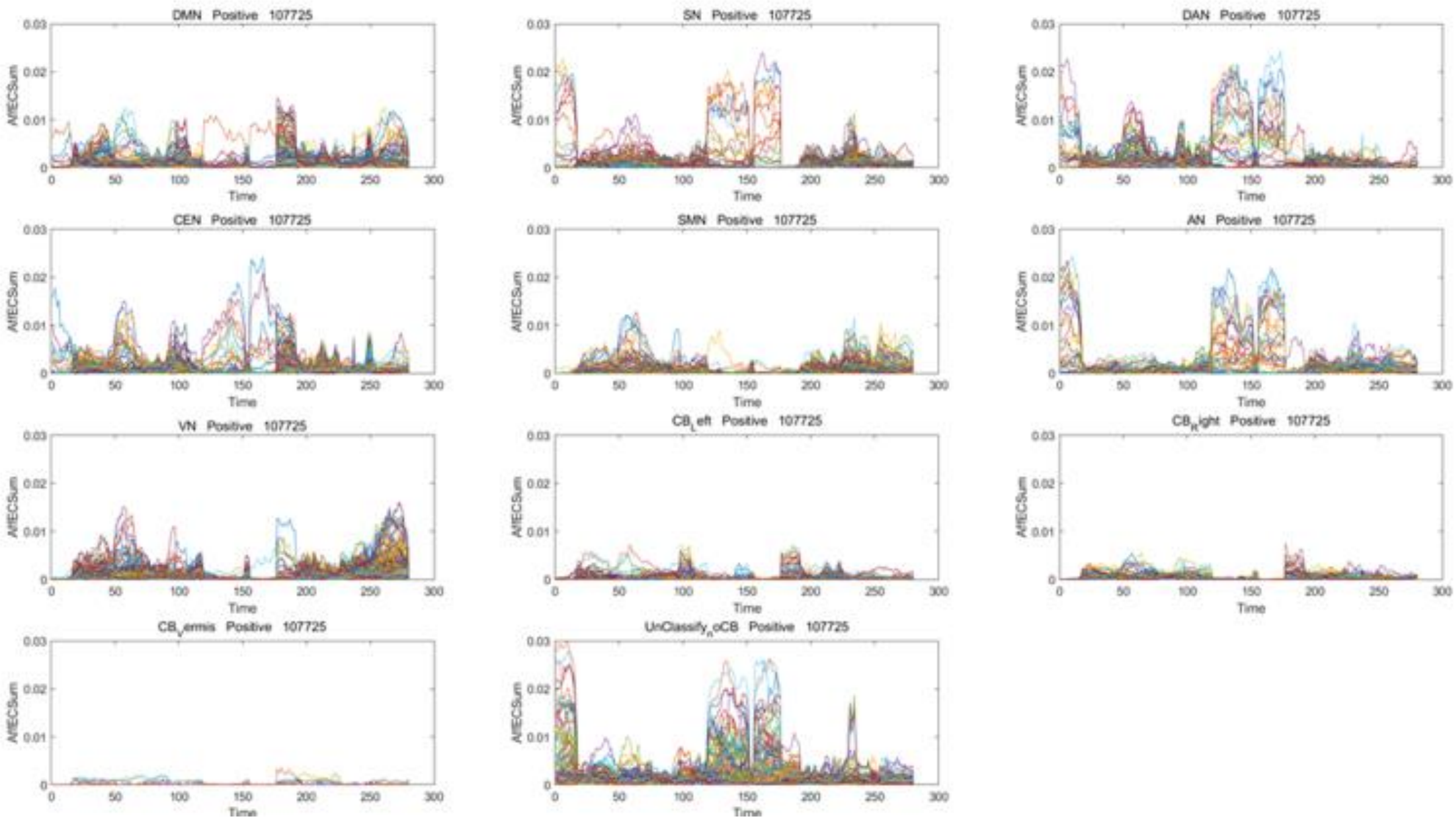

E.

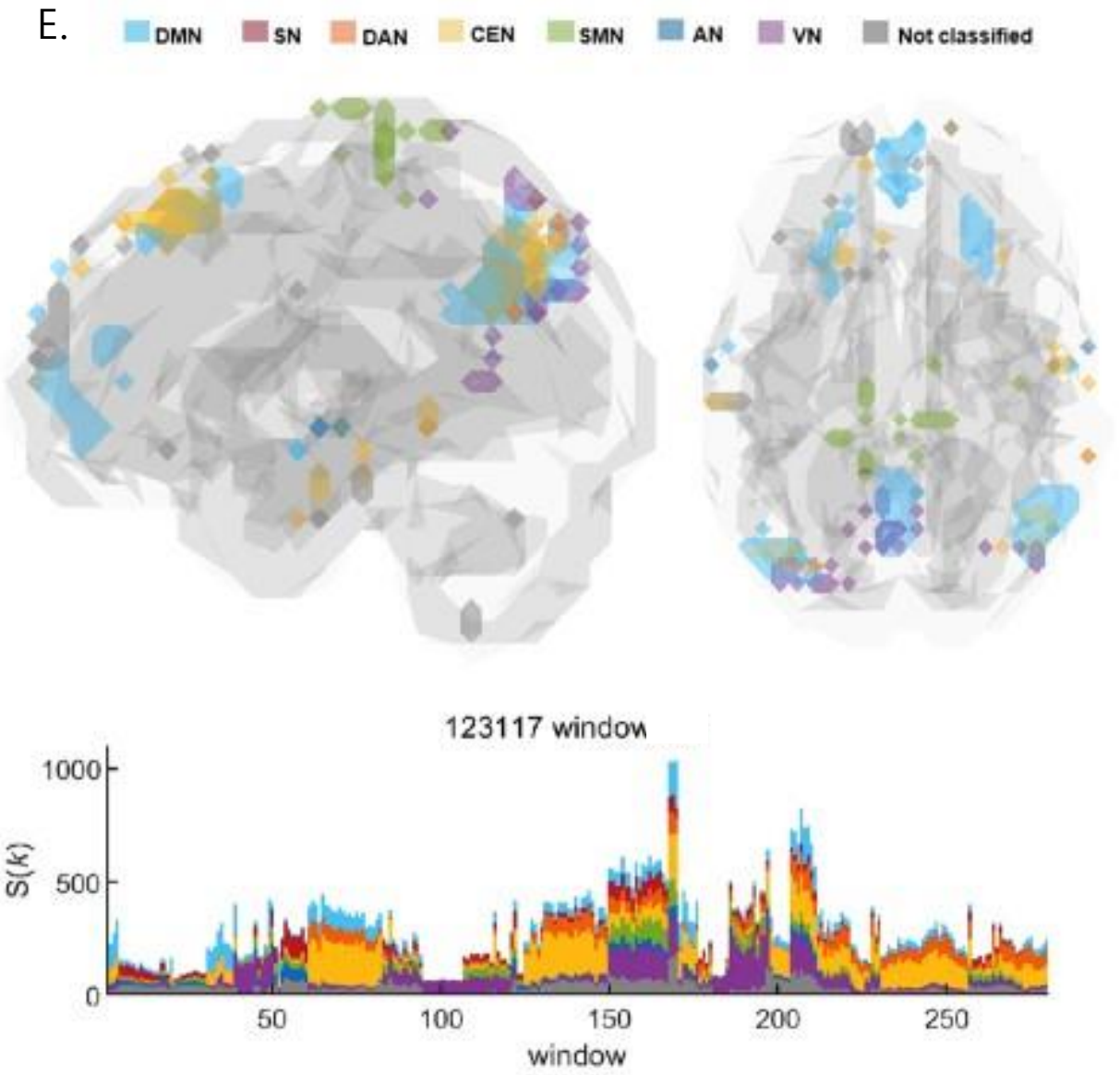

F.

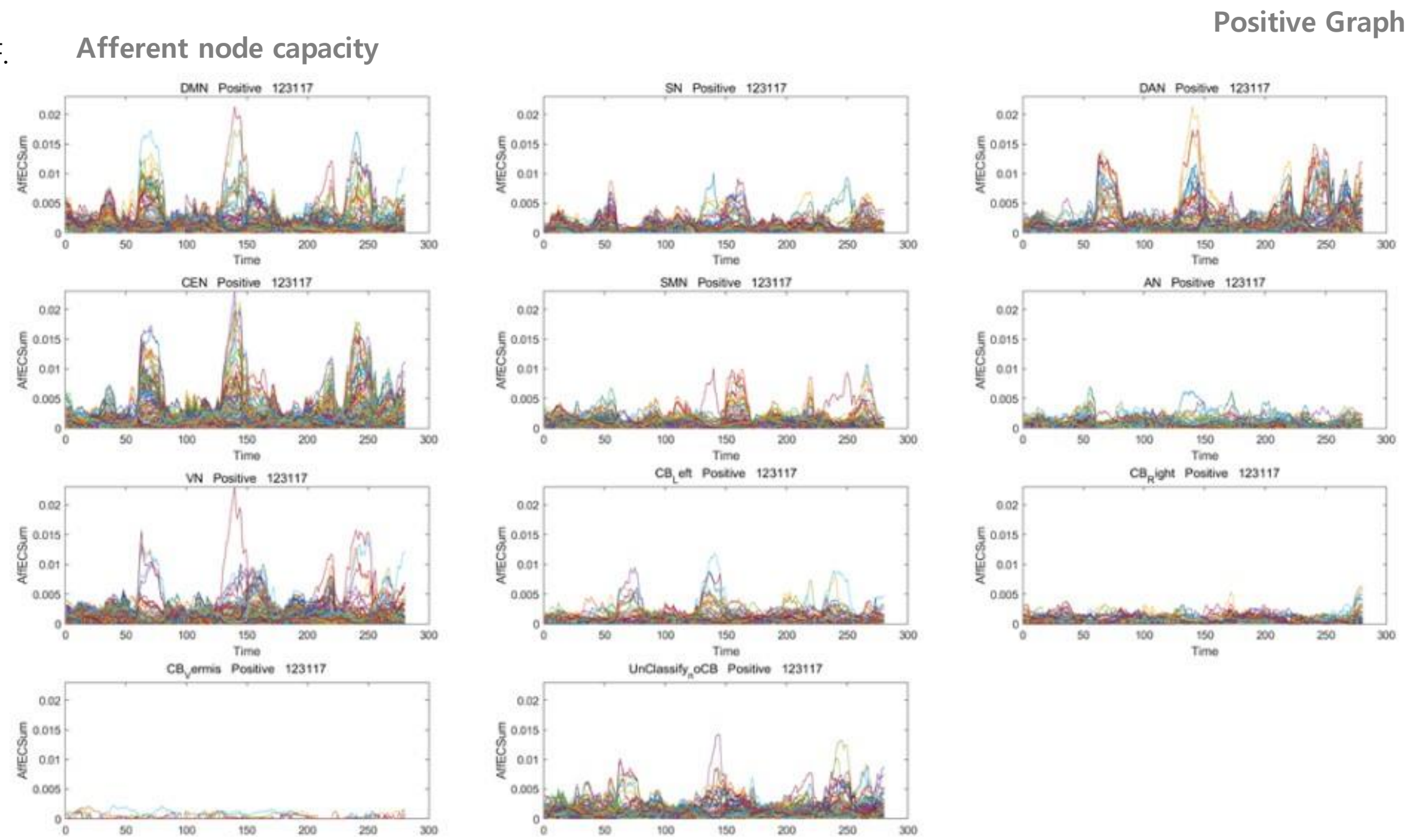

**Supplementary Figure 11. Asymmetry of module composition of states in three subjects showing frontal alternating (A,B), recurrently appearing in left frontal area (C,D), and left cerebellar asymmetry (E,F) patterns.**

A. This individual showed usual pattern of state transition on stacked histogram plots, but on the glass brain images, repeated alternation of right and left DAN/CEN was found.

B. Afferent node capacity showed DMN/DAN/CEN modules and independent SMN, SN, AN and the unclassified, which made rainbow-type distributed modules on $k_{max}$core plots at the later part of the plots.

C. In this extraordinary individual also presented in Figure 3B, $k_{max}$core plots showed ordinary state transition from DMN/SN/DAN/CEN/AN dominance via DMN/CEN dominance to rainbow distributed dominance etc. On glass brain images, however, left dorsal frontal areas of DAN/CEN/SN/AN took the hierarchical supremacy three times at the start, $50^{th}$ time bin and period between $120^{th}$ and $170^{th}$ bins until it yielded to the following DMN/CEN and then VN with other smalls.

D. On afferent node capacity plots, during the period between $120^{th}$ and $170^{th}$ time bins, SN/DAN/CEN/AN/Unclassified co-modules presided as dominating modules with the void VN/SMN/cerebellums

E. This individual was one of the typical examples of state transition. Peculiar finding was in the asymmetry of cerebellum. DMN/CEN, sole VN, DMN/DAN/CEN, rainbow distributed all-ICs took turns, to-and-fro with clear state transitions.

F. Afferent node-voxel capacity showed two similar co-modules, i.e., one DMN/CEN/VN and another SMN/SN/AN/Unclassified. Interestingly, DAN mostly mimicked SMN/SN/AN except for a period around $150^{th}$ time bin, when it adopted the feature of DMN/CEN/VN at that time bin. Left cerebellum made modules and followed the track DMN/DAN/CEN modules, while right cerebellum did not.

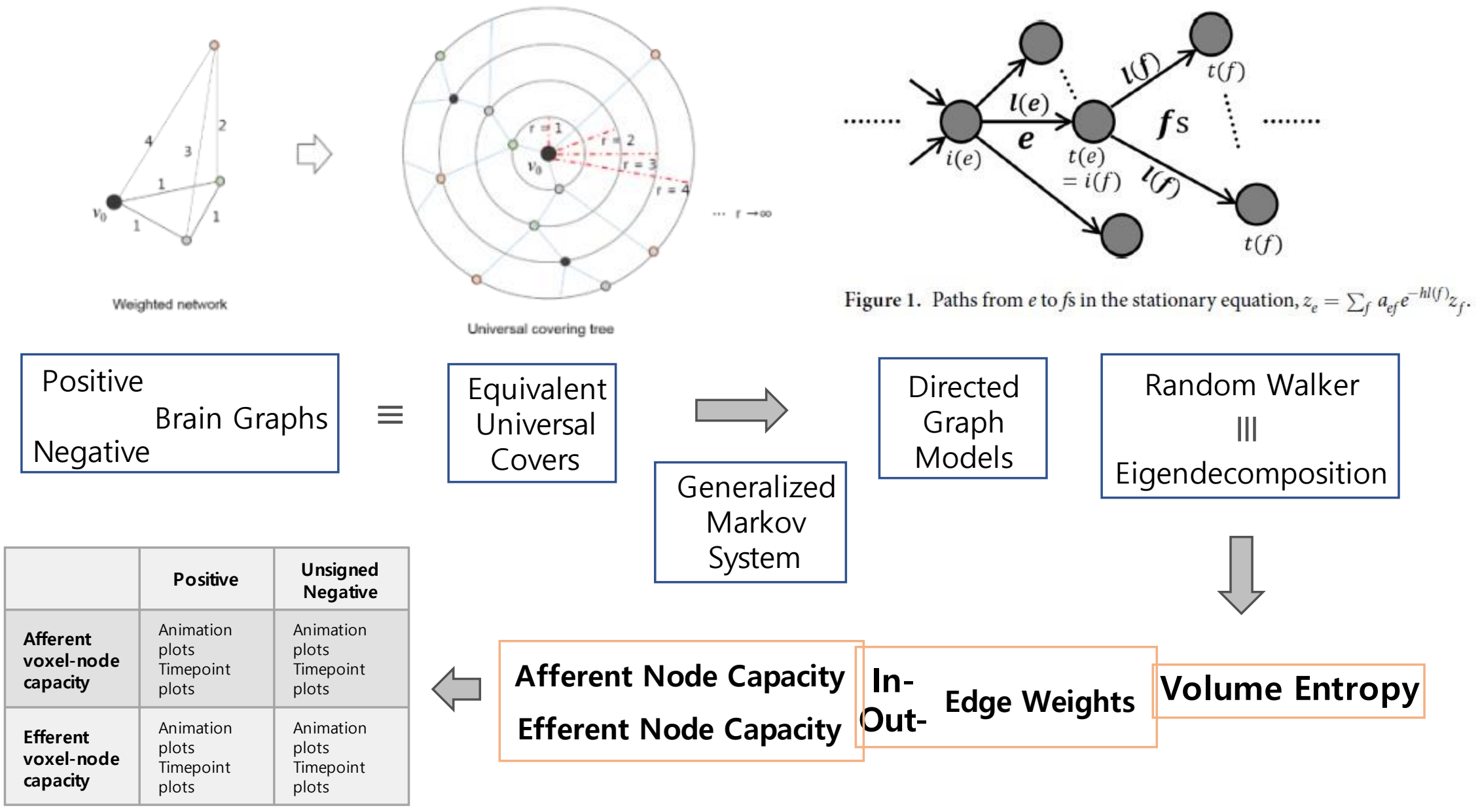

**Supplementary Figure 12. Scheme to estimate volume entropy and afferent/efferent node capacity by making directed weighted graphs from the observed pairwise intervoxel undirected amplitude correlations after thresholding separately on positive and unsigned negative graphs.** The graphs we acquired from resting-state fMRI brain imaging were, after thresholding, put in to the calculation of volume entropy reported by Lee et al. (38) and Lim et al. (40). An undirected graph was transformed to universal cover which is equivalent to the brain graph of interest. By modeling the graph geodesic configuration search with generalized Markov system, the edges of the universal cover were traced by a random walker on the surface (N-1 dimensional) of the N-dimensional ball to the infinite. Th random walk is asymptotically yielding a volume bounded by the maximum and minimum (40). This volume was equivalent to geodesic topological invariant of (N-1) dimensional surface of N-ball, called volume entropy. While volume entropy defines the total information flow of the graph of interest, the edge lengths of the universal cover at the limit to the infinite for the radius of the N-ball, once cropped as a matrix, yielded edge matrix. As we used 1,489 nodes, 1,107,816 undirected edge-matrix before modeling was supposed to become 2,217,121-element asymmetric matrix. Equivalence of random walker on the universal cover in edge-matrix and Eigendecomposition was adopted to calculate edge matrix and volume entropy (38), and Matlab function eigs was used for the exact calculation. Finally, we could produce the edge length (capacity) matrix and thus we summed up all the incoming edge weights to a node, to call it afferent node capacity, and all the outgoing edge weights from that node, to call it efferent node capacity (39). The final products of afferent/efferent node capacity were the normalized ones and thus timepoint plots could be drawn for afferent and efferent node capacities, separately for positive and negative graphs of an individual.

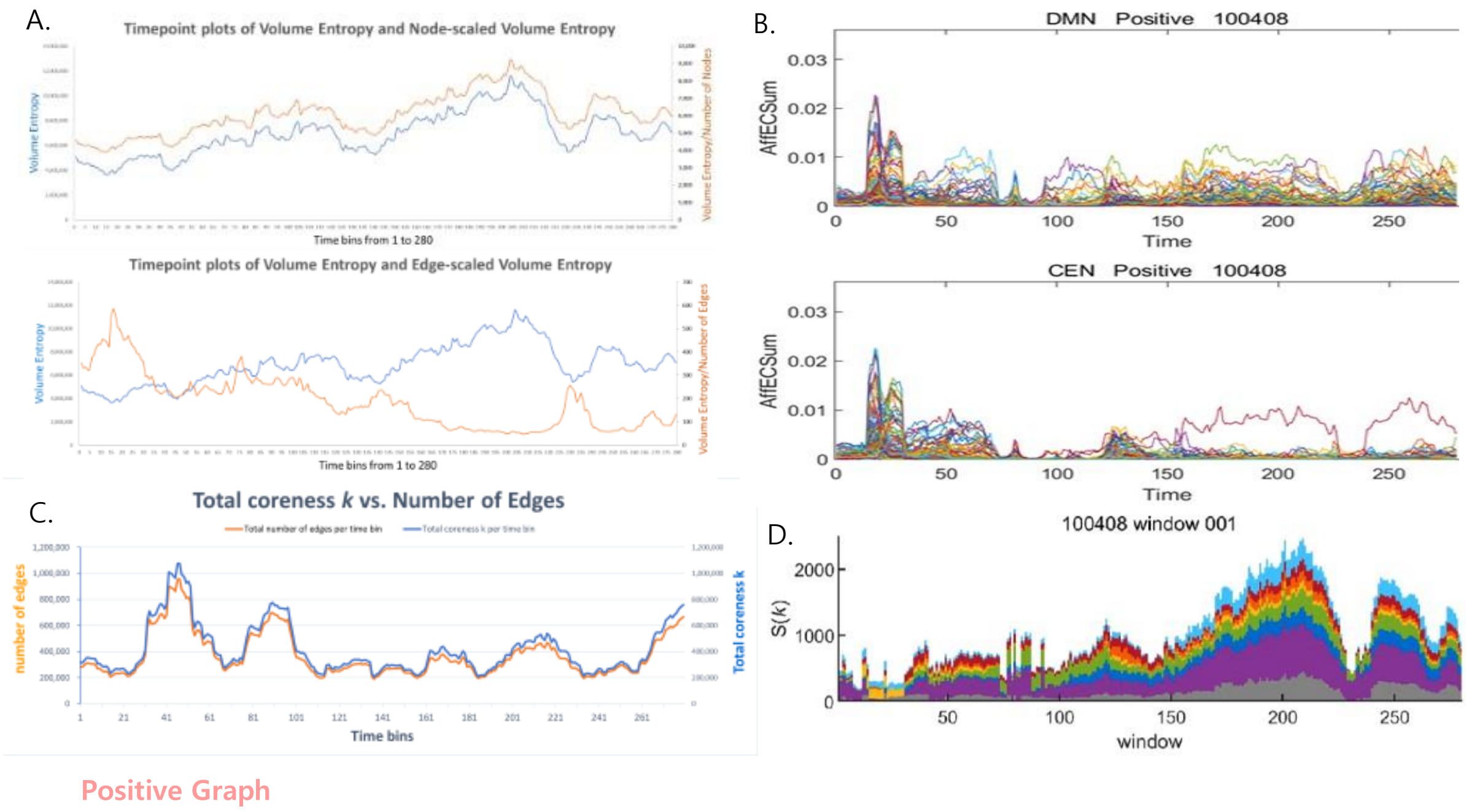

**Supplementary Figure 13. Volume entropy (A) and afferent node capacity (B) with timepoint plots and their corresponding coreness k (C) and $k_{max}$core timepoint plots (D).** All these measures were from positive graphs of an individual (#100408). A. Volume entropy was normalized using (divided by) number of nodes and the curves were exactly the same. In contrast, edge-scaled volume entropy was coarsely inverse of the original curve of volume entropy per se. B. Voxels belonging to DMN and CEN were visualized for afferent node capacity. C. Timepoint plots of total coreness k were plotted together with the total number of edges per time bins, which revealed exactly the same time course. D. Timepoint plots of $k_{max}$core showed initial state transitions. $10^{th}$ to $30^{th}$ time bins showed dominance of voxels belonging to DMN/CEN, which was mimicked by the module formation of and exchange on the afferent node capacity.

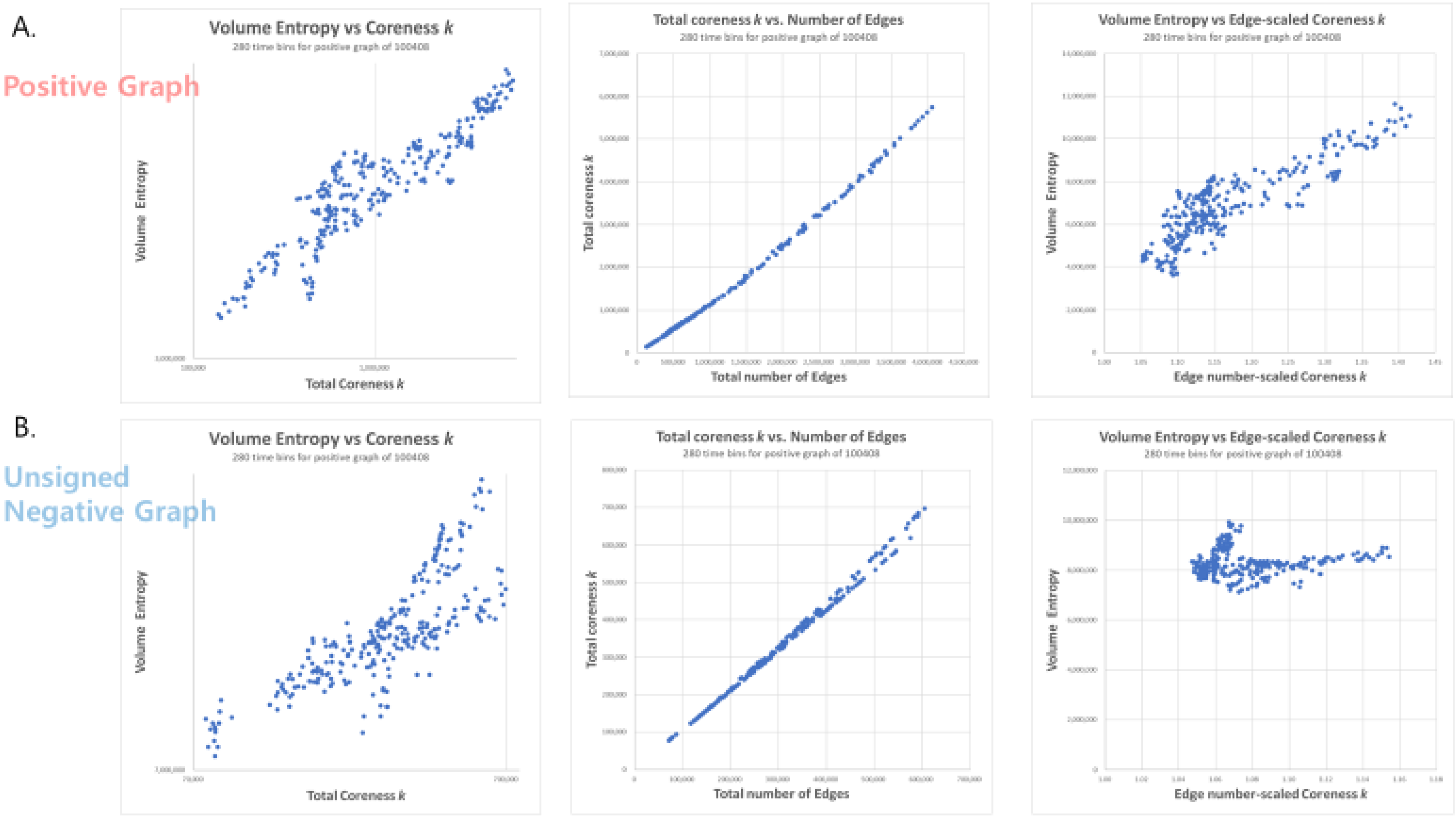

**Supplementary Figure 14. The relationships between the volume entropy and total coreness k of the time bins of the positive and negative graphs of the individuals are presented in Suppl. Fig. 13.**

First column shows the apparent correlation between volume entropy and total coreness k per time bin graphs. It is noted that those two measures are global ones for the graph as a total. On the second column, the expected 1:1 relationship between total edge number and total coreness k, and thus on the third column, relationships between volume entropy and edge-scaled total coreness k were shown. Correlations seemed to exist between volume entropy and total coreness k on the first column plots, however, on the third column especially in negative graphs, the relationship became scattered and were then assumed to have depended on the confounding effect of varying edge numbers.

A.

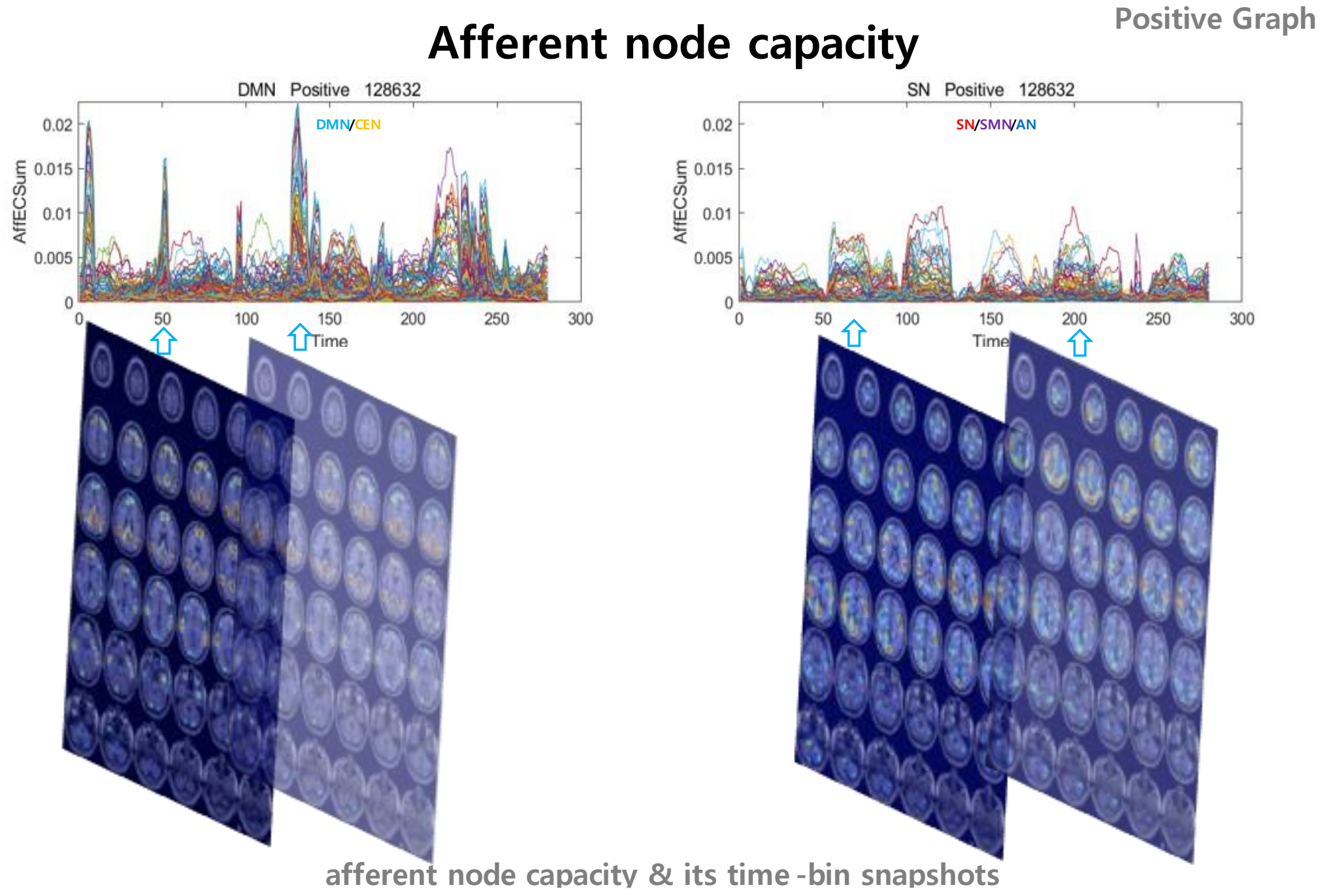

B.

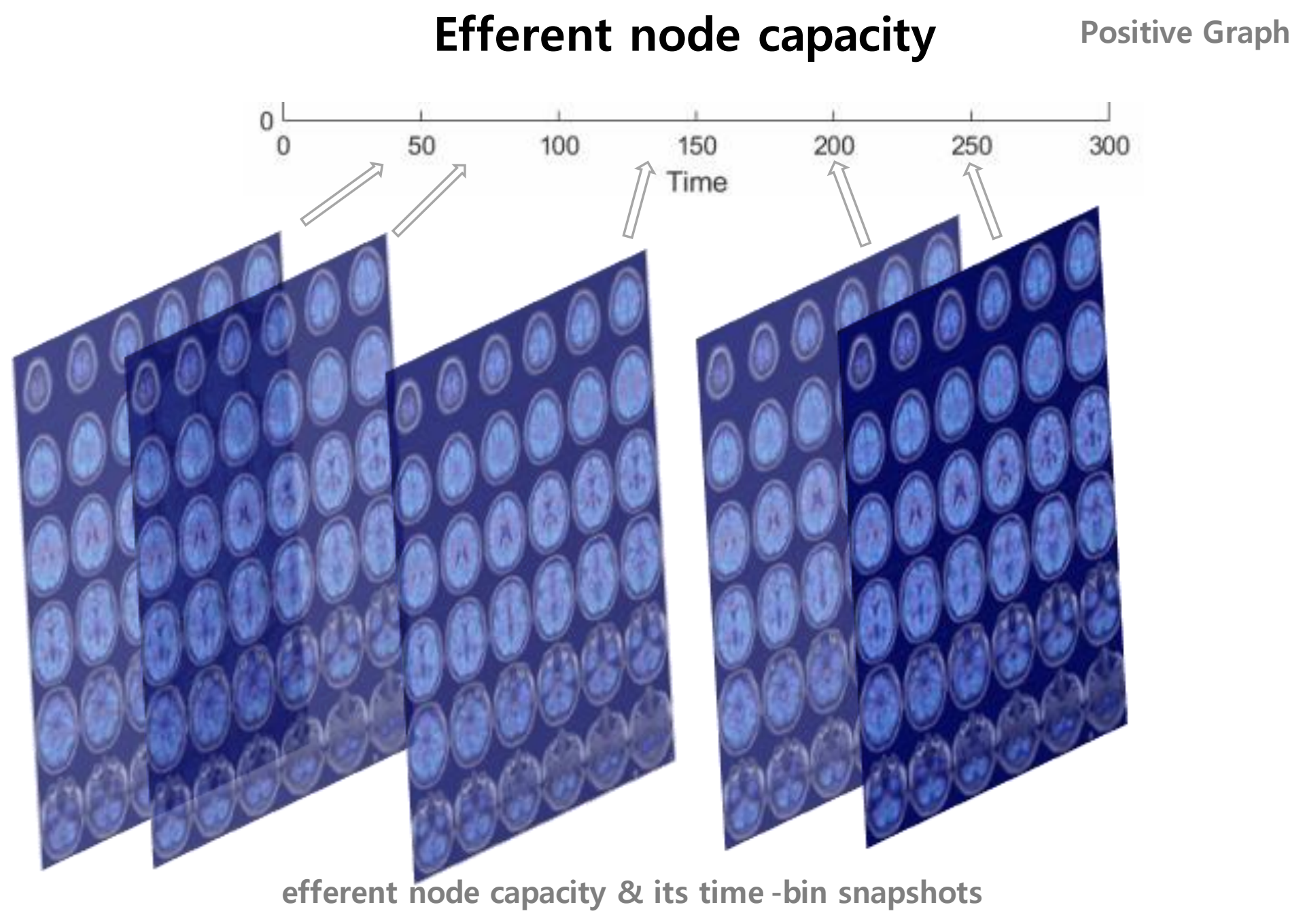

**Supplementary Figure 15. Example of voxels/IC composition timepoint plots and their afferent and efferent node capacity animation maps of positive graphs in an individual (#128632).** The snapshot images of animation maps were presented for immediate visual recognition.

A. Timepoint plots of afferent node capacity of the positive graph of an individual. Snapshots

of animation of afferent node capacity accompanied the timepoint plots at 50th and 130th time bins for voxels of DMN and at 70th and 200th time bins for voxels of SN.

B. Timepoint plots of efferent node capacity of the positive graphs were presented as a line with markers. Snapshots of animation of efferent node capacity were attached at 50th, 70th, 130th, 200th and 250th time bins. It was noted that efferent node capacity was homogeneous with flickering all over the gray matter voxels, and thus timepoint plots were unremarkable even not to make any discernable collective modules.

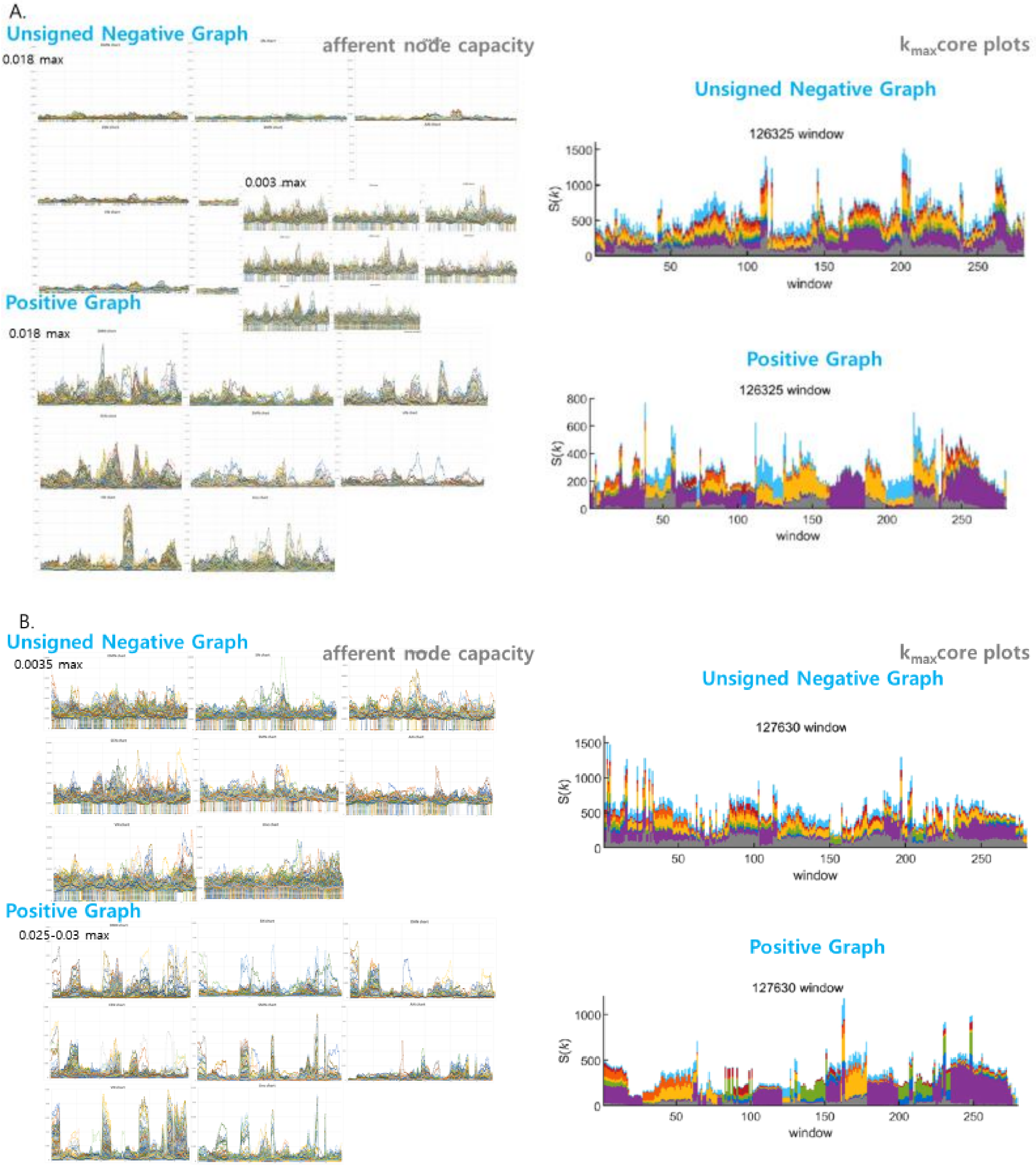

C.

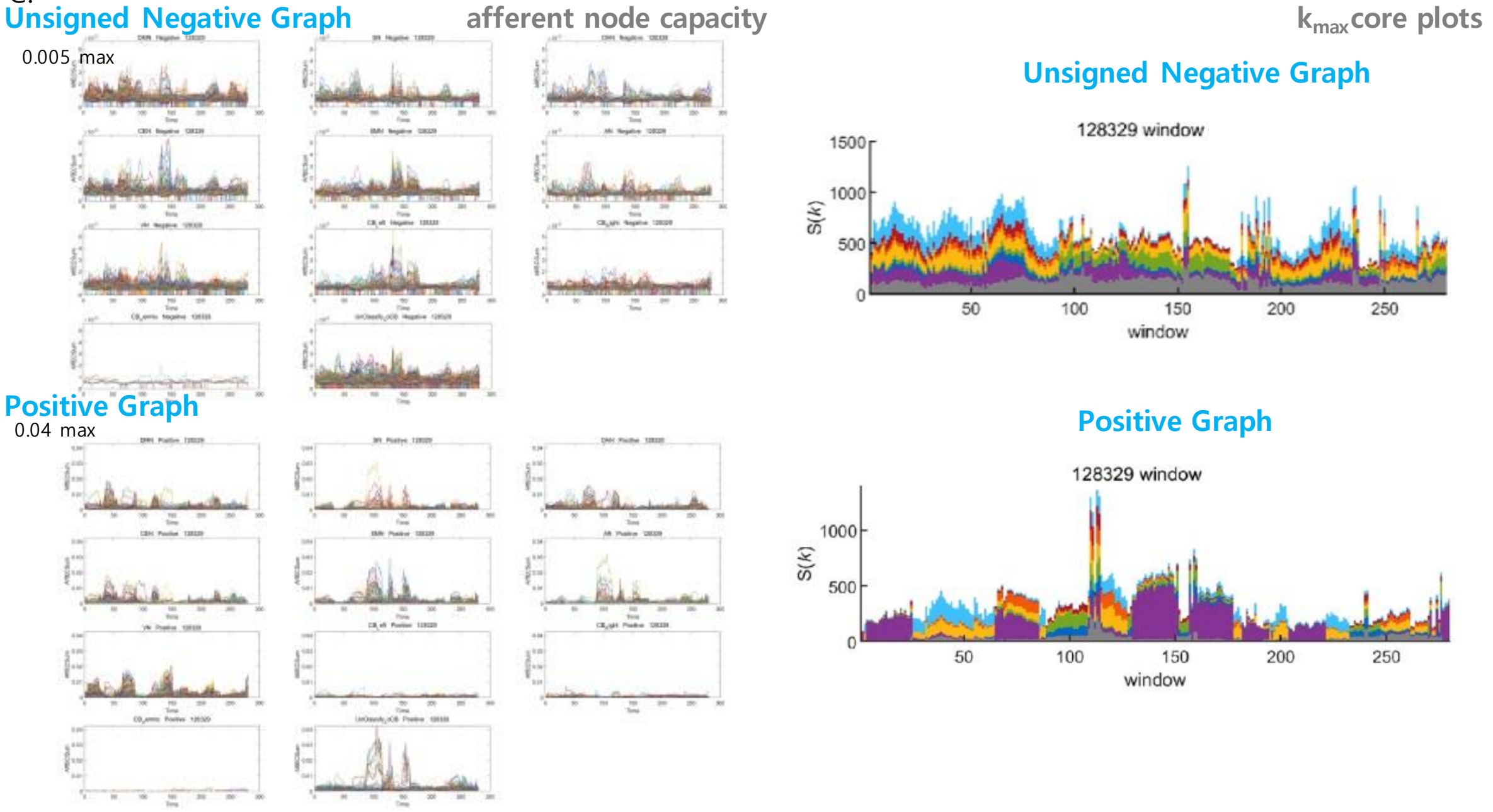

D.

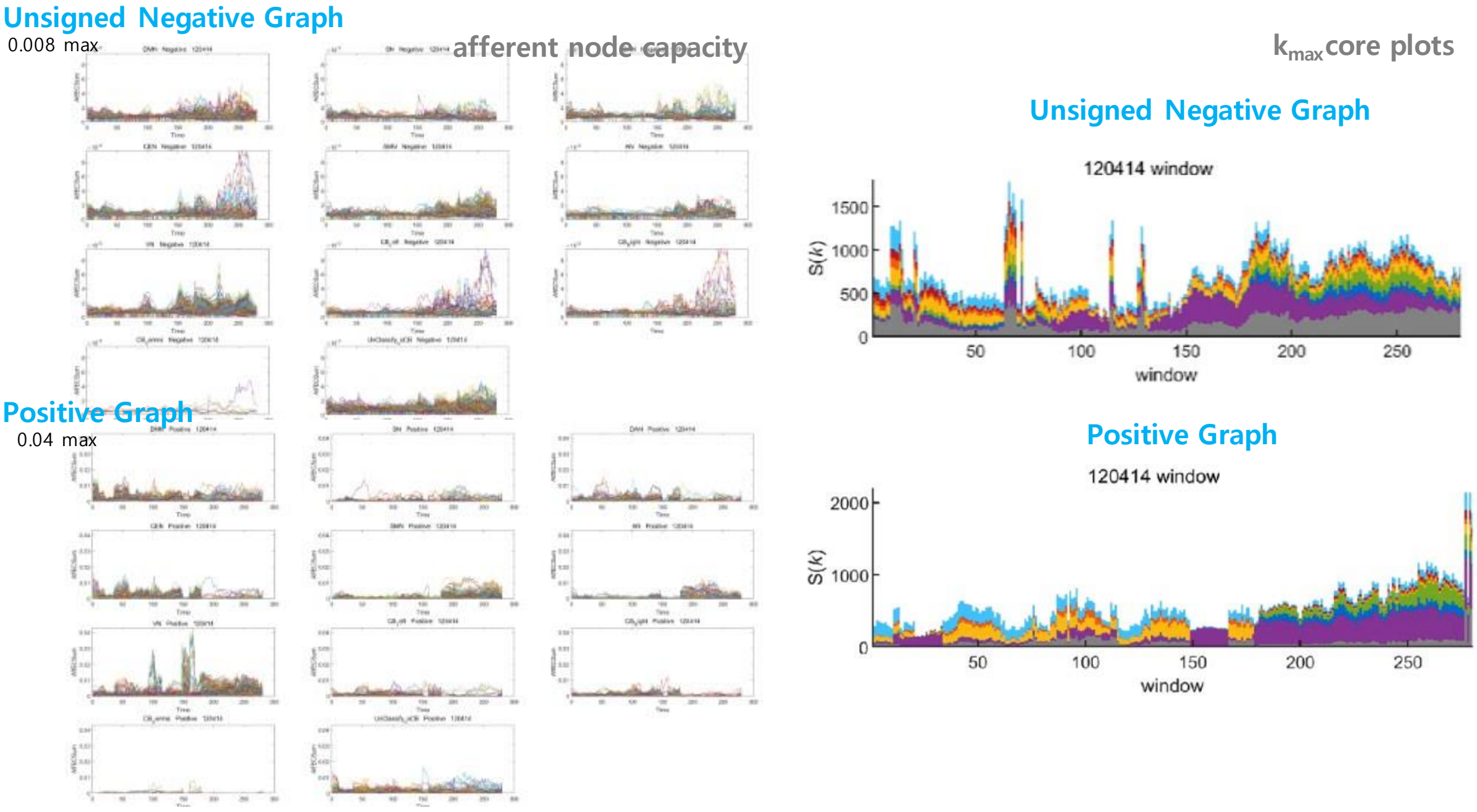

E.

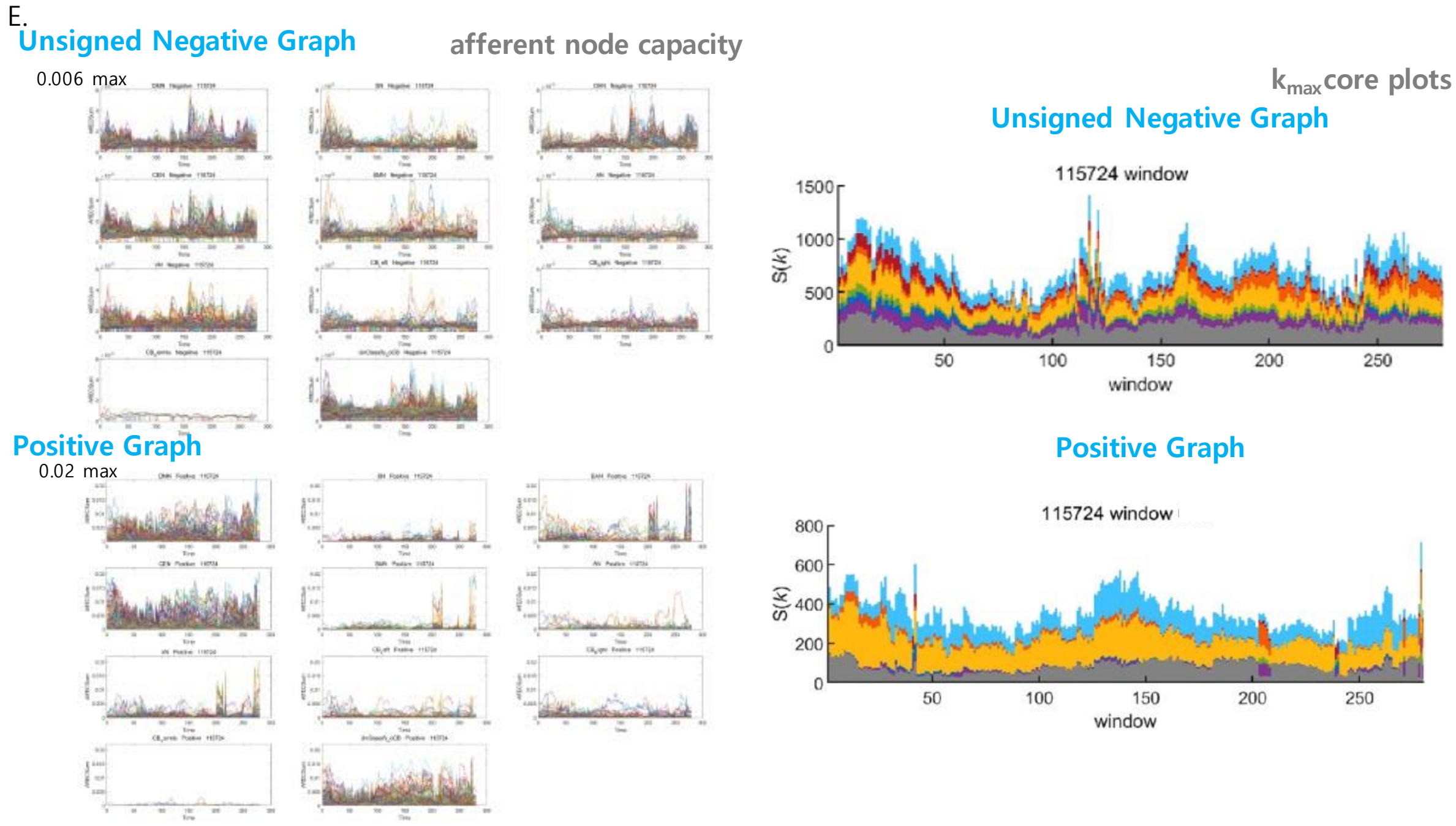

F.

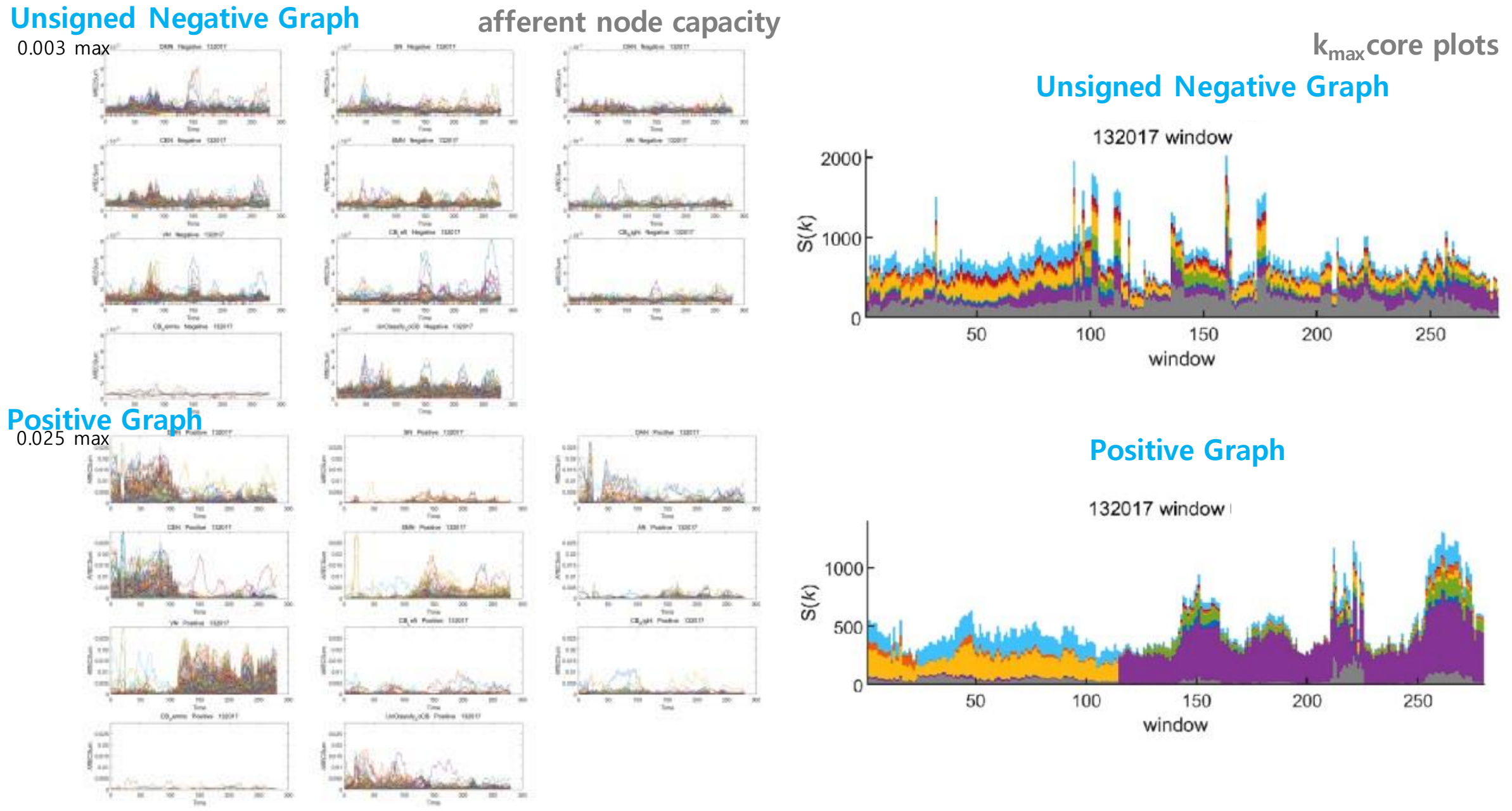

G.

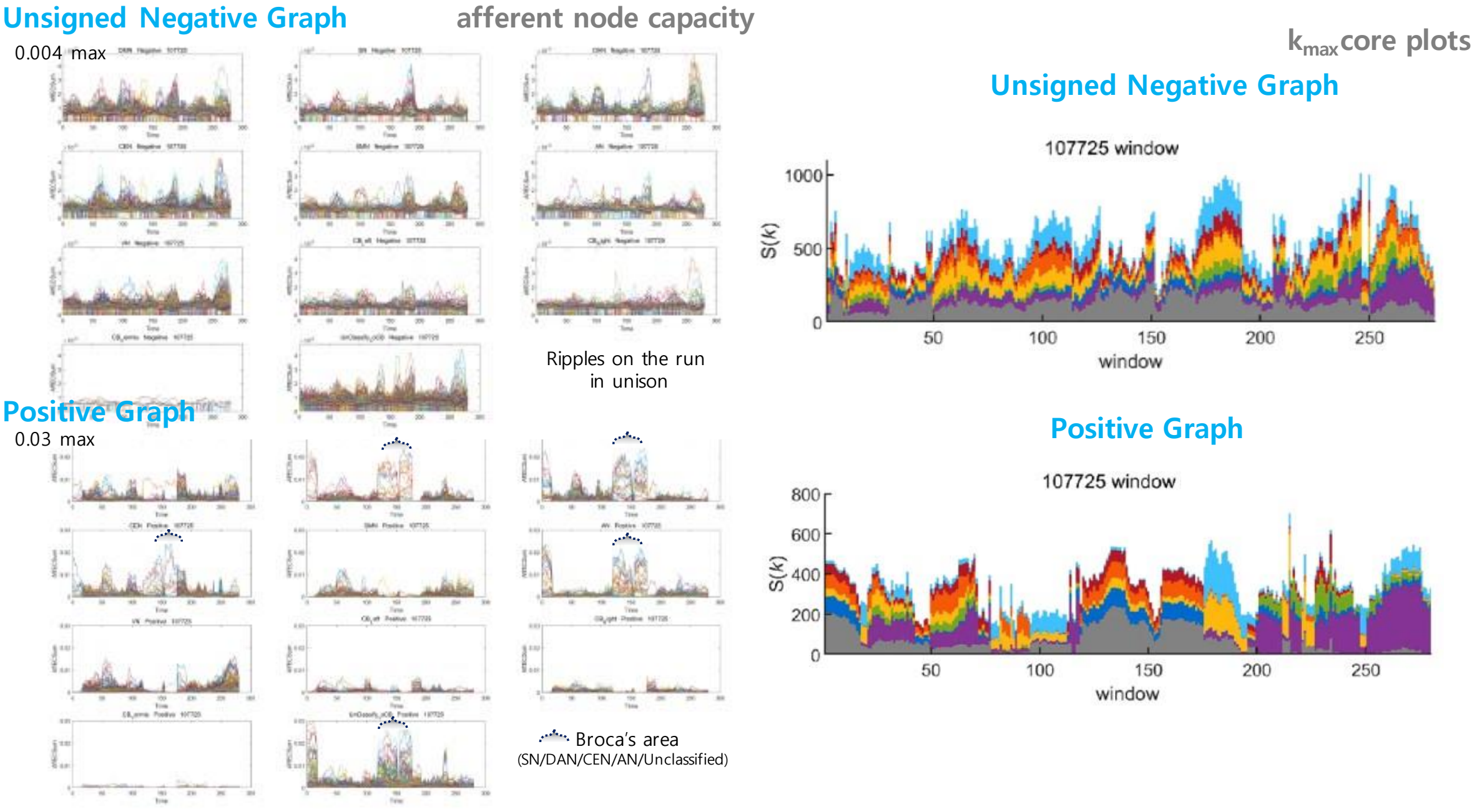

H.

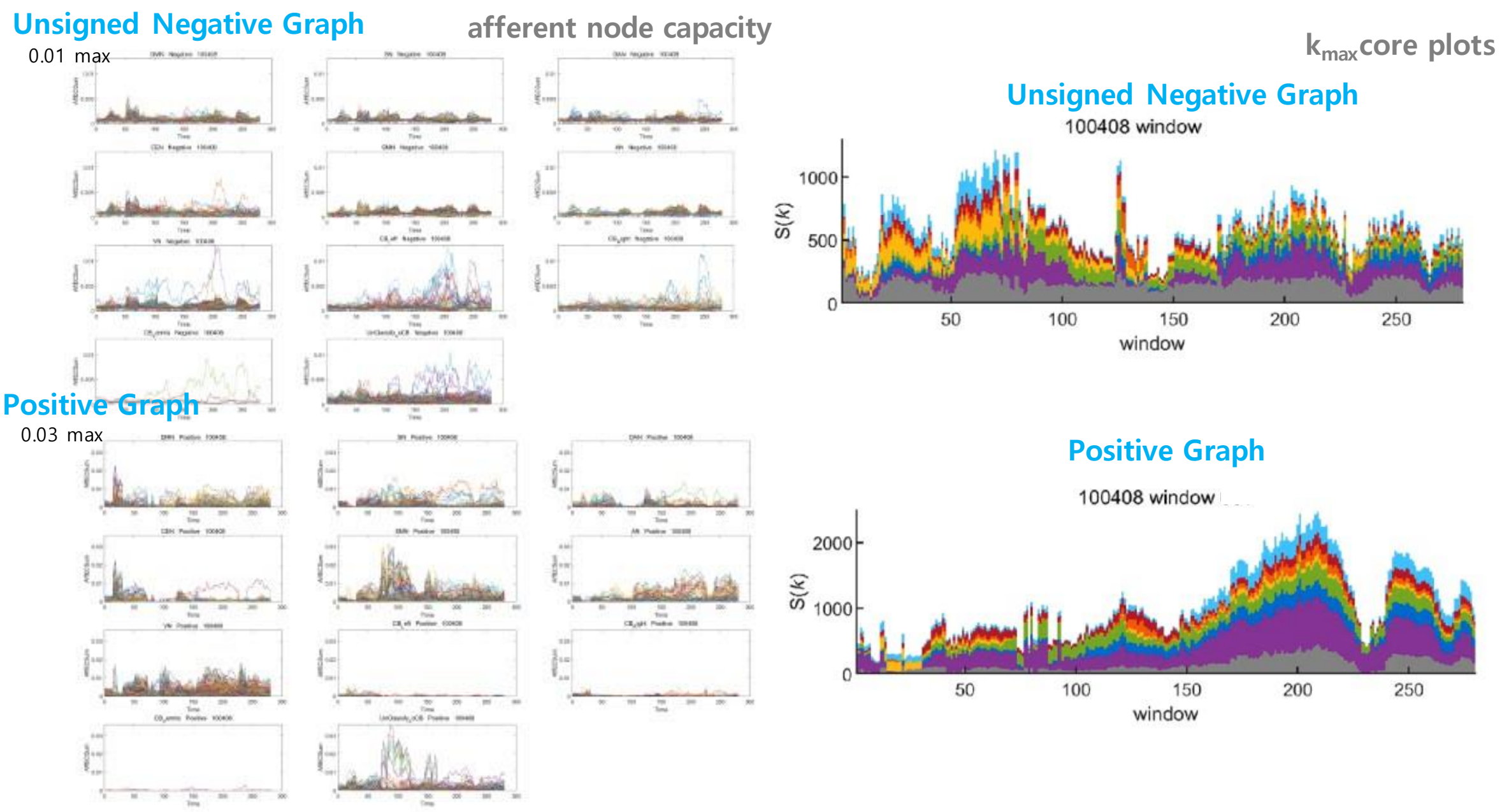

**Supplementary Figure 16. Back-to-back representation of afferent node capacity of voxels/IC timepoint plots of unsigned negative graphs and their corresponding positive graphs in representative individuals, followed by their matching $k_{max}$core stacked histogram timepoint plots.** In the following 8 example showcases, in negative graphs unlike in positive graphs, afferent node capacity plots did rarely show modules nor module

exchange.

A. Afferent edge capacity of negative graphs were much smaller (1/6$^{th}$ ) than those of positive graphs. Unlike afferent node capacity of positive graphs, those of negative graphs did not show evidence of module formation and when the maximum of the ordinate was lowered from 0.018 to 0.003, ripples of collective voxel trajectories rose and fell just in the small and fragmented waves in unison among the ICs. On $k_{max}$core stacked histogram plots, state transition was not prominent in negative graphs, which was prominent in positive graphs.

B. Height of ripples of afferent node capacity of negative graphs was 1/10$^{th}$ of that of positive graphs. Once amplified, the ripples and loose threads were noted in negative graphs. A few state transitions were there on the stacked histogram plots of the negative graphs but not as much explicitly and frequently as those of the positive graphs.

C. Timepoint plots of afferent node capacity showed ripples with much lower values (1/9$^{th}$) in negative graphs than in positive graphs. In contrast, stacked histogram plots of $k_{max}$core showed at least two state transition at 100$^{th}$ time bin and 180$^{th}$ time bin. For these module exchanges of negative graphs, all the ICs constituted the $k_{max}$core at 100$^{th}$ time bin until DMN voxels left and SMN voxels joined and stayed till 180$^{th}$ time bin. Then the voxels/ICs composition returned to the previous one.

D. In this case, 1/5$^{th}$ of height (compared to that of positive graph) of afferent node capacity of negative graph, was dedicated to later surge of trajectory of DMN/CEN/DAN/SMN/AN and cerebellums. Stacked histogram plots of negative graph also showed different pattern between the first and the second halves on $k_{max}$core stacked histogram plots. Afferent node capacity and stacked histogram of positive graphs showed the concordant changes of DMN/CEN and VN/SMN/AN modules.

E. 1/3 height (compared to that of positive graph) of afferent node capacity of negative graph showed ripples in unison, which was also manifest in $k_{max}$core stacked histogram of negative graph. Interestingly, afferent node capacity of positive graph was dominated by the participation of voxels of DMN/CEN/unclassified. This was also corroborated by the same findings of $k_{max}$core stacked histogram plots of positive graph.

F. 1/8$^{th}$ height (compared to that of positive graph) of afferent node capacity of negative graph showed ripples in unison and threads. Stacked histogram plots of $k_{max}$core could be divided to 3 to 5 segments with different voxels/IC composition in negative graph. In contrast, afferent node capacity and $k_{max}$core results were concordant in that the DMN/CEN first and then VN major in the following. Modules of longer duration on afferent node capacity plots were compatible with the top tier voxels/IC on stacked histogram plots in positive graphs.

G. While 1/8$^{th}$ height (compared to that of positive graph) of afferent node capacity showed ripples in unison and matching top tier voxels on the stacked histogram plots of $k_{max}$core in negative graphs, afferent node capacity plots showed typical module formation and exchange in positive graph. Stacked histogram of $k_{max}$core also showed typical state transition in positive graphs. This case was the one in Figure 3B and Suppl. Fig. 11CD. Afferent node capacity matched $k_{max}$core of negative graph as well as in positive graph. Unique in this case was the finding that we needed MRI-overlayed voxel coreness k animation plot to find the Broca area and asymmetry of voxels/IC contribution to modules formation/exchange.

H. 1/3$^{th}$ height (compared to that of positive graph) of afferent node capacity of negative graph showed unimpressive small waves but with much loose threads in VN, left and right cerebellum and the unclassified. Stacked histogram of $k_{max}$core of negative graph also showed progressive changes of IC composition but not with exact state transition. In this case, the stacked histogram of $k_{max}$core of positive graph also showed homogenous progress of totally one state except for the initial transition to DMN/CEN and then back to 'VN and all the others' state. Afferent node capacity plots of positive graph showed very early DMN/CEN module and then the modules were scattered over VN/SMN/SN and AN.

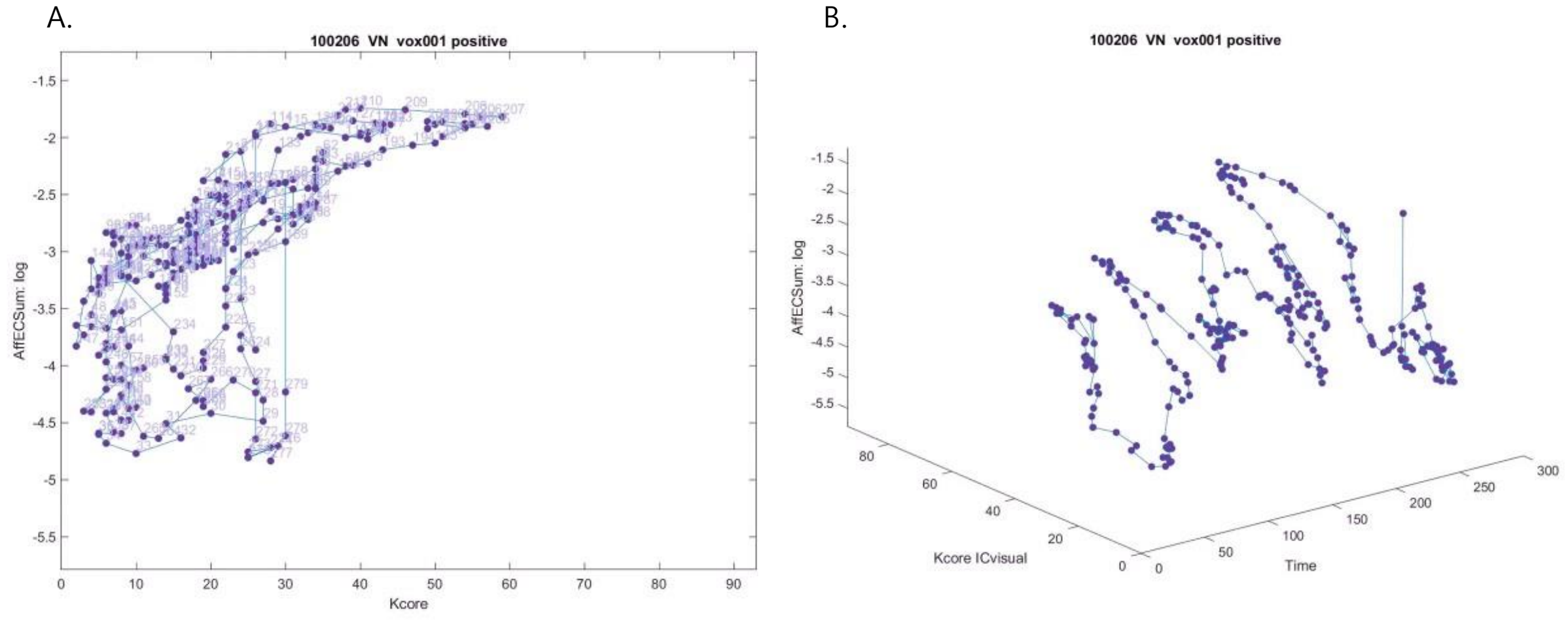

**Supplementary Figure 17. Trajectory tracing of a voxel along the time-bin progress of its own coreness k values and afferent node capacity.** This was the snapshot of animation plots showing all the voxels from an IC (VN, n=247 voxels). It is to be remarked that coreness k value was derived from k core percolation of undirected graph of 5,937 or 1,489 voxels data. For this comparison, we performed another k core percolation for positive graph of #100206 using 10x10x10 $mm^3$ matrix (voxel number of 1,489) with the same threshold of 0.65. 247 voxels were given their coreness k value and afferent node capacity derived from positive graph. 280 time-bin data was plotted in two ways.

A. Logarithmic representation on the ordinate and linear one on the abscissa box was filled with 280 time-bins data of voxel 1 among 247 voxels. In fact, this data was also in animation plot in which we could see the pattern of all the 247 voxels. On the readout of animation plot, frog-like expansion and shrinkage was observed both horizontally and in up-and-down fashion.

B. 280 time-bin data were expanded with the help of Matlab. And animation plot was reviewed. We now could see the collective frog-like motions of each time bin altogether to reach the top tier of hierarchy with the help of afferent node capacity of each voxel at that moment.